\documentclass[longauth]{aa}  

\usepackage{grffile, graphicx}
\usepackage[draft]{hyperref}
\usepackage[T1]{fontenc}
\usepackage{lmodern}
\usepackage{txfonts}
\def\gjb{GJ~504b}
\def\MEarth{M_{\mathrm{Earth}}}
\def\MJ{M_{\mathrm{Jup}}}

\def\Sinit{S_{\mathrm{init}}}
\def\Lsun{L_{\odot}}
\def\Mc{M_{\mathrm{core}}}
\def\kB{k_{\mathrm{B}}}
\def\tKuehl{t_{\mathrm{cool}}}

\begin{document}

   \title{The GJ~504 system revisited}

   \subtitle{Combining interferometric, radial velocity, and high contrast imaging data\thanks{Based on observations collected at the European Organisation for Astronomical Research in the Southern Hemisphere under ESO programs 093.C-0500, 095.C-0298, 096.C-0241, and 198.C-0209, and on interferometric observations obtained with the VEGA instrument on the 
CHARA Array.}}

   \author{M. Bonnefoy\inst{1},  K. Perraut\inst{1}, A.-M. Lagrange\inst{1}, P. Delorme\inst{1}, A. Vigan\inst{2}, M. Line\inst{3}, L. Rodet\inst{1}, C. Ginski\inst{4}, D. Mourard\inst{5}, G.-D. Marleau\inst{6}, M. Samland\inst{7}, P. Tremblin\inst{8}, R. Ligi\inst{9}, F. Cantalloube\inst{7}, P. Molli\`{e}re\inst{4},  B. Charnay\inst{10}, M. Kuzuhara\inst{11, 12}, M. Janson\inst{13}, C. Morley\inst{14}, D. Homeier\inst{15}, V. D'Orazi\inst{16}, H .Klahr\inst{7}, C. Mordasini\inst{6}, B. Lavie\inst{17, 6}, J.-L. Baudino\inst{9,18}, H. Beust\inst{1}, S. Peretti\inst{17}, A. Musso Bartucci\inst{7}, D. Mesa\inst{16, 19}, B. B\'{e}zard\inst{10}, A. Boccaletti\inst{10}, R. Galicher\inst{10}, J. Hagelberg\inst{17, 20}, S. Desidera\inst{16}, B. Biller\inst{7, 21}, A.-L. Maire\inst{7}, F. Allard\inst{15}, S. Borgniet\inst{10}, J. Lannier\inst{1}, N. Meunier\inst{1}, M. Desort\inst{1}, E. Alecian\inst{1}, G. Chauvin\inst{1, 22}, M. Langlois\inst{15}, T. Henning\inst{7}, L. Mugnier\inst{23}, D. Mouillet\inst{1}, R. Gratton\inst{16}, T. Brandt\inst{24}, M. Mc Elwain\inst{25}, J.-L. Beuzit\inst{1}, M. Tamura\inst{11,12, 26}, Y. Hori\inst{11,12}, W. Brandner\inst{7}, E. Buenzli\inst{7}, A Cheetham\inst{17}, M. Cudel\inst{1}, M. Feldt\inst{7}, M. Kasper\inst{1, 27}, M. Keppler\inst{7}, T. Kopytova\inst{1, 3}, M. Meyer\inst{28, 29}, C. Perrot\inst{10}, D. Rouan\inst{10}, G Salter\inst{2}, T. Schmidt\inst{10}, E. Sissa\inst{16}, A. Zurlo\inst{2, 30, 31}, F. Wildi\inst{17}, P. Blanchard\inst{2}, V. De Caprio\inst{16}, A. Delboulb\'{e}\inst{1}, D. Maurel\inst{1}, T. Moulin\inst{1}, A. Pavlov\inst{7}, P.  Rabou\inst{1}, J. Ramos\inst{7}, R. Roelfsema\inst{32}, G. Rousset\inst{10}, E. Stadler\inst{1}, F. Rigal\inst{32}, L. Weber\inst{17}}

   \institute{\inst{1} Univ. Grenoble Alpes, CNRS, IPAG, 38000 Grenoble, France \\
\inst{2} Aix Marseille Univ, CNRS, CNES, LAM, Marseille, France \\              
\inst{3} School of Earth \& Space Exploration, Arizona State University, Tempe AZ 85287, USA \\              
\inst{4} Leiden Observatory, Leiden University, PO Box 9513, NL-2300 RA Leiden, Netherlands \\
\inst{5} Laboratoire Lagrange, UMR 7293 UNS-CNRS-OCA, Boulevard de l’Observatoire, BP 4229, 06304 Nice Cedex 4, France \\
\inst{6} Physikalisches Institut, Universit\"{a}t Bern, Gesellschaftsstrasse 6, 3012 Bern, Switzerland \\
\inst{7} Max Planck Institute for Astronomy, K\"{o}nigstuhl 17, D-69117 Heidelberg, Germany \\ 
\inst{8} Maison de la Simulation, CEA, CNRS, Univ. Paris-Sud, UVSQ, Universite\'{e} Paris-Saclay, F-91191 Gif-sur-Yvette, France \\
\inst{9} INAF-Osservatorio Astronomico di Brera, Via E. Bianchi 46, I-23807 Merate, Italy\\
\inst{10} LESIA, Observatoire de Paris, PSL Research University, CNRS, Sorbonne Universit\'{e}s, UPMC Univ. Paris 06, Univ. Paris Diderot, Sorbonne Paris Cit\'{e}, France\\
\inst{11} Astrobiology Center of NINS, 2-21-1, Osawa, Mitaka, Tokyo, 181-8588, Japan\\
\inst{12} National Astronomical Observatory of Japan, 2-21-1, Os- awa, Mitaka, Tokyo, 181-8588, Japan\\
\inst{13} Department of Astronomy, Stockholm University, AlbaNova University Center, 106 91 Stockholm, Sweden\\
\inst{14} Harvard University, Cambridge, MA, USA\\
\inst{15} CRAL, UMR 5574, CNRS, Universit\'{e} de Lyon, Ecole Normale Sup\'{e}rieure de Lyon, 46 All\'{e}e d'Italie, F-69364 Lyon Cedex 07, France\\
\inst{16} INAF - Osservatorio Astronomico di Padova, Vicolo dell Osservatorio 5, 35122, Padova, Italy \\
\inst{17} Geneva Observatory, University of Geneva, Chemin des Mailettes 51, 1290 Versoix, Switzerland \\
\inst{18} Department of Physics, University of Oxford, Oxford, UK\\
\inst{19} Instituto de Astronomía y Ciencias Planetarias de Atacama, Copayapu 485, Copiapó, Atacama, Chile\\
\inst{20} Institute for Astronomy, University of Hawaii, 2680 Woodlawn Drive, Honolulu, HI 96822, USA\\
\inst{21} SUPA, Institute for Astronomy, The University of Edinburgh, Royal Observatory, Blackford Hill, Edinburgh, EH9 3HJ, UK\\
\inst{22} Unidad Mixta Internacional Franco-Chilena de Astronom\'{i}a, CNRS/INSU UMI 3386 and Departamento de Astronom\'{i}a, Universidad de Chile, Casilla 36-D, Santiago, Chile\\
\inst{23} Office National d'Etudes et de Recherches A\'{e}rospatiales (ONERA), Optics Department, BP 72, 92322 Cha\^{a}tillon, France\\
\inst{24} Astrophysics Department, Institute for Advanced Study, Princeton, NJ, USA\\
\inst{25} Exoplanets and Stellar Astrophysics Laboratory, Code 667, Goddard Space Flight Center, Greenbelt, MD 20771, USA\\
\inst{26} Department of Astronomy, The University of Tokyo, 7-3-1, Hongo, Bunkyo-ku, Tokyo, 113-0033, Japan\\
\inst{27} European Southern Observatory (ESO), Karl-Schwarzschild-Str. 2, 85748 Garching, Germany\\
\inst{28} Institute for Particle Physics and Astrophysics, ETH Zurich, Wolfgang-Pauli-Strasse 27, 8093 Zurich, Switzerland\\
\inst{29} Department of Astronomy, University of Michigan, 1085 S. University Ave, Ann Arbor, MI 48109-1107, USA \\
\inst{30} N\'ucleo de Astronom\'ia, Facultad de Ingenier\'ia y Ciencias, Universidad Diego Portales, Av. Ejercito 441, Santiago, Chile\\
\inst{31} Escuela de Ingenier\'ia Industrial, Facultad de Ingenier\'ia y Ciencias, Universidad Diego Portales, Av. Ejercito 441, Santiago, Chile\\
\inst{32} NOVA Optical Infrared Instrumentation Group, Oude Hoogeveensedijk 4, 7991 PD Dwingeloo, The Netherlands\\}

   \date{Received March 5, 2018; Accepted June 28, 2018}

 
  \abstract
   {The G-type star GJ504A is known to host a 3 to 35 $\mathrm{M_{Jup}}$ companion whose temperature, mass, and projected separation all contribute to making it a test case for planet formation theories and  atmospheric models of giant planets and light brown dwarfs.}
   {We aim at revisiting the system age, architecture, and companion physical and chemical properties using new complementary interferometric, radial-velocity, and high-contrast imaging data.}
   {We used the CHARA interferometer to measure GJ504A's angular diameter and obtained an estimation of its radius in combination with the Hipparcos parallax. The radius was compared to evolutionary tracks to infer a new independent age range for the system. We collected dual imaging data with IRDIS on VLT/SPHERE to sample the NIR (1.02-2.25$\mu$m) spectral energy distribution (SED) of the companion. The SED was compared to five independent grids of atmospheric models (\texttt{petitCODE}, \texttt{Exo-REM}, \texttt{BT-SETTL}, Morley et al., and \texttt{ATMO}) to infer the atmospheric parameters of GJ 504b and evaluate model-to-model systematic errors. In addition, we used a specific model grid exploring the effect of different C/O ratios. Contrast limits from 2011 to 2017 were combined with radial velocity data of the host star through the \texttt{MESS2} tool to define upper limits on the mass of additional companions in the system from 0.01 to 100 au. We used an MCMC fitting tool to constrain the companion's orbital parameters based on the measured astrometry, and dedicated formation models to investigate  its  origin.}
   {We report a radius of $\mathrm{1.35 \pm 0.04\:R_{\odot}}$ for GJ504A. The radius yields isochronal ages of $21\pm 2$ Myr or $4.0\pm1.8$ Gyr for the system and line-of-sight stellar rotation axis inclination of $162.4_{-4.3}^{+3.8}$ degrees or $18.6_{-3.8}^{+4.3}$ degrees. We re-detect the companion in the Y2, Y3, J3, H2, and K1 dual-band images.  The complete 1-4 $\mu$m SED shape of GJ504b is best reproduced by T8-T9.5 objects with intermediate ages ($\leq1.5$Gyr), and/or unusual dusty atmospheres and/or super-solar metallicities. All atmospheric models yield $\mathrm{T_{eff}=550 \pm 50}$K for GJ504b and point toward a low surface gravity (3.5-4.0 dex).  The accuracy on the metallicity value is limited by model-to-model systematics; it is not degenerate with the C/O ratio. We derive $\mathrm{log\:L/L_{\odot}=-6.15\pm0.15}$ dex for the companion from the empirical analysis and spectral synthesis. The luminosity and $\mathrm{T_{eff}}$ yield  masses of $\mathrm{M=1.3^{+0.6}_{-0.3}M_{Jup}}$ and $\mathrm{M=23^{+10}_{-9} M_{Jup}}$ for the young and old age ranges, respectively.  The semi-major axis (sma) is above 27.8 au and the eccentricity is lower than 0.55. The posterior on GJ~504b's orbital inclination suggests a misalignment with the  rotation axis of GJ~504A. We exclude additional objects (90\% prob.) more massive than 2.5 and 30 $\mathrm{M_{Jup}}$ with semi-major axes in the range 0.01-80 au for the young and old isochronal ages, respectively.}
  {The mass and semi-major axis of GJ~504b are marginally compatible with a formation by disk-instability if the system is 4 Gyr old. The companion is in the envelope of the population of planets synthesized with our core-accretion model. Additional deep imaging and spectroscopic data with SPHERE and JWST should help to confirm the possible spin-orbit misalignment and refine the estimates on the companion temperature, luminosity, and atmospheric composition.}


\titlerunning{The GJ504 system revisited}
\authorrunning{Bonnefoy et al.}
\maketitle
   
%

\section{Introduction}
	
	The most recent formation and dynamical evolution models of the solar system \citep[e.g.,][]{2011Natur.475..206W, 2017Icar..297..134R} propose that the wide-orbit giant planets  (Jupiter, Saturn) have largely influenced the composition and/or the architecture of the inner solar system.  Those models are  guided by the population of exoplanets established below $\sim$8 au mainly through transit and radial velocity surveys \citep[e.g., ][]{2007ARA&A..45..397U, 2008PhST..130a4001M, 2009ApJ...693.1084W, 2016ApJS..224...12C, 2016ApJS..226....7C, 2016ApJ...822...86M}.  Several pieces of evidence support the universality of the core-accretion \citep[CA;][]{1996Icar..124...62P, 2004A&A...417L..25A}  formation scenario in this separation range \citep[e.g.,][]{2009A&A...501.1161M, 2010ApJ...709..396B}. Some systems \citep[planets with large sky-projected obliquities; packed systems; see][]{2005ApJ...631.1215W, 2012Sci...337..556C, 2017arXiv171206638B} highlight the dramatic role played by dynamical interactions such as disk-induced migration \citep[for a review, see ][]{2014prpl.conf..667B}, and planet-planet scattering \citep{2008ApJ...678..498N,2008ApJ...686..621F} in stabilizing or (re)-shaping the system architectures in the first astronomical units.  

	Our knowledge of the formation and dynamical evolution of planetary systems at large separation  ($>$8 au) is limited. It relies for the most part on the direct imaging (DI) method whose sensitivity to low-mass companions increases on nearby (d$<150$pc) young systems (age$<$150 Myr). At these ages,  planets can still be hot and self luminous from their formation \citep[depending on the accretion phase, e.g. the so called "hot" and "cold" start conditions;][]{2007ApJ...655..541M, 2017A&A...608A..72M} and be detected at favorable contrasts in the NIR (1-5 $\mu$m).  The implementation of  differential methods \citep{1999PASP..111..587R, 2000PASP..112...91M, 2006ApJ...641..556M} on eight-meter ground-based  telescopes equipped with adaptive optics in the late 2000s led to the breakthrough detections of massive (5-13 $\mathrm{M_{Jup}}$) Jovian planets at short physical separations (9-68 au) around the young ($\sim17-30$ Myr) intermediate-mass (AF) stars HR~8799 \citep{2008Sci...322.1348M, 2010Natur.468.1080M},  $\beta$~Pictoris  \citep{2009A&A...493L..21L, 2010Sci...329...57L},  and HD~95086 \citep{2013ApJ...772L..15R, 2013ApJ...779L..26R}.  Systems  such as HR8799 challenge the CA paradigm whose timescales are too long at large orbital radii compared to the circumstellar disk lifetimes \citep{2001ApJ...553L.153H}. The gravitational instability scenario  \citep[hereafter GI; e.g.,][]{1997Sci...276.1836B, 2013MNRAS.432.3168F}  has been proposed as an alternative to solve that issue. But  the GI model outcomes depend on their sophistication \citep[e.g.,][]{2010ApJ...710.1375K, 2018arXiv180103384M} and some fine tuning is possible \citep[e.g.,][]{2017ApJ...848...40B, 2017ApJ...836...53B} .  
	
	The model development can be guided by the discovery of new systems and by the statistics inferred from the DI surveys  \citep[e.g.,][]{2012ApJ...745....4J, 2017A&A...603A...3V}.   	The  second generation of DI instruments SPHERE \citep{2008SPIE.7014E..18B}, GPI \citep{2008SPIE.7015E..18M}, and SCExAO \citep{2015PASP..127..890J} have been designed to detect fainter companions closer to their stars  ($10^{-6}$ contrasts at 500 mas). Ambitious surveys  such as the SpHere INfrared survey for Exoplanets (SHINE) aim at  building a  meaningful statistics (400-600 stars) on the occurrence and properties of the giant planets from 5 au. These instruments have already detected two more planetary systems around the AF-type stars 51 Eri and HIP~65426 \citep{2015Sci...350...64M, 2017A&A...605L...9C}  and four BD companions around F and G-type stars \citep[][Cheetham et al. 2018, submitted]{2016ApJ...829L...4K, 2017A&A...597L...2M, 2017arXiv171205217C}.   
	
	 The  high-precision astrometry of these instruments  brings constraints on the companion orbital parameters and system achitectures in spite of  the slow orbital motions \citep[][]{2016A&A...587A..57Z, 2016A&A...587A..55V, 2016A&A...587A..56M, 2016ApJ...822L..29R,  2016AJ....152...97W, 2018arXiv180105850C, 2017A&A...608A..79D}.  Stringent detection limits can be derived from these observations at multiple epochs and be combined with radial velocity data of the host star  to  provide insightful constraints on the masses of undetected companions \citep{2017A&A...603A..54L, 2018arXiv180105850C} over all possible semi-major axes. 

	SPHERE and GPI  have extracted  high-quality low-resolution (R$\sim$30-300) NIR (1-2.5$\mu$m) spectra of most of the known substellar companions found at  projected separations below 100 au \citep[e.g.,][]{2014A&A...567L...9B, 2015ApJ...805L..10H, 2016ApJ...824..121D, 2016A&A...587A..57Z, 2017A&A...603A..57S, 2017A&A...608A..79D, 2017arXiv171205828M, 2017AJ....153..182C}. In addition, SPHERE uniquely allows for dual-band imaging of the coolest companions in narrow-band filters  sampling the  $\mathrm{H_{2}O}$ and $\mathrm{CH_{4}}$ absorptions appearing in their SEDs  \citep{2010MNRAS.407...71V, 2016A&A...587A..55V}.  
	
	An empirical understanding of the companions' nature can be achieved through the comparison of their spectra and photometry to those of the many ultracool  dwarfs found in the field \citep[e.g.,][]{2013ApJS..205....6M, 2015ApJ...814..118B, 2016ApJ...830..144R} or in young clusters \citep[e.g.,][]{2017ApJ...837...95B,  2018MNRAS.473.2020L}. Most young planet and BD companions studied so far have spectral features characteristic of M- and L-type objects  with hot atmospheres $\mathrm{1000\leq T_{eff} \leq 3000K}$. Some peculiar features appear such as the red spectral slopes and shallow molecular absorption bands that might be caused by the low surface gravity of the objects \citep[e.g.][]{2016A&A...587A..58B, 2017A&A...608A..79D}. 
	
	Only three companions (51 Eri b, GJ 758b, HD 4113C)  with $\mathrm{T_{eff} \leq 800K}$ and noticeable methane absorptions typical of T-type dwarfs have been detected and/or characterized  with the planet imaging instruments so far \citep{2016A&A...587A..55V, 2017A&A...603A..57S, 2017AJ....154...10R, 2017arXiv171205217C}.  The companions 51 Eri b and GJ 758b exhibit peculiar colors  \citep{2016A&A...587A..55V, 2017ApJ...838...64N, 2017A&A...603A..57S, 2017AJ....154...10R} that do not match any known object.  Both the low surface gravity (e.g., 51 Eri b) and non-solar atmospheric abundances might explain these spectrophotometric properties. Chemical enrichments are indeed predicted to happen at formation \citep[e.g.,][]{2016ApJ...831L..19O, 2016ApJ...832...41M, 2017A&A...603A..57S}. The empirical understanding of these objects is limited by the small number of young T-type objects identified to date \citep{2007ApJ...654..570L, 2014ApJ...787....5N, 2015ApJ...808L..20G, 2017ApJ...841L...1G, 2018arXiv180200493G} or found in metal-rich environments \citep{2008A&A...481..661B}.   
		
 Atmospheric models aim at providing a global understanding of the physical, chemical, and dynamical processes at play in planetary and BD atmospheres.  Models face difficulties matching the NIR colors (J-K, J-H) of  objects at the so-called T/Y transition corresponding to a $\mathrm{T_{eff}}$ of around 500K  \citep[e.g.,][]{2011AJ....142..169B}, but promising new ingredients have been introduced to solve this issue. One is the formation of a cloud deck made of  alkali salts and sulfides \citep{2012ApJ...756..172M} whose impact peaks at $\mathrm{T_{eff}= 500-600K}$.  Another group  chose, rather, to introduce a modification of the temperature gradient caused by fingering convection \citep{2015ApJ...804L..17T, 2016ApJ...824....2L}. The effect of the fingering instability on the thermal gradient, however, has recently been questioned \citep{2018ApJ...853L..30L}. The few detected companions at the T/Y transition are precious benchmarks for atmospheric models because of the known ages and distances of the host stars. \\
	
	A  faint companion was resolved in 2011 at  2.5" projected separation (43.5 au) from the nearby \citep[$\mathrm{17.56\pm0.08pc}$; ][]{2007A&A...474..653V} G0-type star GJ~504 \citep{Kuzuhara2013} in the course of the “Strategic Exploration of Exoplanets and Disks with Subaru” (SEEDS) survey \citep{2009AIPC.1158...11T}.  The companion mass was estimated to be $\mathrm{4^{+4.5}_{-1.0} M_{Jup}}$, making it the first jovian exoplanet resolved around a solar-type star. This mass estimate is nonetheless tied to the  host star age of $160^{+350}_{-60}$ Myr inferred from gyrochronology and activity indicators.  Some tension existed between this age  and the one derived from evolutionary tracks \citep{Kuzuhara2013}, but the authors argued that a reliable isochronal age could not be inferred  because it would have relied on $\mathrm{T_{eff}}$ measurements of the star for which inconsistent values exist in the literature \citep[e.g.,][]{2005ApJS..159..141V, 2012A&A...542A..84D}. \cite{Fuhrmann2015} derived their own $\mathrm{T_{eff}}$ estimate from the modeling of a high-resolution optical spectrum of the star. They found an isochronal age of $4.5^{+2.0}_{-1.5}$ Gyr, implying a mass of  $\mathrm{\sim24M_{Jup}}$ for the companion. \cite{2017A&A...598A..19D} made a strictly differential (line-by-line) analysis of GJ 504A spectra to derive new atmospheric parameters and abundances. They  confirmed that the star has a metallicity above solar ($\mathrm{[Fe/H]=0.22\pm0.04}$) and inferred an isochonal age of $2.5^{+1.0}_{-0.7}$Gyr, leaving GJ~504b in the brown-dwarf mass regime. 
	 
	The companion has NIR broad-band  photometry (J, H, $\mathrm{K_{s}}$, L') similar to late T-type objects \citep{Kuzuhara2013}. \cite{2013ApJ...778L...4J} obtained differential imaging data that showed a strong methane absorption at 1.6$\mu$m which confirms the cool atmosphere of GJ~504b. Complementary observations \citep{2016ApJ...817..166S} were obtained with LBT/LMIRCam at wavelengths of 3.71, 3.88, and 4.00 $\mu$m.  \cite{2016ApJ...817..166S}  estimate a $\mathrm{T_{eff}=543\pm11 K}$  consistent with an object close to the T/Y transition.  The analysis also reveals that the companion might be enriched in metals with respect to GJ~504A. They also find a low surface gravity which is more consistent with the age estimated by \cite{Kuzuhara2013}. However, they did not study the effect of possible systematics related to the choice of the atmospheric models used to interpret the companion photometry.  
	
	GJ~504A is bright  \citep[V=5.19;][]{2009yCat..35040681K} and observable from most northern and southern observatories (dec=+09.42$^{\circ}$). Consequently, the system is suitable to observations with an array of techniques. This paper aims at revisiting the system properties based on interferometric measurements,   high-contrast imaging observations, and existing and new radial-velocity (RV) data. We present the observations and the related data processing in Section \ref{sec:obs}. We derive a new age estimate for the system in Section \ref{sec:age}. We analyze the companion photometric properties following an empirical approach (Section \ref{sec:emp_analysis}) and using atmospheric models (Section \ref{Section:atmomodels}).  Section \ref{sec:mass} summarizes the mass estimates of GJ504b that can be inferred from the analysis presented in the previous sections. In section \ref{Section:archi} we exploit the companion astrometry,   the RV measurements, and the interferometric radius of GJ~504A to study the system architecture. We discuss our results in Section \ref{sec:discussion} and summarize  them in Section \ref{sec:conclusion}.


\section{Observations}
\label{sec:obs}
	\subsection{SPHERE high-contrast observations}
	\label{Section:dataSPHERE}
We observed GJ~504  on seven different nights with the SPHERE instrument mounted on the VLT/UT3 (Table \ref{tab:obs}) as part of the guaranteed time observation (GTO) planet search survey SHINE \citep{2017A&A...605L...9C}. All the observations were acquired in pupil-tracking mode with the 185mas-diameter apodized-Lyot coronograph \citep{2011ExA....30...39C, 2011ExA....30...59G}. 

The target was observed on  May 6, 2015, June 3, 2015, March 29, 2016, and February 10, 2017 with the IRDIFS  mode of SPHERE. The mode enables operation of the IRDIS instrument \citep{2008SPIE.7014E..3LD} in dual-band imaging mode \citep[DBI; ][]{2010MNRAS.407...71V} with the H2H3 filters (Table \ref{Tab:flux-density}), and the  integral field spectrograph \citep[IFS; ][]{2008SPIE.7014E..3EC}  in Y-J (0.95-1.35$\mu$m, $R_{\lambda}\sim40$) mode in parallel.  The companion lies inside the circular   field of view (FOV) of  $\sim$5'' radius. It is however outside of the 1.7''$\times$1.7'' IFS FOV.  

We obtained additional observations with the IRDIFS\_EXT mode on June 5, 2015. The mode enables  DBI with the K1K2 filters (Table	 \ref{tab:obs}) and the simultaneous use of the IFS in the Y-H mode (0.95-1.64$\mu$m, $R_{\lambda} = 30$). GJ~504 was then re-observed on June 6 and 7, 2015 with IRDIS and the DBI  Y2Y3 and J2J3 filters (Table \ref{tab:obs}). 

We collected additional calibration frames with the waffle pattern created by the deformable mirror for the May and June 2015 epochs. Those frames were used to ensure an accurate registration of the star position behind the coronagraph. The waffle pattern was maintained during the whole sequence  of 2016 and 2017 IRDIFS observations to allow a registration of the individual frames along the deep-imaging sequence.    We also collected nonsaturated exposures of the star before and after the sequence of coronographic exposures for astrometric and photometric extraction of point sources.

\begin{table*}
\caption{Log of SPHERE observations}
\label{tab:obs}
\begin{center}
\small
\begin{tabular}{lllllllllll}
\hline
Date	&	UT-Time 	&	Instrument  &  Neutral 	&  $\mathrm{DIT \times NDIT \times N_{EXP}}$ & $\Delta$PA & $<$Seeing$>$ &  Airmass	&	$\tau_{0}$	&	Notes 	\\
		&	(hh:mm)	&					 &	density			&		(IRDIS/IFS)					&					($^{\circ}$)	&	('')	& &	(ms)	& \\
\hline\hline
06-05-2015	&	02:28	&	IRDIFS	&		ND\_3.5	&	$8/16s\times 8/4 \times 1/1$	&	0.46	&	1.63	&	1.22	&	0.9	&	unsat \\	
06-05-2015	&	02:39	&	IRDIFS	&		none		&	$4/16s\times 2/2 \times1/1 $ & 0.07	&	1.72	&	1.21	&	0.9	&	waffles \\	
06-05-2015	&	02:41	&	IRDIFS	&		none		& $4/16s\times 56/16 \times 16/16 $	&	29.32	&	0.89	&	1.21	&	1.9	&	\\	
06-05-2015	&	03:59	&	IRDIFS	&		none		& $4/16s\times 2/2 \times1/1 $	&	0.07	&	0.83	&	1.24	&	1.9	&	waffles	\\	
06-05-2015	&	04:00	&	IRDIFS	&		ND\_3.5		& $8/16s\times 8/4 \times 1/1$	&	0.43	&	0.71	&	1.24	&	2.2	&	unsat \\	
\hline
03-06-2015	&	00:32	&	IRDIFS	&		ND\_2.0 	&	$0.84/2s \times 16/8 \times 1/1$	&	0.18	&	1.53	&	1.23	&	2.9	&	unsat \\	
03-06-2015	&	00:33	&	IRDIFS	&		none		&	$16/16s \times 2/2 \times 1/1$ & 0.23	&	1.67	&	1.23	&	2.7	&	waffles \\	
03-06-2015	&	00:34	&	IRDIFS	&		none		& $16/16s \times 16/16 \times 16/16$ & 28.77	&	1.30	&	1.21	&	2.8	&	 \\	
03-06-2015	&	01:54	&	IRDIFS	&		none		&	$16/16s \times 2/2 \times 1/1$ & 0.23	 &	1.11	&	1.23	&	4.7	&	waffles \\	
03-06-2015	&	01:56	&	IRDIFS	&		ND\_2.0 	&	$0.84/2s \times 16/8 \times 1$	&	0.19	&	0.85	&	1.23	&	6.1	&	unsat \\	
\hline
05-06-2015	&	00:50	&	IRDIFS\_EXT	&		ND\_2.0 	&	$0.84/2s \times 16/8 \times 1$	&	0.20	&	1.47 &	1.21 &	1.9	&	unsat \\	
05-06-2015	&	00:51	&	IRDIFS\_EXT	& none & $16/16s \times 2/2 \times 1/1$	&	0.23	&	1.49	& 1.21	&	1.8	&	waffles \\
05-06-2015	&	00:54	&	IRDIFS\_EXT	& none &  $16/16s \times 16/16 \times 1/1$	&  27.88	&	1.79	&	1.22	&	1.39	&	\\
05-06-2015	&	02:11	    &	IRDIFS\_EXT	& none & $16/16s \times 2/2 \times 1/1$	&	0.19	&	1.75	& 1.26	&	1.5	&	waffles \\
05-06-2015	&	02:13	&	IRDIFS\_EXT	&		ND\_2.0 	&	$0.84/2s \times 16/8 \times 1/1$	&	0.17	&	1.74 &	1.26 &	1.4	&	unsat \\	
\hline
06-06-2015	&	00:41	&	IRDIS-Y2Y3	&		ND\_3.5 	&	$4s \times 15 \times 1$	&	0.50	&	1.27 &	1.21 &	2.1	&	unsat \\	
06-06-2015	&	00:44	&	IRDIS-Y2Y3	&		none 	&	$2s \times 3 \times 1$	&	0.05	&	1.30 &	1.21 &	2.2	&	waffles \\	
06-06-2015	&	00:45	&	IRDIS-Y2Y3	&		none 	&	$2s \times 40 \times 64$	&	35.17	&	1.34 &	1.23 &	2.2 &	 \\	
06-06-2015	&	01:49	&	IRDIS-Y2Y3	&		none 	&	$2s \times 3 \times 1$	&	0.06	&	1.42 &	1.23 &	2.1	&	waffles \\	
06-06-2015	&	02:18	&	IRDIS-Y2Y3	&		none 	&	$2s \times 3 \times 1$	&	0.05	&	1.21 &	1.28 &	2.6	&	waffles \\	
06-06-2015	&	00:41	&	IRDIS-Y2Y3	&		ND\_3.5 	&	$4s \times 15 \times 1$	&	0.38	&	1.31 & 1.28  &	2.5	&	unsat \\	
\hline
07-06-2015	&	00:56	&	IRDIS-J2J3	&		ND\_2.0 	&	$4s \times 15 \times 1$	&	0.50	&	1.63 &	1.21 &	1.5	&	unsat \\	
07-06-2015	&	00:59	&	IRDIS-J2J3	&		none 	&	$8s \times 3 \times 1$	&	0.19	&	1.42 &	1.21 &	1.7	&	waffles \\	
07-06-2015	&	01:00	&	IRDIS-J2J3	&		none 	&$8s \times 32 \times 16$	&	28.27 &	1.95 &	1.23 &	1.38 &	 \\	
07-06-2015	&	02:21	&	IRDIS-J2J3	&		none 	&	$8s \times 3 \times 1$	&	0.14	&	2.55 &	1.30 &	1.2	&	waffles \\	
07-06-2015	&	02:28	&	IRDIS-J2J3	&		ND\_2.0 	&	$4s \times 15 \times 1$	&	0.35	&	2.33 &	1.32 &	1.3	&	unsat \\	
\hline
29-03-2016	&	05:07	&	IRDIFS	&		ND\_3.5 	&  $8/16s\times 21/11 \times 1/1$	&	1.25	&	1.29	&	1.21	&	1.7	&	unsat \\
29-03-2016	&	05:11		&	IRDIFS	&		none	&	$32/32s\times 4/4 \times 26/26$		& 31.22	&	1.10	&	1.22	&	2.1&	waffles \\
29-03-2016	&	06:07	&	IRDIFS	&		ND\_3.5 	&  $8/16s\times 21/11 \times 1/1$	&	0.19	&	1.12	&	1.22	&	1.8	&	unsat \\
\hline
10-02-2017	&	08:05	&	IRDIFS	&		ND\_3.5 	&  $8/16s\times 21/11 \times 1/1$	&	1.23	&	0.65	&	1.22	&	5.1	&	unsat \\
10-02-2017	&	08:09	&	IRDIFS	&		none	& $32/32s\times 4/4 \times 28/28$		& 31.17	&	0.78	&	1.22	& 3.4	&	waffles \\
10-02-2017	&	09:29	&	IRDIFS	&		ND\_3.5 	&  $8/16s\times 21/11 \times 1/1$	&	1.12	&	0.93	&	1.24	&	2.6	&	unsat \\
\hline
\end{tabular}
\end{center}
\tablefoot{UT-Time at start. The seeing is measured at 0.5~$\muup$m. DIT {(Detector Integration Time)} refers to the individual exposure time per frame. NDIT is the number of individual frames per exposure, $N_{EXP}$ is the number of exposures, and $\Delta$PA to the amplitude of the parallactic rotation.}\\
\end{table*}

	The IRDIS and IFS datasets were reduced at the SPHERE Data Center \citep[DC;][]{2017sf2a.conf..347D} using the SPHERE Data Reduction and Handling (DRH)  pipeline \citep{2008SPIE.7019E..39P}. The DRH carried out the basic corrections for bad pixels, dark current, and flat field. The DC performed an improved wavelength calibration,  a correction of the cross-talk, and  removal of bad pixels for the IFS data \citep{2015A&A...576A.121M}. It also applied the anamorphism correction to the IRDIS data.  We registered the frames fitting a two-dimentional moffat function to the waffles. 
	
	We temporally binned some of the registered cubes of IRDIS frames to ensure we could run the angular differential imaging \citep[ADI;][]{2006ApJ...641..556M} algorithms efficiently (binning factors of 2,  4,  and  8 for the K1K2, J2J3, and Y2Y3 data; factors of 7 and  2 for the May 2015 and June 2015 H2H3 data).  
	We  selected the resulting IFS datacubes based on the ratio of average fluxes in an inner and an outer ring centered on 75 and 597 mas separation to ensure that we kept the frames with the best Strehl ratio (flux ratio $\geq1.3$).  Conversely, we selected 80\% (H2H3, K1K2, J2J3 datasets) to 60\% (Y2Y3 dataset) of the frames with the lowest halo values beyond the AO correction radius where GJ~504b lies (e.g. in a ring located between 19 and 26 full-width-at-half-maxima).
	
		The absolute on-sky orientation of the instrument and the detector pixel scale were calibrated as part of a long-term monitoring conducted during the GTO \citep{2016A&A...587A..56M, 2016SPIE.9908E..34M}. The values are reported in Table \ref{Tab:Astrometry}.
		
		We used the \texttt{Specal} pipeline (Galicher et al., submitted) to apply the ADI steps on the IRDIS data. We applied the Template Locally Optimized Combination of Images  \citep[TLOCI; ][]{2014IAUS..299...48M} algorithm to extract the photometry and astrometry of the companion and to derive detection limits. The algorithm has been shown to extract the flux and position of such companions with a high fidelity (Chauvin et al, in prep). We also used the principal component analysis \citep[PCA; ][]{2012ApJ...755L..28S} implemented in \texttt{Speca}l and ANDROMEDA \citep{Mugnier-a-09, 2015A&A...582A..89C} algorithms to confirm our results. 
		We processed the IFS data with a custom pipeline exploiting the temporal and spectral diversity \citep{2015MNRAS.454..129V}. The pipeline derived detection limits following the estimation of the flux losses based on the injection of fake planets with flat spectra. The sensitivity curves account for the small-number statistics affecting the noise estimates at the innermost working  angles \citep{2014ApJ...792...97M}.
	    
	    \begin{figure}
  \centering
  \includegraphics[width=\columnwidth, angle=0]{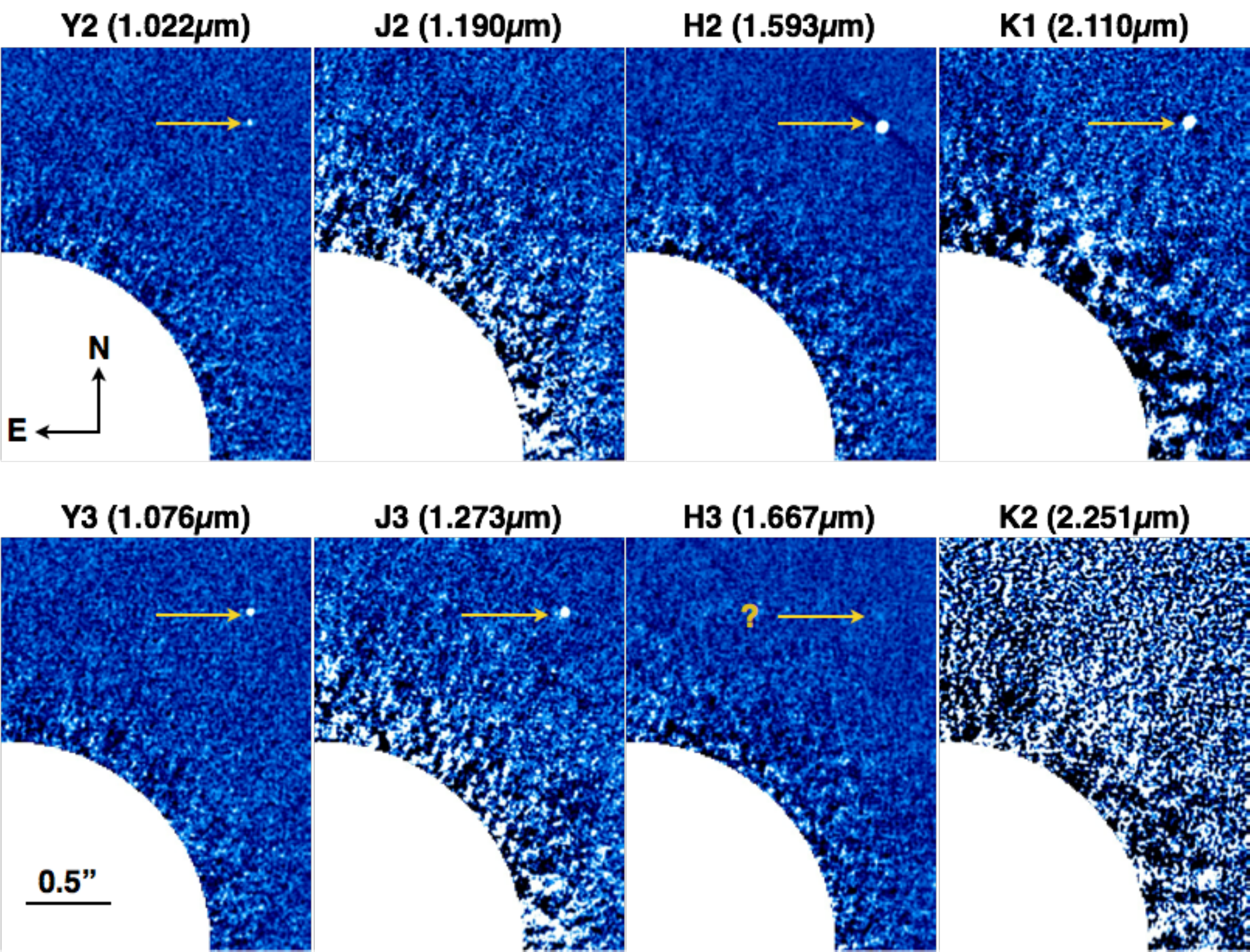}
  \caption{High-contrast images of the immediate environment of GJ~504A obtained with the DBI filters of IRDIS and using the TLOCI angular differential imaging algorithm. The star center is located in the lower-left corner of the images. GJ~504b is re-detected (arrow) into the Y2, Y3, J3, H2,  and K1 bands. The companion is tentatively re-detected in the H3 channel. The H2-H3 images correspond to the March 2016 data.}
  \label{Fig:Vignette}
\end{figure}

		The Y3, J3, H2, and K1 filters sample the main emission peaks  of cold companions ("on-channels") while the central wavelengths of the Y2, J2, H3, and K2 filters are chosen to sample the  molecular  absorptions.  The companion is therefore re-detected in the "on" channels with signal-to-noise ratios (S/Ns) ranging from 10 to 46 (Fig. \ref{Fig:Vignette}). We also re-detect the object in the Y2 ($\Delta Y2=16.71\pm0.16$ mag)  channel at a S/N of 7. To conclude, we also tentatively re-detect the object in the H3 band in the May 2016 data, which are the deepest  obtained on the system with SPHERE.  We considered the photometry extracted from the H3 channel as an upper limit in Sections \ref{sec:emp_analysis} and \ref{Section:atmomodels} to be conservative. We also derive upper limits in the J2 and K2 channels using the injection of artificial planets. 
		
		The PCA and ANDROMEDA photometry confirms the contrasts and astrometry found with the TLOCI algorithm.  Table \ref{Tab:Astrometry} summarizes the astrometry extracted from the data using TLOCI. The June 2015 astrometry obtained with the different filter pairs on consecutive days are consistent. We model these measurements in Section \ref{Section:orbit}.  The final contrasts were converted to apparent magnitudes (Table \ref{Tab:flux-density}) using the star photometry estimated for the SPHERE/IRDIS pass-bands (Appendix \ref{sec:AppA}). 
		  
\begin{table*}
 \centering
  \caption{GJ~504b astrometry.}
\small
  \begin{tabular}{llllllll}
  \hline   
	\hline
Date	&	Instrument	&	Filter		&	Platescale		&	True North	&	Sep		&	PA	 \\
	&	&	&	(mas/pixel)	&	(deg)	&	(mas)		&	(deg) 	\\
	\hline
26/03/2011 	&	HiCIAO	&	H	&	$9.500\pm0.005$ &$0.35\pm0.02$	&	$2479\pm16$		&	$327.94	\pm0.39$  \\
22/05/2011	&	HiCIAO	&	H	&	$9.500\pm0.005$ &$0.35\pm0.02$	&		$2483 \pm 8$		& $327.45 \pm	0.19$	 \\
12/08/2011	&	IRCS		&	L'	&	$20.54\pm0.03$ &$0.28\pm0.09$	&	$2481 \pm	33$	&	$326.84 \pm	0.94$	 \\
28/02/2012	&	HiCIAO		&	$K_{s}$	&$9.500\pm0.005$&$0.35\pm0.02$	&		$2483 \pm	15$	&	$326.46 \pm	0.36$ \\
12/04/2012	&	HiCIAO	&	J	&	$9.500\pm0.005$&$0.35\pm0.02$	&		$2487 \pm	8$	&	$326.54 \pm	0.18$	 \\
25/05/2012	&	IRCS		&	L'	&	$20.54\pm0.03$ &$0.28\pm0.09$ &		$2499 \pm	26$	&	$326.14 \pm 	0.61$	\\
05/05/2015	&	SPHERE	&	H2	&	$12.255\pm0.009$ &	$1.712\pm0.063$	&		$2491\pm3$	&	$323.46	 \pm 0.07$			\\ 
03/06/2015	&	SPHERE	&	H2	&	$12.255\pm0.009$ &	$1.712\pm0.063$	&	 $2496	\pm	3$	&	$323.50		\pm	0.07$		\\ 
05/06/2015	&	SPHERE	&	K1	&	$12.267\pm0.009$ &	$1.712\pm0.063$& $2497 \pm 4$	 & $323.60 \pm 0.10$	 \\ 
06/06/2015	&	SPHERE	&	Y2	&	$12.283\pm0.009$ &	$1.712\pm0.063$& $2495 \pm 5$	 &	$323.50 \pm 0.14$	 \\  
06/06/2015	&	SPHERE	&	Y3	&	$12.283\pm0.009$ &	$1.712\pm0.063$	& $2501 \pm	3$	 &	$323.49 \pm 0.07$ \\  
07/06/2015	&	SPHERE	&	J3		&	$12.261\pm0.009$ &	$1.712\pm0.063$	&	$2499 \pm 6$ &	$323.40	\pm 0.14$	 \\ 
29/03/2016	&	SPHERE	&	H2	&	$12.255\pm0.009$ &	$1.78\pm0.08$	& $2495	\pm	2$		&	$322.48	 \pm	0.05$	\\ 
29/03/2016	&	SPHERE	&	H3$^{a}$	&	$12.255\pm0.009$ &	$1.78\pm0.08$	& $2493 \pm 12$	&	$ 322.83 \pm 0.32$ \\  
10/02/2017	&	SPHERE	&	H2 &	$12.255\pm0.009$ &	$1.719\pm0.056$	&	$2493	\pm	3$		&	$321.74	 \pm  0.08$	\\ 
\hline
 \end{tabular}
 \tablefoot{HiCIAO and IRCS astrometry from \cite{Kuzuhara2013}. $^{a}$Tentative re-detection at H3.}
\label{Tab:Astrometry}
\end{table*}

		We converted the SPHERE apparent magnitudes of GJ~504b to flux densities using a spectrum of Vega \citep{1985IAUS..111..225H, 1985A&A...151..399M}, the filter passbands\footnote{http://www.eso.org/sci/facilities/paranal/instruments/sphere/inst/\-filters.html}, and atmospheric extinction curves computed with the SKYCALC tool for our observing conditions \citep{2012A&A...543A..92N, 2013A&A...560A..91J}. 
	We followed this procedure to convert the J, H, K, L', CH4S, and L photometry from \cite{Kuzuhara2013} and \cite{2013ApJ...778L...4J}\footnote{We considered    Mauna Kea transmissions for an airmass of 1.0  and a water vapor column of 3mm (https://www.gemini.edu/sciops/telescopes-and-sites/observing-condition-constraints/ir-transmission-spectra). The transmission has a negligible impact on the central values ($\leq1$\%) with respect to our error bars.}. Finally, we  directly used the zero points and magnitudes reported in \cite{2016ApJ...817..166S} to compute the L\_NB6, L\_NB7, and L\_NB8  flux densities. Table \ref{Tab:flux-density} summarizes the companion apparent magnitudes and flux densities used in this study. 

\begin{table*}
 \centering
  \caption{Apparent magnitudes and flux densities of GJ~504b. The J2 and K2 upper-limit magnitudes correspond to the 3$\sigma$ detection level.}
  \begin{tabular}{ccccccccc}
  \hline   
	\hline
Filter		&	$\lambda_c$	&	$\Delta \lambda$	&	Mag	&	Uncertainty 	&	Flux	&	1$\sigma$ lower limit &	1$\sigma$  upper-limit	&	Ref. \\
	&	($\mathrm{\mu}$m) &	($\mathrm{\mu}$m)	&	(mag)	&	(mag)	&	($\mathrm{W.m^{-2}.\mu m^{-1}}$) &	($\mathrm{W.m^{-2}.\mu m^{-1}}$)	&	($\mathrm{W.m^{-2}.\mu m^{-1}}$)	&	\\
	\hline
Y2	&	1.022	&	0.049 	&	20.98	&	0.20	&	 2.325e-17	&	1.934e-17	&	2.795e-17	&	This work	\\
Y3	&	1.076	&	0.050	&	20.14	&	0.09	&	  4.237e-17	&	3.900e-17	&	4.603e-17	&	This work \\
J2		&	1.190	&	0.042	&	21.28	&	\dots	&			\dots		&		\dots		&	1.078e-17	&	This work \\
J3		&	1.273	&	0.046	&	19.01	&	0.17	&	6.705e-17		&	5.733e-17	&	7.841e-17	&	This work \\
H2	&	1.593	&	0.052	&	18.95	&	0.30	&	3.260e-17		&	2.473e-17	&	4.297e-17	&	This work \\
H3$^{a}$	&	1.667	&	0.054	&	 21.81	&	0.35	&	1.990e-18		&	1.442e-18	&	2.747e-18	&	This work \\
K1	&	2.110		&	0.102	&	18.77	&	0.20	&	1.423e-17		&	1.184e-17	&	1.711e-17	&	This work \\
K2	&	2.251	&	0.109	&	$\ge$19.96	&	\dots	&			\dots		&		\dots		&	3.690e-18	&	This work \\
J		&	1.252	&	0.152	&	19.78	&	0.10	&	3.555e-17		&	3.243e-17	&	3.898e-17	&	Janson+13\\
H		&	1.633	&	0.288	&	20.01	&	0.14	&	1.131e-17		&	9.944e-18	&	1.287e-17	&	Janson+13\\
$\mathrm{K_{s}}$	&	2.139	&	0.312&	19.38	&	0.11	&	7.591e-18	&	6.860e-18	&	8.401e-18	&	Janson+13\\
$\mathrm{CH_{4}S}$ &	1.551	&	0.139&	19.58	&	0.13	& 1.974e-17	&	1.752e-17	&	2.226e-17 &	Janson+13\\
$\mathrm{CH_{4}L}$	& 1.719	&	0.142&	$\ge$20.63	&	\dots	&		\dots			&		\dots	 &	5.360e-18 &	Janson+13\\
L'		&	3.770	&	0.700	&	16.70	&	0.17	&	1.093e-17	&	9.344e-18	&	1.278e-17	&	Kuzuhara+13\\ 
L\_NB6	&	3.709	&	0.188	&	17.59	&	0.17	&	5.154e-18	&	4.407e-18	&	6.028e-18	&	Skemer+16\\
L\_NB7	&	3.875	&	0.234	&	16.47	&	0.19	&	1.229e-17	&	1.032e-17	&	1.464e-17	&	Skemer+16\\
L\_NB8	&	4.000	&	0.068	&	15.85	&	0.17	&	1.920e-17	&	1.641e-17	&	2.245e-17	&	Skemer+16\\
 \hline
 \end{tabular}
 \tablefoot{$^{a}$Tentative re-detection at H3. The photometry corresponds to the one extracted from \texttt{Specal}. We considered it as an upper limit for the empirical and atmospheric model analysis}
\label{Tab:flux-density}
\end{table*}

	\subsection{Radial velocity}
		We obtained 38 spectra between March 31, 2013, and May 23, 2016, with the SOPHIE spectrograph \citep{2006tafp.conf..319B} mounted on the OHP 1.93m telescope. The spectra cover the 3872-6943 \AA~range with a R$\sim$75 000 resolution. The data were reduced using the Software for the Analysis of the Fourier Interspectrum Radial velocities \citep[SAFIR, ][]{2005A&A...443..337G}.  From the fit of the cross-correlation function, we derive a $v \cdot sin\:i$ of $6.5\pm1$ km/s, in agreement with the value ($6\pm1$ km/s) reported in  \cite{2017A&A...598A..19D}.
			The data reveal radial-velocity variations with amplitudes greater than 100m/s that we model in Section \ref{subsub:polspot}. The SOPHIE data are not enough to precisely measure the period of the variations but they are compatible with the star rotation period measured by \cite{1996ApJ...466..384D}. 
			To complement the SOPHIE data, we also used 57 archival RV data points from the long-term monitoring  of the star obtained as part of  the Lick planet search survey. They span from June 12, 1987 to February 2, 2009 \citep{2014ApJS..210....5F}.
			
	\subsection{Interferometry}
	\label{subsec:interf}
We observed GJ504 on 2017 23, 24, and 25  June 2017 with the VEGA instrument \citep{vega, 2013JAI.....240003L} at the CHARA interferometric array \citep{chara}. We used the VEGA medium spectral resolution mode ($\sim$6000) and selected three spectral bands of 20~nm centered at 550, 710 and 730~nm. We recorded seven datasets with the E2W1W2 telescope triplet, allowing us to reach baselines spanning from about 100 to 220~m. Each target observation of about 10 minutes is interspersed with observations of reference stars to calibrate the instrumental transfer function. We used the JMMC SearchCal\footnote{www.jmmc.fr/searchcal} service \citep{searchcal} to select calibrators that are bright and small enough, and close to the target: HD~110423 (whose uniform-disk angular diameter in R band equals 0.250~$\pm$~0.007~mas according to \citet{JSDC2}) and HD~126248 (0.362~$\pm$0.011~mas). 

We used the standard VEGA data-reduction pipeline \citep{vega} to compute the calibrated squared visibility of each measurement. Those visibilities were fitted with the  LITpro\footnote{$\textrm{\tiny www.jmmc.fr/litpro\_page.htm}$} tool to determine a uniform-disk angular diameter $\theta_{UD}$~=~0.685~$\pm$~0.019~millisecond of arc (mas). We used the Claret tables \citep{Claret2011} to determine the limb-darkened angular diameter $\theta_{LD}$~=~0.71~$\pm$~0.02~mas using a linear limb-darkening law in the R band for an effective temperature ranging from 6000 and 7000~K (limb-darkening coefficient of 0.44). Assuming a parallax of 56.95~$\pm$~0.26~mas \citep{2007A&A...474..653V}, we deduced a radius of $R_{\bigstar}$~=~1.35~$\pm$~0.04~$R_\odot$ for GJ~504A. 

\section{Revised stellar properties}
\label{sec:age}

\begin{figure}
  \centering
  \begin{tabular}{c}
    \includegraphics[width=\columnwidth]{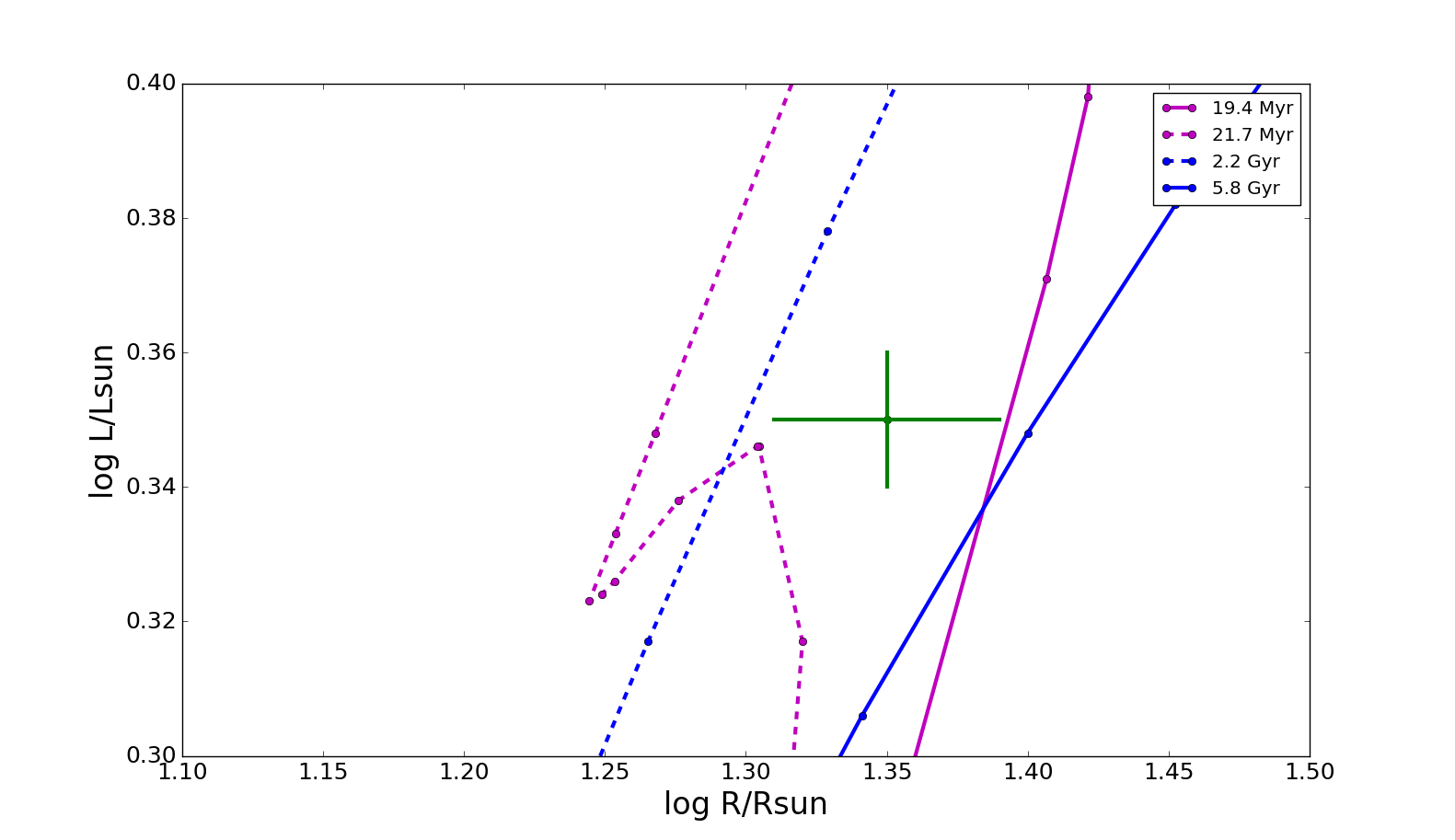} \\
  \end{tabular}
  \caption{Position of GJ504 in the Hertzsprung-Russell diagram. The constraints on the fundamental parameters are indicated by the 1$\sigma$-error box ($\log (L/L_\odot)$, $R_{\bigstar}$). PARSEC isochrones for [Fe/H] = $0.22\pm0.08$ dex (Z = 0.024, Y = 0.29) are overplotted in blue lines for the old age solution, and in purple for the young age solution.}
  \label{fig:HR_ages}
\end{figure}

We compared the radius and the star luminosity derived in  Appendix \ref{sec:AppA} to the PARSEC isochrones \citep{Bressan2012} for a Z~=~0.024 (Fig.~\ref{fig:HR_ages}) corresponding to  the [Fe/H]=$0.22\pm0.04$ dex  of GJ~504A \citep{2017A&A...598A..19D}. The tracks were generated using the CMD3.0 tool\footnote{$\textrm{\tiny http://stev.oapd.inaf.it/cgi-bin/cmd}$}.  The 1-$\sigma$ uncertainty on $L$ and $R$ are consistent with two age ranges for the system: $21\pm 2$ Myr and $4.0\pm1.8$ Gyr, according to these models. We also infer a new mass estimate of 1.10-1.25 M$_{\odot}$ for the star. We find similar solutions using the DARTMOUTH  models \citep{2008ApJS..178...89D}. These isochronal ages are inconsistent with the intermediate age reported in \cite{Kuzuhara2013}. The old age range overlaps with the one reported in \cite{Fuhrmann2015} and \cite{2017A&A...598A..19D}. The young age estimate had been neglected in  \cite{Fuhrmann2015} and was not discussed further in \cite{2017A&A...598A..19D}.  We re-investigate below how our isochronal age estimates fit with the other age indicators in light of the measured metallicity of the host-star \citep{2017A&A...598A..19D} and recent work on clusters.\\ 
 
 The Barium abundance is known to decrease with  stellar age \citep[e.g., ][]{2009ApJ...693L..31D, 2017A&A...605A..66B}. The value for GJ~504A \citep[$\mathrm{[Ba/Fe]=-0.04\pm0.01\pm0.03}$dex;][]{2017A&A...598A..19D} is compatible with those of thin-disk stars \citep{2017A&A...606A..94D}. It is clearly at odds with the one derived for 10-50 Myr-old stars in associations and clusters \citep{2009ApJ...693L..31D, 2013MNRAS.431.1005D, 2015MNRAS.454.1976R, 2017A&A...598A..86D}.  The kinematics of GJ~504 is also known to be inconsistent with young moving groups (YMG) or any known young open clusters \citep{Kuzuhara2013, 2017A&A...598A..19D} which are the only groups of young stars with distances compatible with that of GJ~504A.  Stars from young nearby associations  and from young clusters ($<$150 Myr) are generally restricted to  solar metallicity values while GJ~504A has a super solar metallicity \citep[e.g. ][]{2009A&A...501..553D, 2012MNRAS.427.2905B, 2017A&A...601A..70S, 2017A&A...605A..66B}.  The Hyades super-cluster is the closest group of  metal-rich stars to GJ~504A. But the kinematics of GJ~504A is incompatible with these stars, in particular the V heliocentric space velocity \citep{2001MNRAS.328...45M} and the ages of these clusters are in any case at odds with those inferred from the tracks. The \texttt{BANYAN} $\mathtt{\Sigma}$ tool \citep{2018arXiv180109051G} yields a null probability of  membership to the 27 nearby  ($\leq 150$ pc)  associations (NYA; including the Hyades), considered, and predicts the system to belong to the field (99.9\% probability). 

\cite{2017A&A...598A..19D} report stellar ages of 440 Myr and 431 Myr from the  log R'$_{HK}$ and log L$_{X}$/$L_{bol}$ of GJ~504A using the \cite{2008ApJ...687.1264M} calibrations.  The R'$_{HK}$ index of GJ~504A \citep[-4.45 dex;][]{1998ApJS..118..239R} is in fact still compatible with those of some late-F/early-G stars (HIP 490, HIP 1481) from the Tucana-Horologium association \citep[$45\pm4$ Myr][]{2008ApJ...687.1264M, 2015MNRAS.454..593B} and may also reside within the envelope of values of Sco-Cen stars \citep[11-17 Myr;][]{2011ApJ...738..122C, 2012ApJ...746..154P}. The  R'$_{HK}$ is also compatible with an age younger  than 1.45 Gyr  set by the stellar activity in the open cluster NGC 752. That upper limit is not consistent with the old isochronal age  of GJ~504A \citep[Fig. 2 of][]{2013A&A...551L...8P}, but it does not account for the possible impact of GJ504 enhanced metallicity \citep{1998MNRAS.298..332R} and for the possible long-term activity cycles ($>$ 30 years) of the star whose existence has not been investigated thus far. \cite{Kuzuhara2013} argued that the X-Ray activity  of GJ~504A  \citep[$L_{x}/L_{bol}=-4.42$ dex; ][]{1999A&AS..135..319H} is less reliable than the R'$_{HK}$ index because of the temporal baseline which is much shorter than the one of the Calcium line measurement  \citep[while the two age indicators are correlated; ][]{1997AJ....114.1673S}.  We do not discuss this indicator any further. 

The Lithium line of GJ~504A  has previously been used by \cite{Kuzuhara2013} to infer an age range of 30-500 Myr. In fact, different values for the abundance and equivalent widths have been reported for the star \citep[equivalent width  ranging from 81 m\AA~to 83.1m\AA; A(Li)=2.74--2.91][]{1990ApJ...354..310B, 1996A&A...311..951F, 2005PASJ...57...45T, 2010ApJ...724..154G, 2012ApJ...756...46R}. The spread is  likely related to the uncertainty in the line-fitting method,  atmospheric parameter uncertainties, and atmospheric models used \citep{2015PASJ...67...85H}.   Lithium is also known to be a crude age estimator at the intrinsic mass and $\mathrm{T_{eff}}$ of the star \citep{Kuzuhara2013}.  The Li abundance of GJ~504A is in fact still compatible with the values reported for the Sco-Cen stars \citep{2011ApJ...738..122C}, but, conversely, it is consistent with some 1.1-1.3 $M_{\odot}$ stars of the well-characterized solar-metallicity cluster NGC 752 \citep[Fe/H=$+0.01\pm0.04$;][]{2004A&A...426..809S, 2016A&A...590A..94C} and of the metal-enriched $\sim$3 Gyr old cluster NGC 6253 \citep[Fe/H=$+0.43\pm0.01$;][]{2010AJ....139.2034A, 2012AJ....144..137C}.

\begin{figure*}
  \centering
  \begin{tabular}{cc}
  \includegraphics[width=9cm]{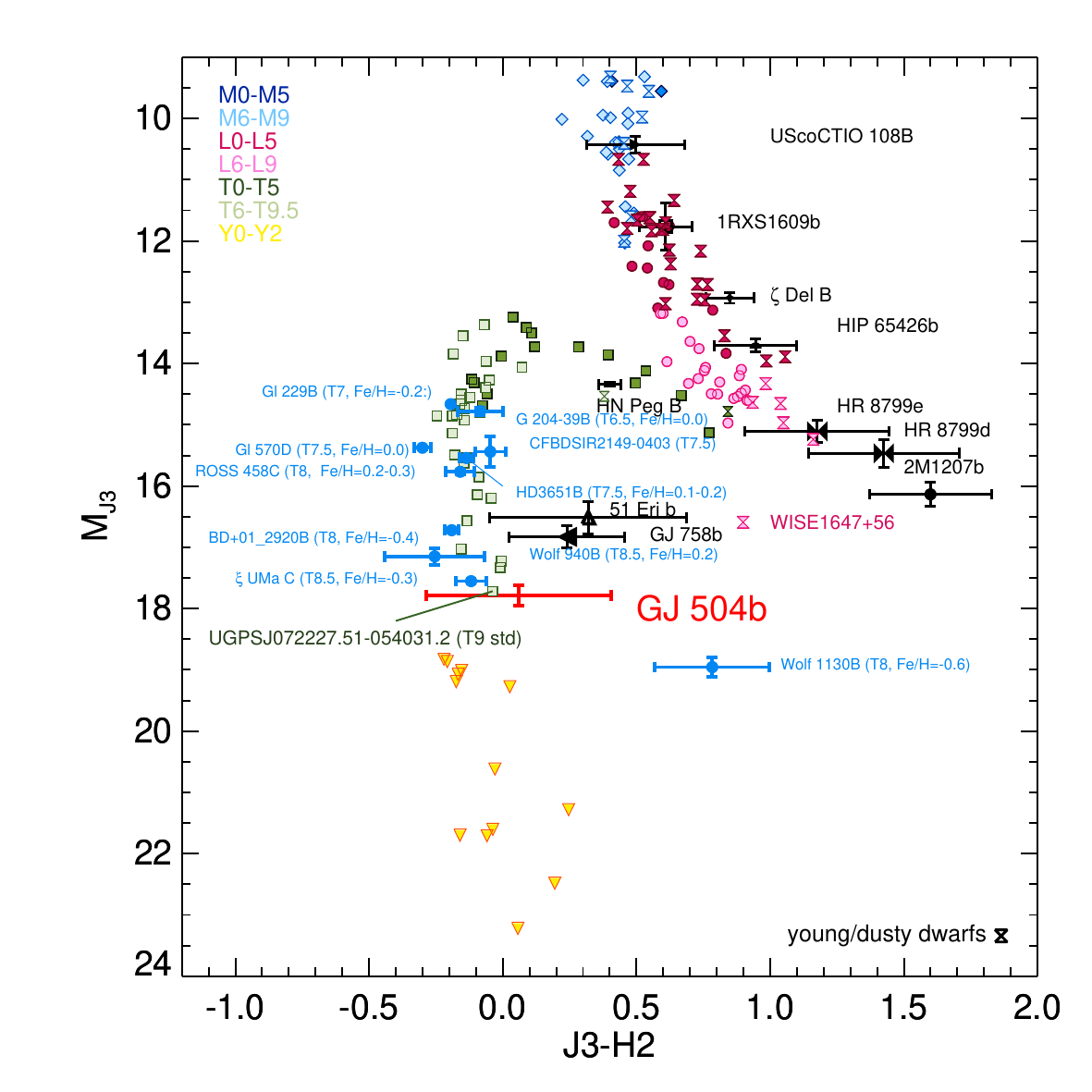}  &
     \includegraphics[width=9cm]{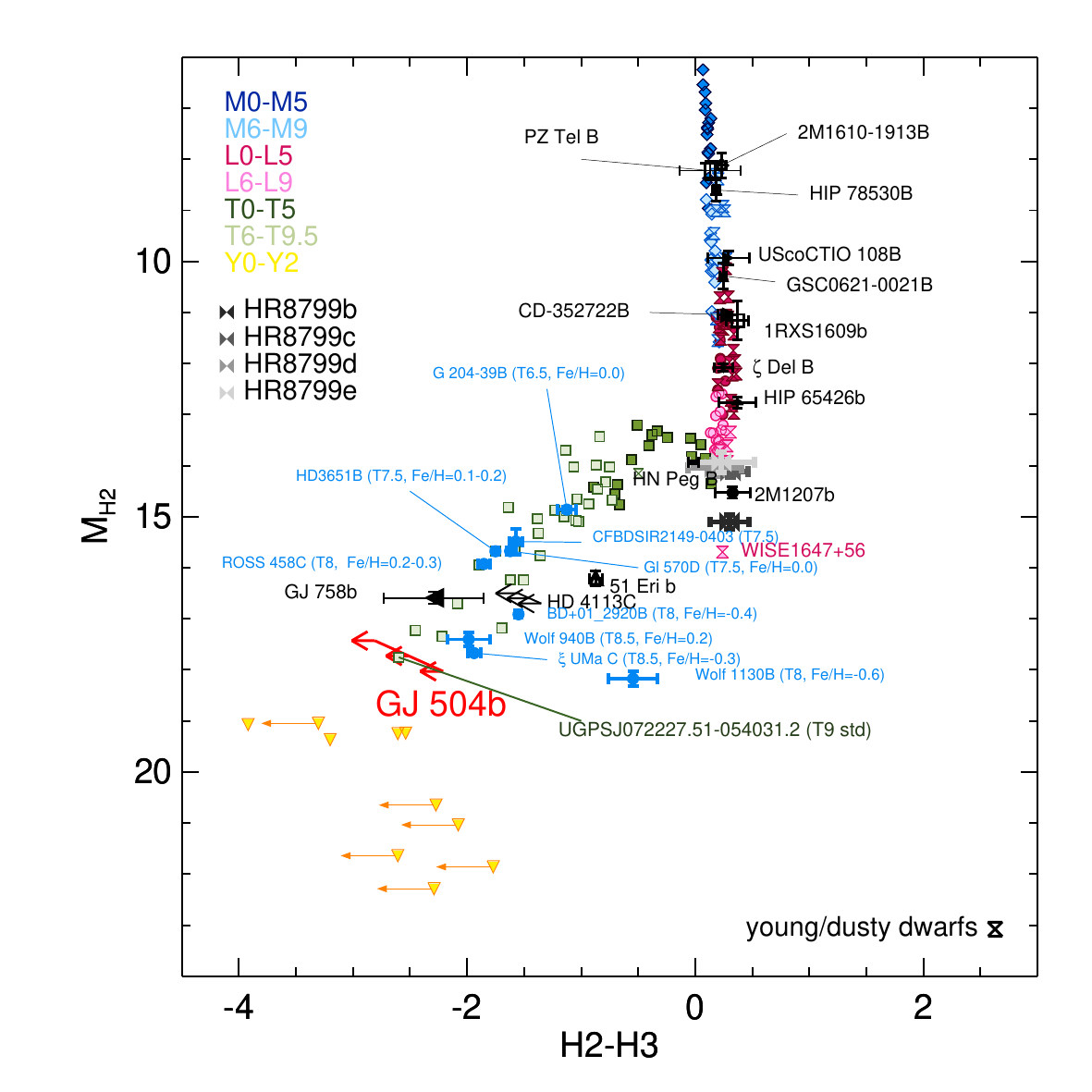}\\
 \end{tabular}
  \caption{Color-magnitude diagrams for the SPHERE/IRDIS photometry. The benchmark T-type companions are overlaid (full blue symbols). Their properties are summarized in Appendix \ref{Append:B}} 
  \label{fig:CMDs}
\end{figure*}

\cite{Kuzuhara2013}  derive an age of $160_{-60}^{+70}$ Myr for the system using the rotation period and various  gyrochronology relations \citep{2008ApJ...687.1264M, 2007ApJ...669.1167B, 2009ApJ...695..679M}.   It is possible to derive the age of stars with a convective envelope from a measured rotation period only  if they belong to the "I sequence" of slow rotators.  These relations are well established and robust for such solar-type stars. With a rotation period of 3.33 days for a spectral type of G0, GJ504 is a fast rotator, and therefore belongs to the "C sequence" of fast rotators as defined in \cite{2003ApJ...586..464B},  or has just reached the  "I sequence". The significant probability that GJ504 is a fast rotator means the calibrated gyrochronological relations used to directly measure its age with associated error bars are not reliable.  This is confirmed by observations and model realizations \citep[e.g., ][]{2013A&A...556A..36G, 2015A&A...577A..98G} that show that G stars with a period of 3.3 days can have any age between 1 and ~200Myr.  Conversely, gyrochronology provides a very robust upper limit on the age of such objects at the border between the I  and C sequences, which by design have to be younger than the age at which fast rotators of a given mass have all converged toward the "I sequence" of slow rotators. \cite{2003ApJ...586..464B} and \cite{2009ApJ...695..679M} show that G-type star convergence time is typically $\sim$150Myr. Close inspection of the M34 rotation sequence derived by \cite{2011ApJ...733L...9M} shows that all G stars of this cluster have turned into slow rotators. This means that if the rotation period of GJ504A derived by \cite{1996ApJ...466..384D} is correct, then the star is probably younger than 150 Myr and the age of M34 ($\sim$ 220Myr) is a conservative upper limit. 

\begin{table}
 \centering
  \caption{Summary of the different diagnostics on the age of GJ~504A}
  \begin{tabular}{ccc}
  \hline   
	\hline
Indicator			&	Age range \\
 \hline
Isochrones			&	$21 \pm 2$ Myr or 	$4.0\pm1.8$ Gyr \\
Barium		&	$\gg$ 1 Gyr \\
Activity	&	$\leq$ 1.45 Gyr \\
Rotation	&	$\leq$ 220 Myr \\
Lithium	&	$\lesssim$ 3 Gyr \\
 \hline
 \end{tabular}
\label{Tab:ageind}
\end{table}

Table \ref{Tab:ageind} summarizes the ages derived from the different indicators.  None of the two possible isochronal age ranges can be firmly excluded. Asteroseismology might disentangle between our solutions \citep[e.g.,][]{2015MNRAS.452.2127S}. We will consider both age ranges in the following sections. In Section \ref{subsec:agerec} we discuss two scenarios to explain the divergent conclusions from the age indicators. 

\section{Empirical analysis of GJ~504b photometry}
\label{sec:emp_analysis}
The  SPHERE photometry more than doubles the number of photometric data points sampling the NIR (1-2.5 $\mu$m)  SED \citep{Kuzuhara2013, 2013ApJ...778L...4J} of GJ~504b. The  H2-H3 color confirms the detection of a 1.6$\mu$m  methane absorption in GJ~504b's atmosphere \citep{2013ApJ...778L...4J}.  The Y2-Y3 color of GJ~504b is modulated by the red wing of the potassium doublet at 0.77 $\mu$m \citep{2007A&A...474L..21A}. The J2-J3 and K1-K2  colors indicate that the companion has strong additional methane and water bands at 1.1 and 2.3 $\mu$m. The IRDIS photometry allows for a detailed comparison of GJ~504b  to the large set of  brown dwarf and young giant planets for which  NIR spectra are available. 

\
\begin{figure*}
  \centering
  \begin{tabular}{cc}
  \includegraphics[width=9cm]{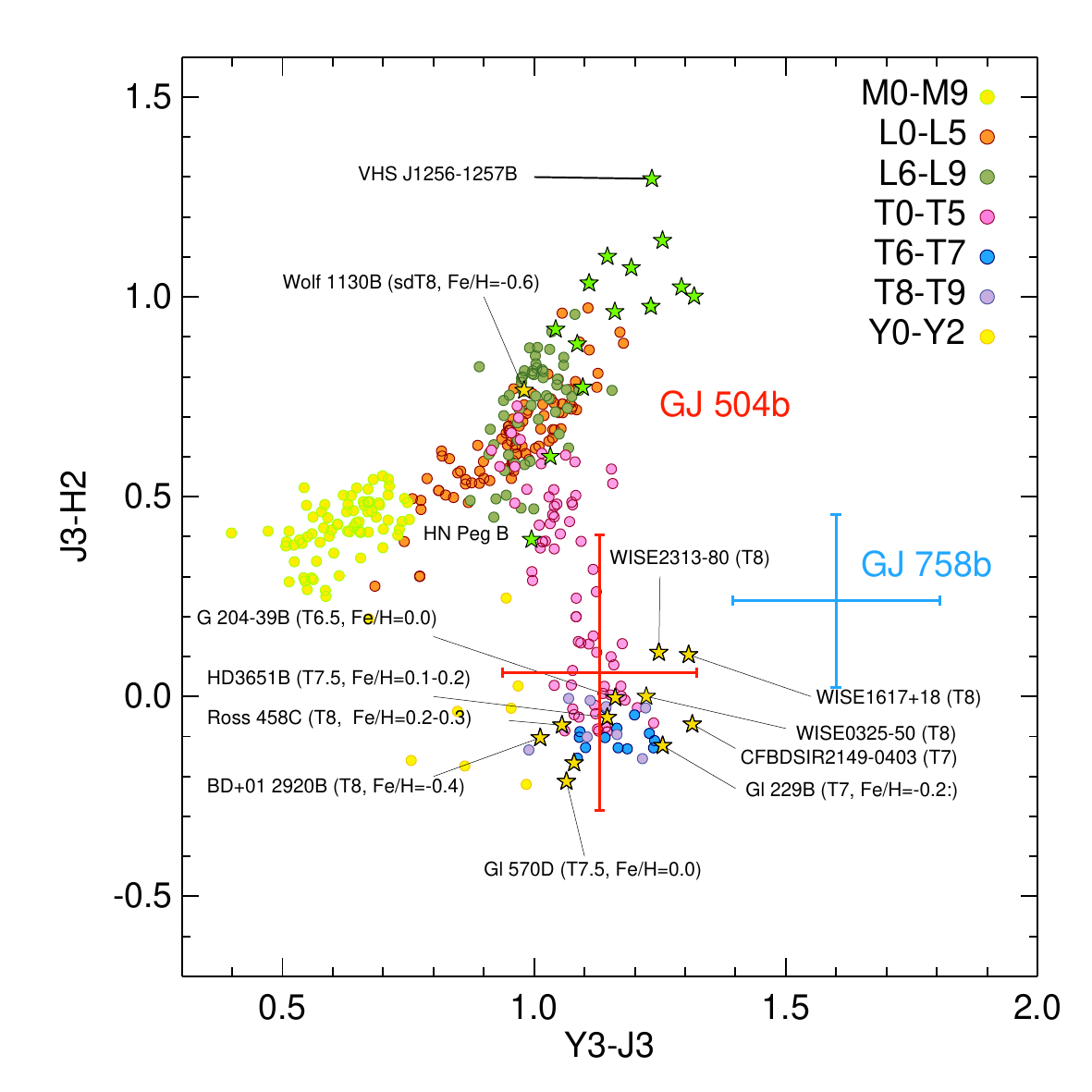} &
  \includegraphics[width=9cm]{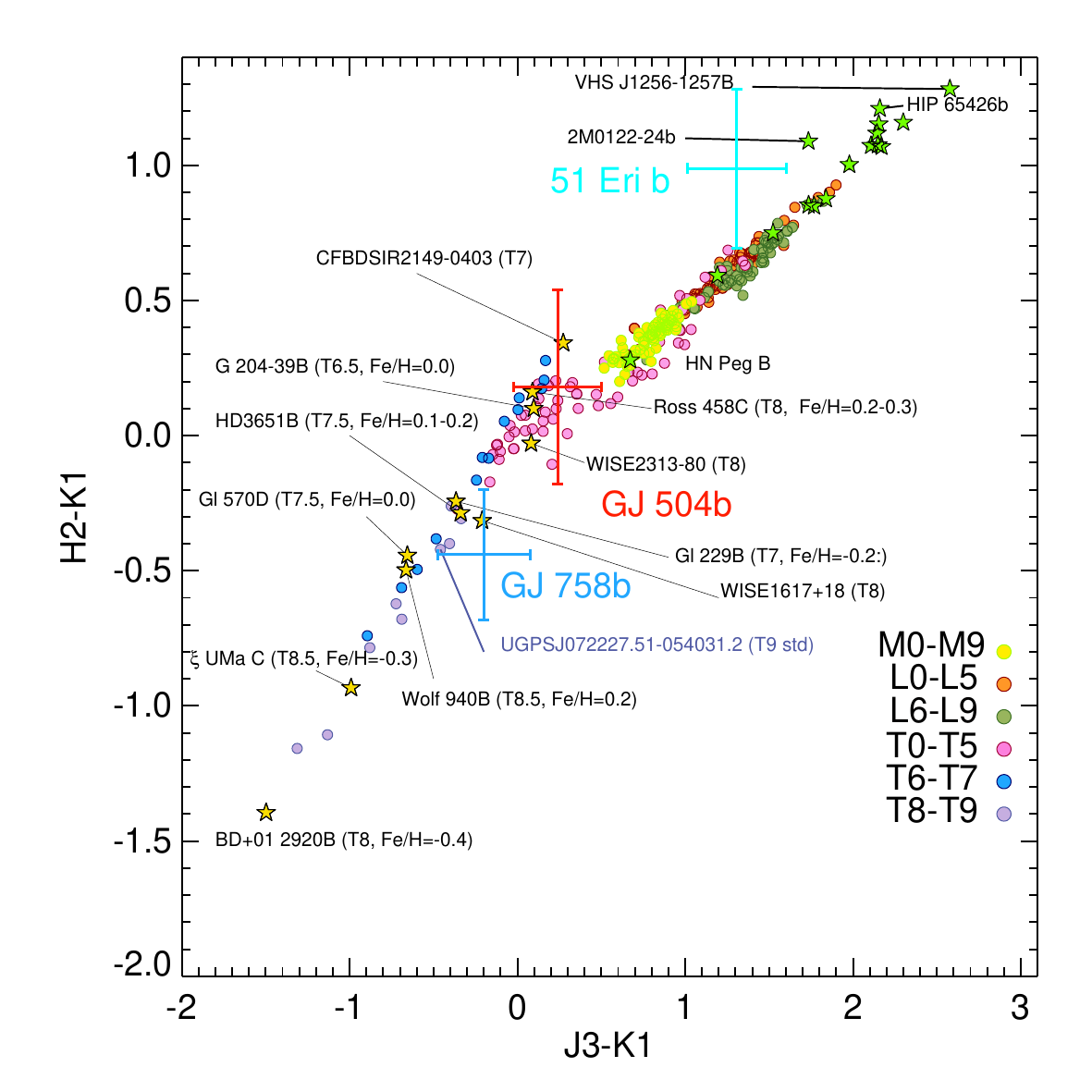} \\
  \end{tabular}
  \caption{Color-color diagram using the SPHERE/IRDIS photometry. The green stars correspond to dusty and/or young dwarfs at the L/T transition. The yellow stars correspond to the benchmark T-type companions and isolated objects listed in Table \ref{Tab:AppB}.}
  \label{ref:color-colordiag}
\end{figure*}


Fig. \ref{fig:CMDs} shows GJ~504b photometry in two selected color-magnitude diagrams (CMDs) exploiting the IRDIS photometry. Appendix \ref{App:C} details how the CMDs are created.   Late T-type companions with some knowledge on their metallicity are shown for comparison (light blue squares, see Appendix \ref{Append:B}).   GJ~504b has a similar Y, J, H, and K-band luminosity  and Y3-Y2, Y3-J3, J3-H2, and Y3-H2 colors as those of T8.5-T9 objects. The companion $\xi$ UMa C  has the closest  absolute J3 and H2 magnitude to GJ~504b, but the latter has redder H2-H3 colors indicative of a suppressed 1.6$\mu$m CH$_{4}$ absorption that might be related to sub-solar metallicity.  GJ 504b J and H-band luminosity are consistent with those of the  T9 standard UGPSJ072227.51-054031.2 \citep{2010MNRAS.408L..56L, 2011ApJ...743...50C}. The upper limits on the J2-J3,  H2-H3, and K1-K2 colors are close to those of late-T dwarfs.

	We overlay GJ~504b IRDIS photometry in color-color diagrams (CCD; see Appendix \ref{App:C}  for details) corresponding to the SPHERE filter sets (Fig. \ref{ref:color-colordiag}). The late T-type benchmark objects (Appendix \ref{Append:B}) are packed in the J3-H2/Y3-J3 CCD despite the different metallicity of these objects.  GJ~504b has a placement compatible with those objects; it has redder colors than most early Y dwarfs. Conversely, the benchmark companions with sub-solar metallicities have bluer colors in the J3-K1/H2-K1 CCD diagram than those with solar-metallicities for a given spectral type.  The K1-band colors are indeed expected to be modulated by the pressure-induced absorptions of H$_{2}$ which is in turn related to the metallicity and gravity.  GJ~504b has redder colors than the T9 standard UGPSJ072227.51-054031.2 despite the fact that the two objects share the same luminosity (see below). It has a similar placement to the T8 companion Ross 458C whose host star  is sharing the same metallicity range as GJ~504A but has an age \citep[150-800 Myr,][]{2010ApJ...725.1405B} intermediate between the two age ranges derived in Section \ref{sec:age}.  Three other late-T objects have similar deviant colors:  WISEP J231336.41-803701.4 \citep{2011ApJ...735..116B}, CFBDSIR J214947.2-040308.9  \citep{2012A&A...548A..26D}, and 51 Eri b \citep{2015Sci...350...64M}. CFBDSIR2149-04 is possibly younger than the field and/or metal enriched \citep{2017A&A...602A..82D}.   The planet 51 Eri b is orbiting a young star \citep{2015ApJ...813L..11M} and is proposed to be metal-enriched \citep{2017A&A...603A..57S}.  Those objects confirm that the gravity and/or the metallicity induces a shift toward redder colors in that CCD.
	
\begin{figure}
  \centering
\begin{tabular}{c}
  \includegraphics[width=\columnwidth]{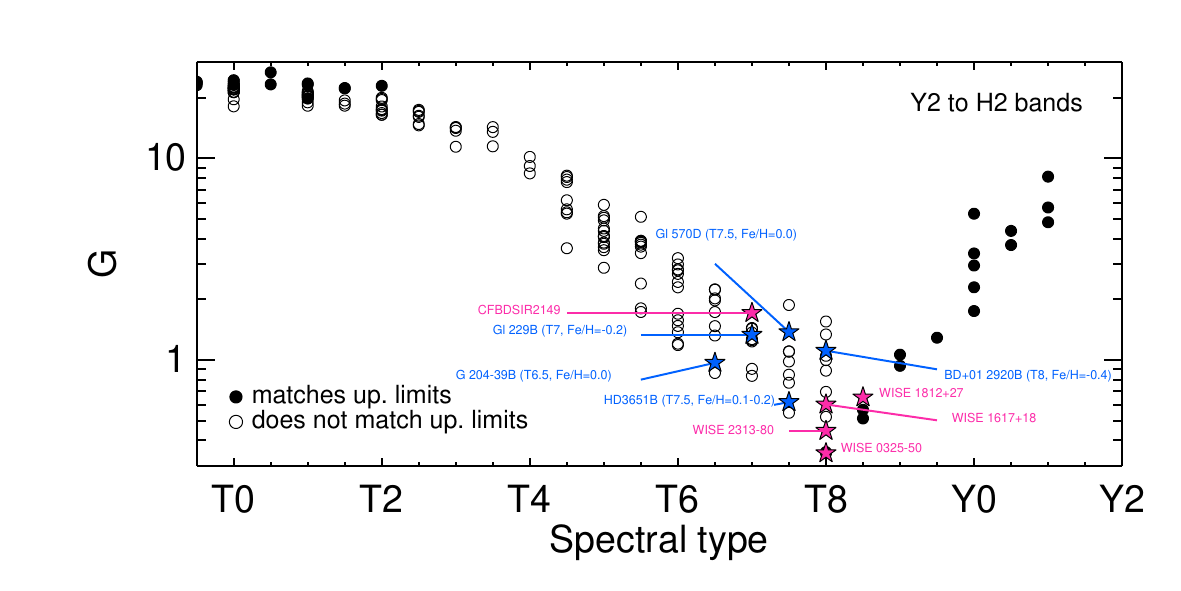} \\
  \includegraphics[width=\columnwidth]{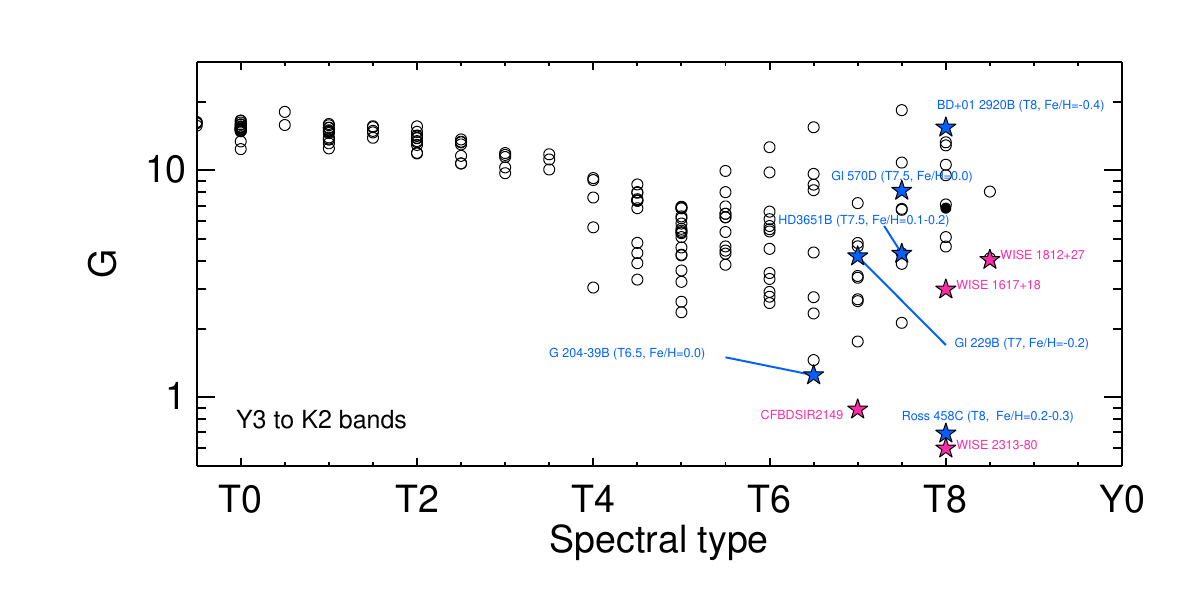} \\
\end{tabular}
   \caption{Goodness-of-fits (G) corresponding to the comparison of GJ504b photometry to those of empirical objects in the Y2 to H2 bands (top) and from the Y3 to K2 bands (bottom). The blue stars correspond to benchmark T-type companions while the pink ones correspond to peculiar free-floating T-type objects (see Appendix \ref{Append:B}).}
  \label{fig:Gemp}
\end{figure}

 	We used the G goodness-of-fit indicator \citep{2008ApJ...678.1372C} to compare the photometry of GJ~504b to those of reference objects (Fig. \ref{fig:Gemp}).   
 	
 	\begin{equation}
		G_{k}=\sum_{i=1}^{n} w_{i} \left ( \frac{f_{i} - \alpha_{k}\textsl{F}_{k,i}}{\sigma _{i}} \right )^{2}
	\end{equation}  

  where $f$ and $\sigma $ are the observed photometry of GJ~504b and associated error, and $w$ are the filter widths. $\textsl{F}_{k}$ corresponds to the photometry of the template spectrum $k$. $\alpha_{k}$ is a multiplicative factor between the companion photometry and the one of the template which minimizes $G_{k}$.

 	The exclusion of the K-band photometry from the fit allows the comparison to be extended to the Y dwarf domain where the K band flux of those objects is fully suppressed. The reference photometry is taken from the SpeXPrism library \citep{2014ASInC..11....7B} in addition to \cite{2014AJ....147..113C}, \cite{2013ApJS..205....6M}, and \cite{2015ApJ...804...92S}. We also added the photometry of peculiar late-T dwarfs described in Appendix \ref{Append:B}.  Figure \ref{fig:visemp} provides a visual comparison of the fit for some objects of interest.  We confirm that the overall NIR luminosity of the companion is best represented by the T9 standard UGPSJ072227.51-054031.2  \citep{2010MNRAS.408L..56L}. Companions with super-solar metallicity and/or cloudy atmospheres tend to have reduced G values compared to analogs with depleted metals. The T8 dwarf WISEA J032504.52–504403.0 produces the best fit of the YJH band flux; it is estimated to have a 100\% cloudy atmosphere with low surface gravity (log~g=4.0) and be on the younger end of the age range (0.08-0.3 Gyr) of all considered objects in \cite{2015ApJ...804...92S}. The intermediate age and metal-rich companion ROSS 458C produces an excellent fit of the Y- to K-band fluxes of GJ~504b, but it  is clearly more luminous.

\begin{figure}
  \centering
  \includegraphics[width=\columnwidth]{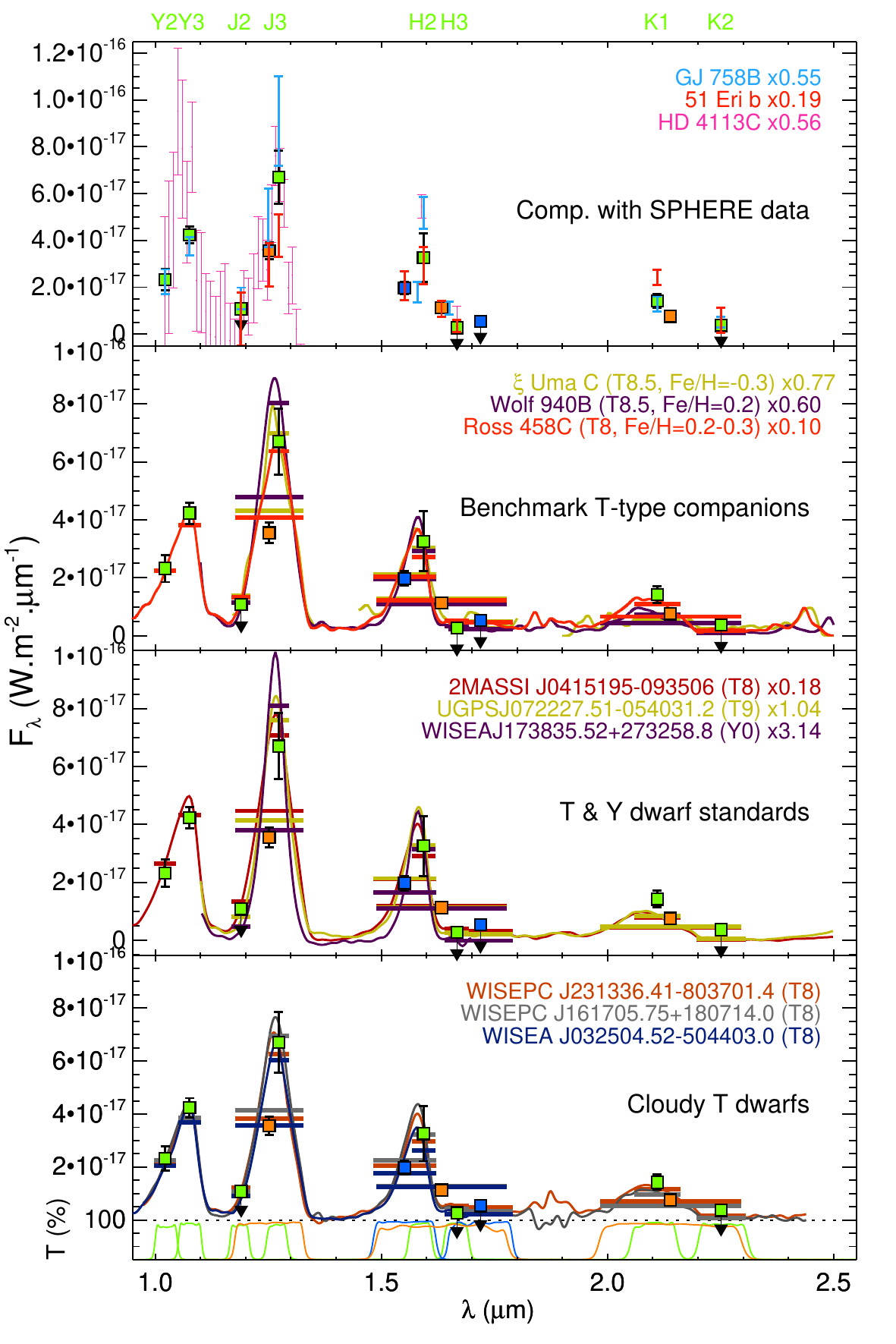}
  \caption{Visual comparison of the SED of GJ~504b (green squares) to that of T-type companions observed with VLT/SPHERE, of  benchmark companions with various metallicities, and of cloudy T dwarfs. The laying bars correspond to the flux of the template spectra averaged over the filter passbands whose transmission is reported at bottom.}
  \label{fig:visemp}
\end{figure}

We conclude that GJ~504b is a T$9^{+0.5}_{-1}$ object with peculiar NIR colors  that could be attributed to low surface gravity and/or enhanced metallicity. We use atmospheric models in the following section to further explore this latter findings. 

Using the $BC_{J}=2.0^{+0.4}_{-0.1}$ mag and  $BC_{H}=1.7^{+0.4}_{-0.2}$ mag of  T$9^{+0.5}_{-1}$ dwarfs from \cite{2013Sci...341.1492D}, we find a $log\:(L/L_{\odot}) = -6.33^{+0.12}_{-0.20}$ and  a $log\:(L/L_{\odot}) = -6.30^{+0.14}_{-0.22}$ for GJ~504b, respectively\footnote{Using $M_{bol, \odot}$=4.74 mag \citep{2016AJ....152...41P}.}.  The bolometric corrections might however not be appropriate for the peculiar SED of GJ~504b because it corresponds to the averaged values for "regular" dwarfs in spectral type bins. Therefore, we considered the $log\:(L/L_{\odot}) = -6.20 \pm 0.03$ of the T9 object UGPS J072227.51-054031.2 \citep{2013Sci...341.1492D} and the flux-scaling factor $\alpha=1.04$ value found above to estimate a  $log\:(L/L_{\odot}) = -6.18 \pm 0.03$ dex for GJ~504b . If the T8.5 companion Wolf 940B is used instead \citep[$log\:(L/L_{\odot})=-6.01\pm0.05$][]{2010ApJ...720..252L}, we find a $log\:(L/L_{\odot}) = -6.23 \pm 0.05$ dex for GJ~504b. 

\section{Atmospheric properties of GJ~504b}
\label{Section:atmomodels}
	\subsection{Forward modeling with the G statistics}
	\subsubsection{Model description}
	We considered five independent grids of synthetic spectra relying on different theoretical models to characterize the atmospheric properties of the companion and to show differences in the retrieved properties related to the model choice. The grid properties are summarized in Table \ref{Tab:atmomodchar}. We provide a succinct description of the  atmospheric models below. 
	
	We used the model grid of the Santa Cruz group (hereafter the "Morley" models). The grid was previously compared to the GJ~504b SED  \citep{2016ApJ...817..166S}.  It explores  the case of metal-enriched atmospheres. These 1D radiative-convective equilibrium atmospheric models are similar to those described in \cite{2012ApJ...756..172M} and \cite{2014ApJ...787...78M}.  They use the ExoMol methane line lists \citep{2014MNRAS.440.1649Y}. The wings of the pressure-broadened  K I and Na I bands  in the optical can extend into the NIR in Y and J bands and are known to affect the modeling of T-dwarf spectra.  In those models, the broadening is treated following \cite{2000ApJ...531..438B}. The models  consider the improved treatment of the collision-induced absorption (CIA) of H$_{2}$ \citep{2012JQSRT.113.1276R}.  They consider chemical equilibrium only, and account for the formation of resurgent clouds at the T/Y transition made of Cr, MnS, Na$_{2}$S, ZnS, and KCl particles. The cloud structure and opacities are computed following \cite{2001ApJ...556..872A}. The clouds are parametrized by the sedimentation efficiency ($f_{sed}$) which represents the balance between the upward transport of vapor and condensate by turbulent mixing in the atmosphere with the downward transport of condensate by sedimentation. Models with low $f_{sed}$ correspond to atmospheres with thicker clouds populated by smaller-size particles.  The grid of models do consider a uniform cloud deck. 

	The \texttt{BT-SETTL} 1D models \citep{2013MSAIS..24..128A} consider a cloud model where the number density and size distribution of condensates are determined following the scheme proposed by \cite{1978Icar...36....1R} as a function of depth, for example by comparing the timescales for nucleation, gravitational settling, condensation, and mixing layer by layer. Therefore, the only free parameters left are the effective temperature $T_{eff}$, the surface gravity log g (cgs), and the metallicity ($[M/H]$) with respect to the Sun reference values \citep{2011SoPh..268..255C}.  The cloud model generates sulfide clouds at the T/Y transition self-consistently.  It accounts for the nonequilibrium chemistry  of CO/CH$_{4}$, CO/CO$_{2}$, and N$_{2}$/NH$_{3}$. The radiative transfer is carried out through the \texttt{PHOENIX} atmosphere code \citep{2012RSPTA.370.2765A}, and uses the ExoMol CH$_{4}$ line list. The pressure-broadened K I and Na I line profiles are computed following \cite{2007A&A...474L..21A}. The grid of models used for GJ~504b analysis was computed to work in the temperature range of late-T/early-Y dwarfs and was previously compared to the SPHERE photometry of GJ~758b \citep{2016A&A...587A..55V}. These models do not explore the impact of the metallicity. 
	
	We used the  \texttt{petitCODE} 1D model atmosphere originally presented in \cite{2015ApJ...813...47M}. The model has been updated to produce realistic  transmission and emission spectra of giant planets  \citep{2016MNRAS.461.1053M, 2016MNRAS.459.1393M, 2017A&A...600A..10M}.  We used the code version  described in \cite{2017A&A...603A..57S}. It  has been vetted on the observations of 51 Eri b and on benchmark brown-dwarf companion spectra (Gl 570D and HD 3651B) whose temperatures fall close to that expected for GJ~504b \citep{2017A&A...603A..57S}. The \texttt{petitCODE} model self-consistently calculates atmospheric temperature structures assuming radiative-convective equilibrium and equilibrium chemistry. The gas opacities are currently taken into account considering the following species: H$_2$O, CO, CH$_4$, CO$_2$, C$_2$H$_2$, H$_2$S, H$_2$, HCN, K, Na, NH$_3$, OH, PH$_3$, TiO and VO. This includes the CIA of H$_2$--H$_2$ and H$_2$--He. The model makes use of the ExoMol CH$_{4}$ line list. The alkali line profiles (Na, K) are obtained from N. Allard \cite[priv com, see also][]{2007A&A...474L..21A} and  are considering a specific modeling \citep[see][]{2015ApJ...813...47M}.	 The  models we use here consider the formation of clouds.  The clouds model follows a modified scheme as presented in \cite{2001ApJ...556..872A}. The mixing length is set equal to the atmospheric pressure scale height in all cases. Above the cloud deck, the cloud mass fraction is parametrized by $f_{sed}$. The atmospheric mixing speed is equal to $K_{zz}/H_{p}$, with $K_{zz}$ the atmospheric eddy diffusion coefficient and  $H_{p}$ the pressure scale height. For the case of 51 Eri b \citep{2017A&A...603A..57S}, models were considering $K_{zz} = 10^{7.5} cm^{2}.s^{-1}$.  The grids have been extended   to the cases of  $K_{zz} = 10^{8.5} cm^{2}.s^{-1}$ and $f_{sed}$=0.5, 1.0\dots 3.0, and  $K_{zz} = 10^{6.5} cm^{2}.s^{-1}$ and $f_{sed}$=2.5 or 3.0.  The cloud model considers the opacities of KCl and Na$_{2}$S, the latter being the most abundant sulfite grain species expected to form in the atmosphere of a companion such as GJ~504b \citep{2012ApJ...756..172M}.

\begin{table*}
\centering
\caption{\label{Tab:atmomodchar} Characteristics of the atmospheric model grids compared to the SED of GJ\,504~b.}
\begin{tabular}{llllll}
 \hline
Parameter	&	\texttt{BT-SETTL}	&Morley 	&	\texttt{ATMO} & \texttt{Exo-REM}	&	\texttt{petitCODE}	\\
\hline 
$\mathrm{T_{eff}}$ (K)	&	200, 220\dots420	&		450,475\dots625	&	400,450\dots700	&	 300,350\dots2000	&	300,350\dots850$^{a}$ \\
									&		450, 500\dots1000	&								&								&								&										\\
log g (dex)						&	3.5,4.0, 4.5 			& 3.5, 4.0\dots5.0 			&	3.5, 4.0, 4.5			&	3.0, 3.1\dots6.0		&	3.0, 3.5\dots5.0			\\
$\mathrm{[M/H]}$ (dex)	&	0 							&	0.0, 0.5, 1.0 				&	0.0, 0.2, 0.5			&		-0.5, 0, 0.5			&	0.0, 0.2\dots1.4			\\
$\mathrm{K_{zz} (cm^{2}.s^{-1})}$			&	\dots						&		\dots						&	$10^{6}$ 				&	\dots						&	$10^{6.5}$, $10^{7.5}$,  $10^{8.5}$ \\		
$\mathrm{f_{sed}}$							&	\dots						&		1, 2, 3, 5, $\inf$		&	\dots						&	\dots						&		0.5, 1.0, \dots3.0$^{b}$ \\				
$\mathrm{f_{cloud}}$		&	1.00						&			1.00						&					1.00			&		0, 0.25\dots1.00	&		1.00 						\\
$\mathrm{\gamma}$		&	\dots						&		\dots				&	0, 1.2, 1.3						&		\dots					&		\dots						\\									
 \hline
\end{tabular}
\tablefoot{$^{a}$restricted to 500--850K for $K_{zz}=10^{7.5}$ and $f_{sed} \leq 2.0$. $^{b}\:f_{sed}$ values of 2.5 and 3.0 only when  $K_{zz}=10^{6.5}$. Additional $f_{sed}$=0.2 when $K_{zz}=10^{7.5}$ and $\mathrm{T_{eff} \geq}$ 500K.}
\end{table*}

	The 1D model \texttt{Exo-REM} (Baudino et al. 2015, 2017) solves for radiative-convective equilibrium, assuming conservation of the net flux (radiative+convective) over the 64 pressure-level grid. The first version of the cloud model of \texttt{Exo-REM} only considered the absorption of iron and silicate particles \citep{2015A&A...582A..83B}. The cloud vertical profile remained fixed \citep{2006ApJ...640.1063B} with the optical depth at some wavelengths being left as a free parameter. In spite of their relative simplicity,  these models were found to reproduce the spectral shape of the planets HR8799cde \citep{2016A&A...587A..58B} and of the late-T companion GJ~758b \citep{2016A&A...587A..55V}, but not necessarily their absolute fluxes. The grids used for GJ~504b correspond to a major upgrade of the models which are valid for planets with $\mathrm{T_{eff}}$ in the range 300-1700K.  This new version of \texttt{Exo-REM}  is described in more detail in \cite{2017arXiv171111483C}.  The radiative transfer equation is solved using the correlated-k approximation and opacities related to the CIA of H$_{2}$-He and to ten molecules (H$_{2}$O, CH$_{4}$, CO, CO$_{2}$ , NH$_{3}$, PH$_{3}$ , Na, K, TiO and VO) as described in \cite{2017ApJ...850..150B}.   The abundances in each atmospheric layer of the different molecules and  atoms are calculated for a given temperature profile assuming thermochemical equilibrium for TiO, VO and PH$_{3}$, and nonequilibrium chemistry for C-, O- and N- bearing
compounds comparing the chemical time constants to the vertical mixing time scales \citep{2014ApJ...797...41Z}. The latter is parametrized through an eddy mixing coefficient $\mathrm{K_{zz}}$ calculated from the mixing length theory and the convective flux from Exo-REM.  The cloud model now includes the formation of iron, silicate, Na$_{2}$S, KCl, and water clouds. The microphysics of the grains (size distribution and populations) is computed self-consistently following  \cite{1978Icar...36....1R}  (similarly to BT-SETTL) by comparing the timescales for condensation growth, gravitational settling, coalescence, and vertical mixing.  \texttt{Exo-REM}  considers the case of patchy atmospheres where the disk-averaged flux $F_{total}$ is a mix of clear regions ($F_{clear}$) and cloudy ones ($F_{cloudy}$) following 
\begin{equation}	
	(1 - f_{cloud}) \times F_{clear} + f_{cloud} \times F_{cloudy},
\end{equation}	
where $f_{cloud}$ is the cloud fraction parameter. In total, those models only leave $\mathrm{T_{eff}}$,  log g,  $\mathrm{[M/H]}$, and $f_{cloud}$ as free parameters.\\ 
	
	While all the previous models account for the formation of clouds,  \cite{2015ApJ...804L..17T} proposes through the \texttt{ATMO} models that this ingredient might not be needed to describe the atmosphere of brown dwarf and giant exoplanets. 	 \texttt{ATMO} is a 1D/2D radiative-convective equilibrium code suited for the modeling of the atmosphere of brown dwarfs, and irradiated and nonirradiated exoplanets \citep{2015ApJ...804L..17T, 2016ApJ...817L..19T, 2016A&A...594A..69D, 2017ApJ...850...46T}. The radiative transfer equation is solved using the correlated-k approximation as implemented in \cite{2014A&A...564A..59A} and \cite{amundsen:2017aa}. It accounts for the CIA of  H$_{2}$-H$_{2}$ and H$_{2}$-He and the opacities of CH$_{4}$, H$_{2}$O, CO, CO$_{2}$, NH$_{3}$, Na, K, TiO, VO, and FeH coupled with the out-of-equilibrium chemical network of \cite{2012A&A...546A..43V}. This nonequilibrium chemistry is directly related to $K_{zz}$ \citep{2007ApJ...669.1248H}.  The methane  opacities are updated with the ExoMol line list. The K I and Na I line profiles are calculated following \cite{2007A&A...474L..21A}.	 The  L/T and T/Y transitions are interpreted in that case as a temperature gradient reduction in the atmosphere coming from the fingering instability of chemical transitions (CO/CH$_{4}$, N$_{2}$/NH$_{3}$). That gradient reduction is parametrized through the adiabatic index $\gamma$ which is left as a free-parameter.   The \texttt{ATMO} models are shown to successfully reproduce the spectra of T and Y dwarfs \citep{2015ApJ...804L..17T, 2017ApJ...842..118L}  and of young  and old objects at the L/T  transition \citep{2016ApJ...817L..19T, 2017ApJ...850...46T}. For the case of GJ~504b, the grids used in \cite{2017ApJ...842..118L} have been extended to higher metallicities to encompass the solutions found by \cite{2016ApJ...817..166S}. We set $K_{zz}=10^{6}cm^{2}.s^{-1}$ to limit the extent of the grid. That value is within the range of expected values found for mature late-T objects   \citep[$10^{4}-10^{6} cm.s^{-2}$; ][]{2006ApJ...647..552S, 2007ApJ...656.1136S, 2009ApJ...695..844G}. But higher values may be needed for the case of GJ~504b (see below).

  \begin{figure}
  \centering
  \includegraphics[width=\columnwidth]{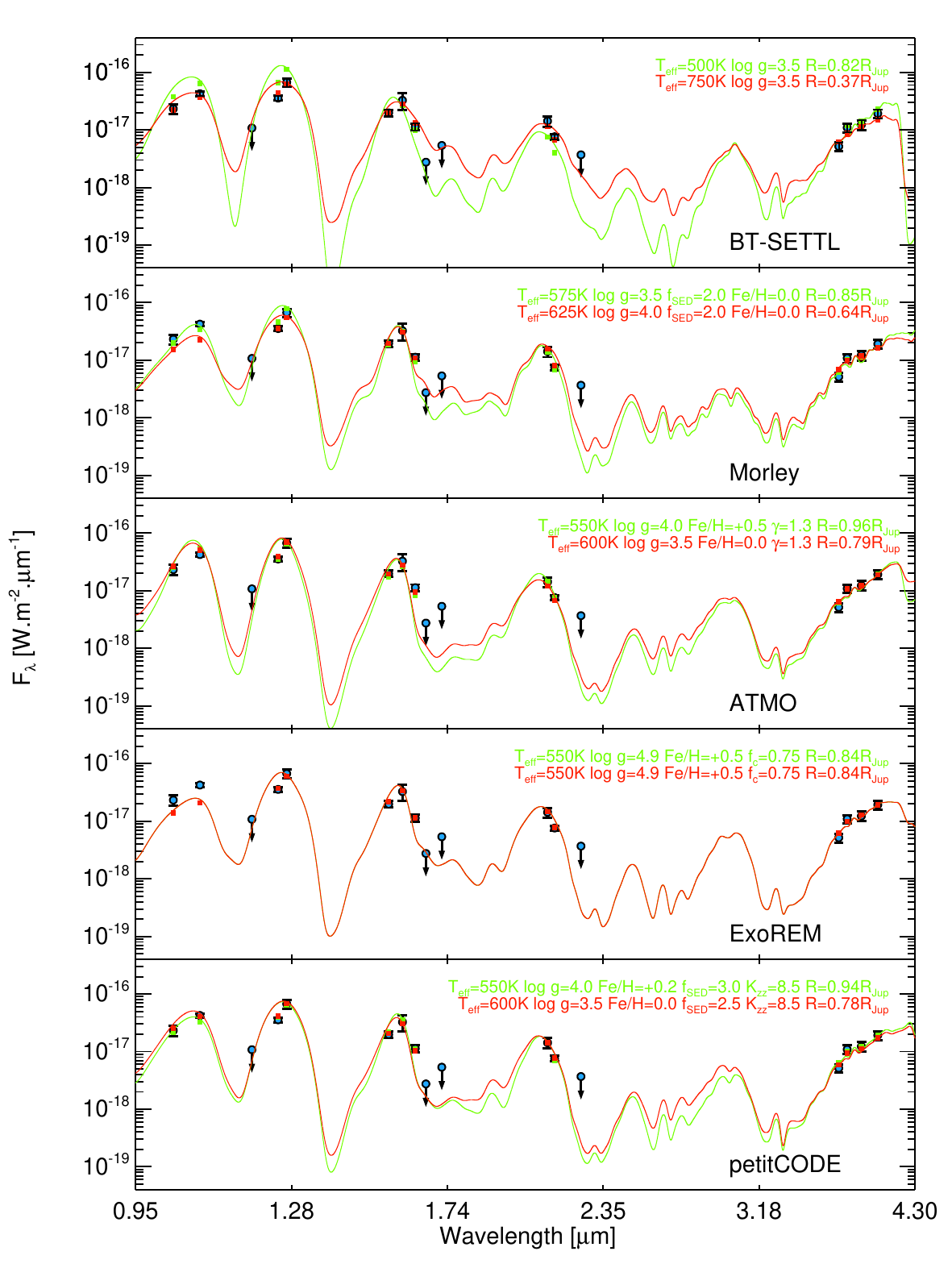}
  \caption{Best-fitting model spectra when using the G statistics. Solutions with some pre-requisite on the object radius are shown in green. The solutions without any constraints on the object radius are shown in red. The GJ~504b's photometry is overlaid as blue dots.}
  \label{fig:bestfitatmo}
\end{figure}

\subsubsection{Results}	
\label{subsubsec:resultsatmo}
We compared the photometry of GJ~504b to the grids of models using the fitting method described in Section \ref{sec:emp_analysis}. The fit is used to determined $\alpha=R^{2}/d^{2}$, with $R$ being the object radius and $d$ the target distance. We allowed the radius to vary in the range 0.82--1.26 $\mathrm{R_{Jup}}$, which corresponds to the radii predicted for the bolometric luminosity (Section \ref{sec:emp_analysis}) and absolute photometry of GJ~504b  in the Y2, Y3, J3, H2, and K1 bands  by the "hot-start" COND evolutionary models for the two age ranges estimated for the system \citep{2003A&A...402..701B}.  We also considered the case where the radius is left unconstrained in the fit. The  solutions minimizing $G$ are reported in Table~\ref{Tab:atmopar} and shown in Fig. \ref{fig:bestfitatmo}. The fitting method does not allow for detailed exploration of the degeneracies in the parameter space of the models, but it does not require any model grid re-interpolations. 


\begin{table}
\centering
\tiny
\caption{\label{Tab:atmopar} Fitting solutions corresponding to the comparison of GJ~504b photometry to atmospheric models using the G goodness-of-fit indicator. The reported masses are derived from the radius and log g.} 
\begin{tabular}{llllll}
\hline
	&	\texttt{BT-SETTL}	&Morley 	&	\texttt{ATMO} & \texttt{Exo-REM}	&	\texttt{petitCODE}	 \\
\hline
		\multicolumn{6}{c}{R constrained ("hot start" models, 19 Myr $\rightarrow$5.8 Gyr)} \\
$\mathrm{T_{eff}}$ (K)	&	550	&		575	&	550	&	 550	&	550 \\ 
log g (dex)						&	3.5		& 3.5			&	4.0	&	4.9		&	4.0		\\ 
$\mathrm{[M/H]}$ (dex)	&	0: 							&	0.0 				&	0.5			&		 0.5			&	0.2	\\ 
$\mathrm{K_{zz} (cm^{2}.s^{-1})}$			&	\dots						&		\dots						&	$10^{6}$: 				&	\dots						&	$10^{8.5}$\\ 
$\mathrm{f_{sed}}$							&	\dots						&		2.0		&	\dots						&	\dots						&		3.0 \\ 
$\mathrm{f_{cloud}}$		&	1:					&				1:					&			\dots			&		0.75	&		1:		\\	
$\mathrm{\gamma}$		&	\dots						&		\dots				&	 1.3						&		\dots					&		\dots				\\		
R ($\mathrm{R_{Jup}}$)		&	0.82					&		0.85			&	0.96					&		0.84					&		0.94	 \\ 
M ($\mathrm{M_{Jup}}$)		&			0.9			&	0.9			&	3.7					&	22.8					&		3.6	 \\ 
$\mathrm{log\:(L/L_{\odot}})$	& -6.25 & -6.14 & -6.11 & -6.23 & -6.13 \\ 
G		&	9.356					&	1.301				&	 0.820						&		1.066					&		0.553		\\
 \hline
  \multicolumn{6}{c}{R unconstrained} \\
$\mathrm{T_{eff}}$ (K)	 &	750$^{a}$	&	625$^{a}$	&	600 &	550 & 600 \\ 
log g (dex)		&	3.5 &	4.0	&	3.5 & 4.9 & 3.5	\\ 
$\mathrm{[M/H]}$ (dex)	&	0.0:	&	0.0	&	0.0 & 0.5 & 0.0	\\ 
$\mathrm{K_{zz} (cm^{2}.s^{-1})}$ &	\dots	& \dots & $10^{6}$: &	\dots & 	 $10^{8.5}$ \\	
$\mathrm{f_{sed}}$	 & \dots	&	2.0 		&	\dots		& \dots	& 2.5		\\ 
$\mathrm{f_{cloud}}$ &	1: &	1:	& \dots  & 0.75 & 1:\\ 
$\mathrm{\gamma}$	 & \dots	& \dots	& 1.3 & \dots & \dots \\ 
R ($\mathrm{R_{Jup}}$)			&	0.37 	&	0.64 & 0.79 & 0.84	& 0.78 \\	
M ($\mathrm{M_{Jup}}$)	 & 	0.2	& 1.7	& 0.8	& 22.8 &  0.8\\	
$\mathrm{log\:(L/L_{\odot}})$	 & -6.40 & -6.25 & -6.13 & -6.23 & -6.14 \\ 
G		& 1.378		& 1.165	& 0.684	& 1.066 & 0.543 \\		
\hline		
 \end{tabular}
\tablefoot{$^{a}$The fitting solutions predict a H3-band flux in disagreement with the upper limit set by the IRDIS observations.}
\end{table}
  
The \texttt{ATMO} and \texttt{petitCODE} models yield the best fit to the companion SED.  The fit converges toward implausibly small radii and higher temperatures when $\alpha$ is left unconstrained. This likely arises from the red colors of GJ~504b which are better represented by hotter atmospheres in spite of the companion's low luminosity, as shown in Section \ref{sec:emp_analysis}.  This problem is amplified when the \texttt{BT-SETTL} models are 
considered. The \texttt{BT-SETTL} fitting solutions are also unable to reproduce the upper limit in the H3 band. Those models also failed to reproduce the absolute fluxes and colors of GJ~758b \citep{2016A&A...587A..55V}.  

When the radius is allowed to vary in the interval 0.82--1.26 $\mathrm{R_{Jup}}$, the fit with the \texttt{BT-SETTL}, \texttt{Exo-REM}, and Morley models tends to converge toward lower $\mathrm{T_{eff}}$ values and the lowest radii in the interval in order to reproduce the object's low luminosity. The low radii are those expected (0.84-0.99 $\mathrm{R_{Jup}}$) for a  "hot-start" object for the old age range of the system. In such a case,  the surface gravity of objects with the observed band-to-band luminosity should be in the range 4.60-5.16 dex. Only the Exo-REM models  yield best fits for  high gravities in agreement with the "hot-start" predictions. However, the evolution of $G$ with $\mathrm{T_{eff}}$ and log g shows that the latter is poorly constrained. If we make the hypothesis of a young age for the system (see below), the COND models predict radii in the range 1.22-1.26 $\mathrm{R_{Jup}}$. That tight constraint on R sets the $\mathrm{T_{eff}}$ of the model fit in the range 450-500K. All but the BT-SETTL models reproduce the SED of GJ~504b  for higher surface gravities  (4.5-4.6 dex). Those high surface gravities are inconsistent with the COND predictions for the young age estimates (3.34-3.61 dex).  However, the relation between the age, mass, and radius also depends on the initial conditions ("warm-start" models) and  the idealized "hot-start" scenario \citep[e.g.,][]{2007ApJ...655..541M, 2013A&A...558A.113M} might not be suitable to  GJ~504b, in particular for the young-age scenario (see also Section \ref{sec:mass}). 

 We then  estimate a $\mathrm{T_{eff}=550\pm50}$K for the companion based on the values found from the fit without any pre-requisite on the radius and excluding the BT-SETTL solutions. The value is consistent with the one found by  \cite{2016ApJ...817..166S} using a subset of  photometric datapoints. We  find a $log(L/L_{\odot})=-6.10\pm0.09$ using the $\mathrm{T_{eff}}$ given in parenthesis in Table \ref{Tab:atmopar} and the radii estimated from the fit. That value is consistent within error bars with the one derived in Section \ref{sec:emp_analysis} and by \cite{2016ApJ...817..166S}.

The \texttt{Exo-REM} grids  with cloudless models (${f_{cloud}=0}$) clearly fail to reproduce the object's SED. The best fit is achieved with models considering  a nonuniform cloud coverage (75\%). This percentage of cloud coverage is consistent with that found for the young exoplanet 51 Eri b \citep{2017AJ....154...10R}. Nevertheless, the \texttt{petitCODE} synthetic spectra considering a uniform cloud cover provide the best fit of all considered models. In addition, the \texttt{ATMO} models which do consider the thermo-chemical instability as an alternative to cloud formation yield $G$ values lower than those of the Exo-REM models. Therefore, additional data are needed to comment on the occurrence of clouds in the atmosphere of GJ~504b (see Section \ref{subsec:JWST}).

\begin{figure*}
  \centering
  \includegraphics[width=14cm]{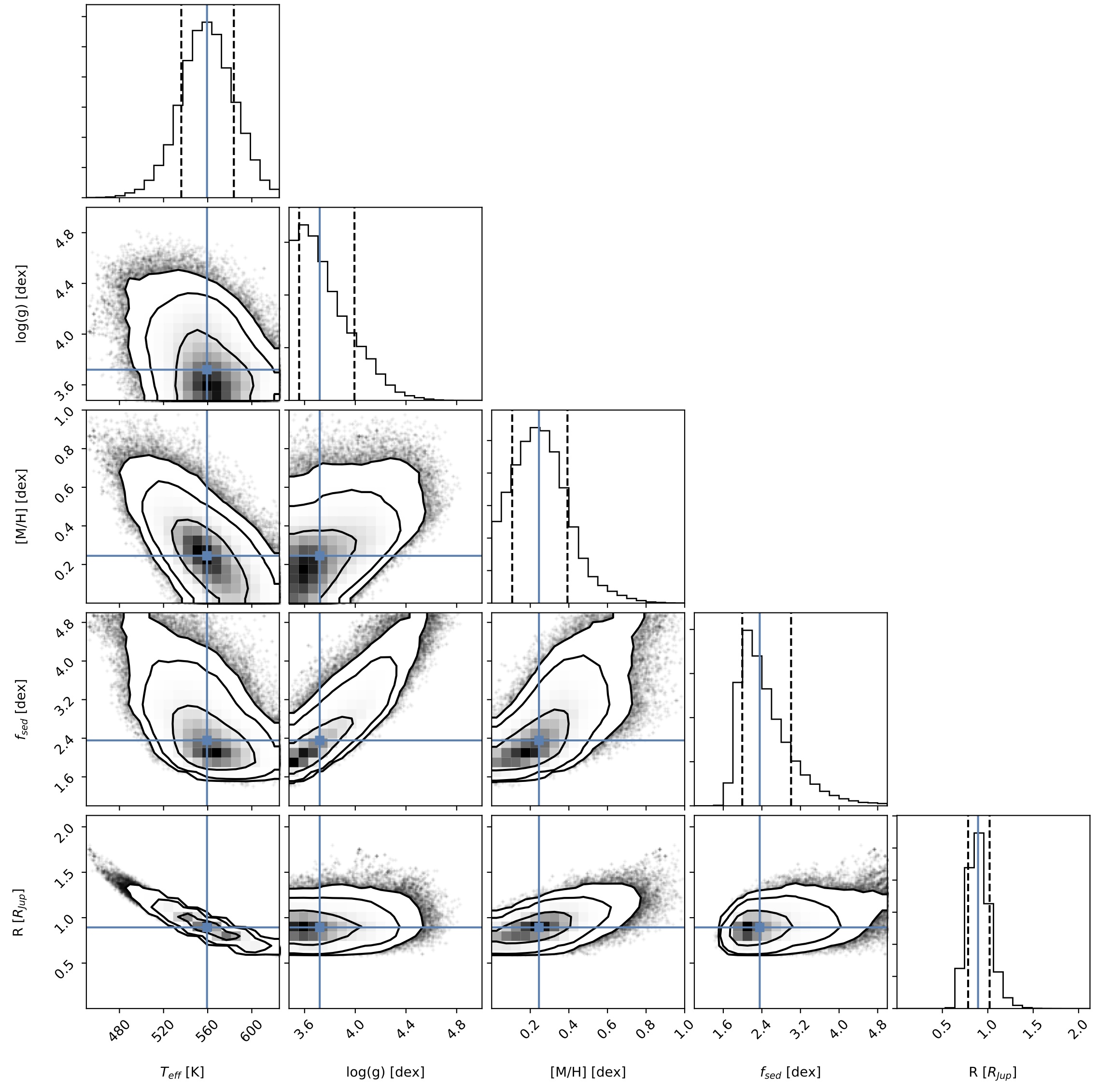}
  \caption{Posterior distributions for GJ~504b atmospheric parameters when the Morley models are considered.}
  \label{fig:MCMCmorley}
\end{figure*}

\begin{figure*}
  \centering
  \includegraphics[width=14cm]{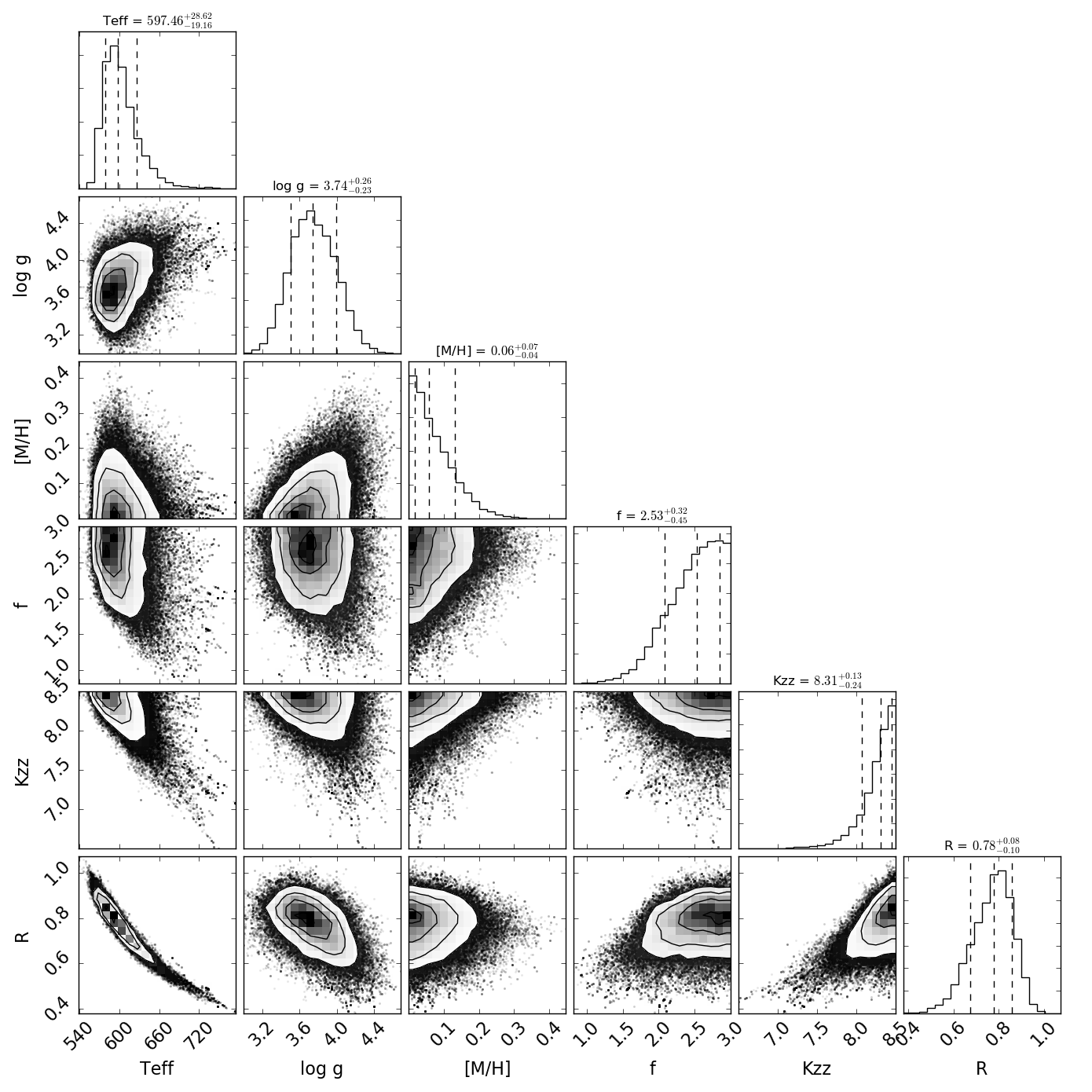}
  \caption{As in Fig. \ref{fig:MCMCmorley} but with the \texttt{petitCODE} atmospheric model used as input.}
  \label{fig:MCMCBACON}
\end{figure*}

Several indications in the fitting solution based on the $G$ statistics  confirm the peculiarity of GJ~504b  atmosphere:
\begin{itemize}

\item All but the Exo-REM models provide a best fit for low surface gravities. The evolution of G with log g indicates that this parameter is well constrained by the Morley, \texttt{ATMO}, and \texttt{petitCODE} grids. This is not the case however for the two other models.  \cite{2011ApJ...735..116B} and \cite{2015ApJ...804...92S} find surface gravities in the same range as GJ~504b for the cloudy T8 objects WISEPC J231336.41-803701.4, WISEA J032504.52-504403.0,  and ROSS 458C.  Our values are also consistent with those found for 51 Eri b \citep{2017A&A...603A..57S, 2017AJ....154...10R}. 

\item The \texttt{petitCODE} and Morley cloudy models find $\mathrm{f_{sed}}$ in the range 2--3. These values are  lower than the ones found for WISEA J032504.52-504403.0  when using models from the Santa-Cruz group \citep{2015ApJ...804...92S}.  They are higher, however, than the one derived with the \texttt{petitCode} models for 51 Eri b \citep[using the SPHERE spectrum;][]{2017A&A...603A..57S}, but are consistent with the $\mathrm{f_{sed}}$ quoted for 51 Eri b using the Morley model grid \citep{2017AJ....154...10R}. Those $\mathrm{f_{sed}}$ values are lower than those found for old late-T objects and consistent with the low surface gravities found. 

\item The \texttt{petitCODE} models favor solutions with high $\mathrm{K_{zz}}$ values ($\mathrm{10^{8.5} cm^{2}.s^{-1}}$).  $\mathrm{K_{zz}}$ enters by setting the cloud particle size (together with $\mathrm{f_{sed}}$) in \texttt{petitCODE}. The solution also corresponds to the largest $\mathrm{f_{sed}}$ values available in the grid. This can be interpreted as a need for models with reduced cloud opacity rather than intense vertical mixing.  The $\mathrm{K_{zz}}$ value of GJ~504b  is well above ($\mathrm{10^{4}-10^{6}  cm^{2}.s^{-1}}$) the one determined for the companion Wolf 940B \citep{2010ApJ...720..252L}. Wolf 940A has the same metallicity ($\mathrm{[M/H] = +0.24 \pm 0.09}$) as GJ~504A. But the Wolf 940 system is clearly old (3-10 Gyr). 

\item The best fit  with the Morley grid corresponds to a model with $\mathrm{[M/H]=0}$.  This is at odds with the conclusions from \cite{2016ApJ...817..166S} found with the same model grid. We discuss the disagreement  below.
\end{itemize}

 We explore in the following section the degeneracies between the free parameters  of the models.

	\begin{figure*}
  \centering
  \includegraphics[width=14cm]{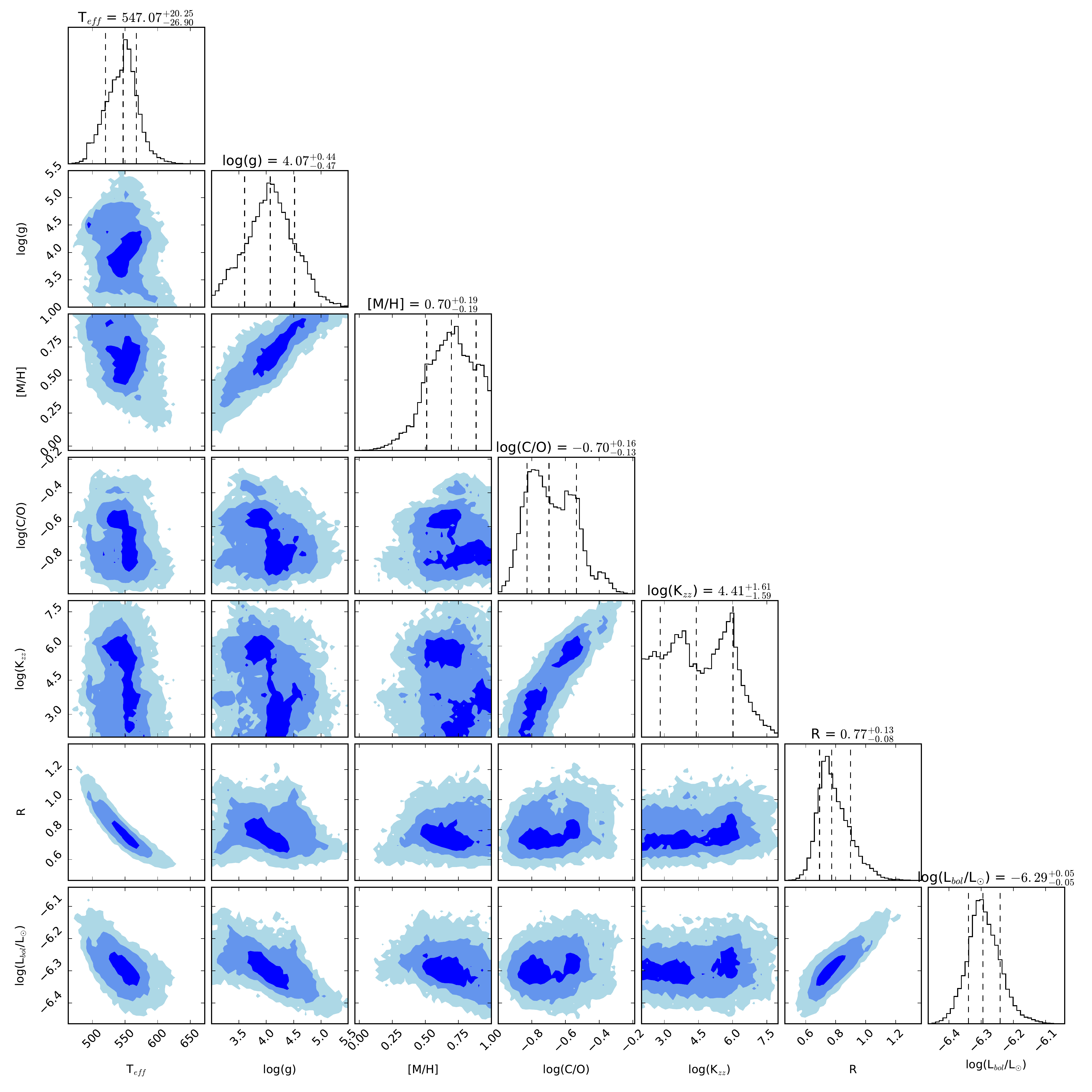}
  \caption{Posterior distribution of atmospheric parameters corresponding to the forward modeling of GJ~504b photometry with cloud-free models exploring different C/O ratios.}
  \label{fig:NSLine}
\end{figure*}

\subsection{Evaluating the degeneracies}
We ran Markov-Chain Monte-Carlo (MCMC) simulations of GJ~504b photometry for the most regular grids (Morley and \texttt{petitCODE}) of models to explore the posterior probability distribution  for each model free parameter, and to evaluate the degeneracies between the different parameters. Each datapoint was considered with an equal weight in the likelihood function. The radius is left to evolve freely during the fit. We used the python implementation of the \texttt{emcee}  package \citep{2010CAMCS...5...65G, 2013PASP..125..306F} to perform the MCMC fit of our data. The convergence of the MCMC chains is tested using the integrated autocorrelation time \citep{2010CAMCS...5...65G}. Each MCMC step required a model to be generated for a set of free parameters that was not necessarily in the original model grid.  We then performed linear re-interpolation of the grid of models in that case. 

We coupled \texttt{emcee} to the Morley grid using a custom code (Vigan et al. in prep). Upper limits are accounted for in the fit as a penalty term in the calculation of the log-likelihood: if the predicted photometry of the model in a given filter is above the upper limit set by the observations, it is taken into account in the calculation of the likelihood; if it is below, it is not taken into account.  We excluded the rained-out models  ($\mathrm{f_{sed}= +\infty}$) beforehand.  The posterior distributions are shown in Fig. \ref{fig:MCMCmorley}. We estimate (1$\sigma$ confidence level) $\mathrm{T_{eff}=559_{-24}^{+25}}$K,  $\mathrm{log\:g=3.72^{+0.27}_{-0.16}}$dex, $\mathrm{[M/H]=0.25\pm0.14}$ dex, $\mathrm{f_{sed}=2.36^{+0.65}_{-0.37}}$, and R=$0.89^{+0.13}_{-0.11}R_{Jup}$. The solution is in good agreement with the one found with the G statistics when R is constrained. The posteriors on $\mathrm{T_{eff}}$,  log g, and $\mathrm{f_{sed}}$  are quite similar to those reported in \cite{2016ApJ...817..166S} using a close MCMC approach and the same model grid. We nonetheless find a lower metallicity. Our value is in excellent agreement with the one determined for GJ~504A. This parameter is correlated with the $\mathrm{T_{eff}}$ and $R$. \cite{2016ApJ...817..166S}  set priors on $R$ corresponding to a range of radii predicted by the "hot-start" evolutionary models.  Adopting a flat prior on the radius in the range 0.82--1.26 $\mathrm{R_{Jup}}$  (see Section \ref{subsubsec:resultsatmo}) does not modify our posteriors significantly. We find  $\mathrm{T_{eff}=552_{-20}^{+16}}$K,  $\mathrm{log\:g=3.72^{+0.28}_{-0.17}}$dex, $\mathrm{[M/H]=0.27^{+0.14}_{-0.13}}$ dex, $\mathrm{f_{sed}=2.40^{+0.66}_{-0.38}}$, and R=$0.93^{+0.11}_{-0.07}R_{Jup}$.  The analysis does not alleviate the correlation between the $\mathrm{f_{sed}}$ and log g values. The radius is more consistent with those of old brown dwarfs. The luminosity is in good agreement with the one determined empirically.

The BACON code used in  \cite{2017A&A...603A..57S} couples the \texttt{petitCODE} grids of models to  \texttt{emcee}.  BACON has been validated on the benchmark T-type companions Gl 570D and HD 3651B  \citep{2017A&A...603A..57S}. We used it on GJ~504b photometry. The posterior distributions are shown in Fig. \ref{fig:MCMCBACON} and confirm the fitting solutions with the G statistics when R is unconstrained. However most of the solutions are  found for unphysical radii which are highly correlated to $\mathrm{T_{eff}}$. Moreover  the $\mathrm{[M/H]}$ determination is degenerate with the cloud parameters ($\mathrm{K_{zz}}$ and $\mathrm{f_{sed}}$).  The posteriors on $\mathrm{[M/H]}$ might be extended to higher values if  the grids of models were created for higher $\mathrm{K_{zz}}$ and $\mathrm{f_{sed}}$ values, as it is the case (for $\mathrm{f_{sed}}$) in the Morley grid. The upper limits were not taken into account in the fit.

GJ~504A has a C/O ratio\footnote{estimated from the abundances reported in Table 4 of \cite{2017A&A...598A..19D}.} of $0.56^{+0.26}_{-0.18}$, close to the  value for the Sun \citep[$\mathrm{C/O_{\odot} = 0.55 \pm 0.10}$;][]{2009ARA&A..47..481A, 2008A&A...488.1031C}. The atmospheric models used for GJ~504b  assume a solar C/O value. Nevertheless, this might not be the case if GJ~504b formed in a disk \citep[see][]{2011ApJ...743L..16O, 2016ApJ...831L..19O}. In such a case, one needs to investigate how a different C/O ratio  could bias the atmospheric parameter determination. Atmospheric retrieval is a powerful method to estimate the abundances of individual molecules carrying C and O.  We attempted a retrieval of the abundances of H$_{2}$O, CO$_{2}$, CO, and CH$_{4}$ with the HELIOS-R  \citep{2017AJ....154...91L} and NEMESIS \citep{Irwin2008} codes.  We obtained flat distributions because of the limited number of photometric data points used as inputs and the uncertainties on the data.

\begin{figure}
  \centering
  \includegraphics[width=\columnwidth]{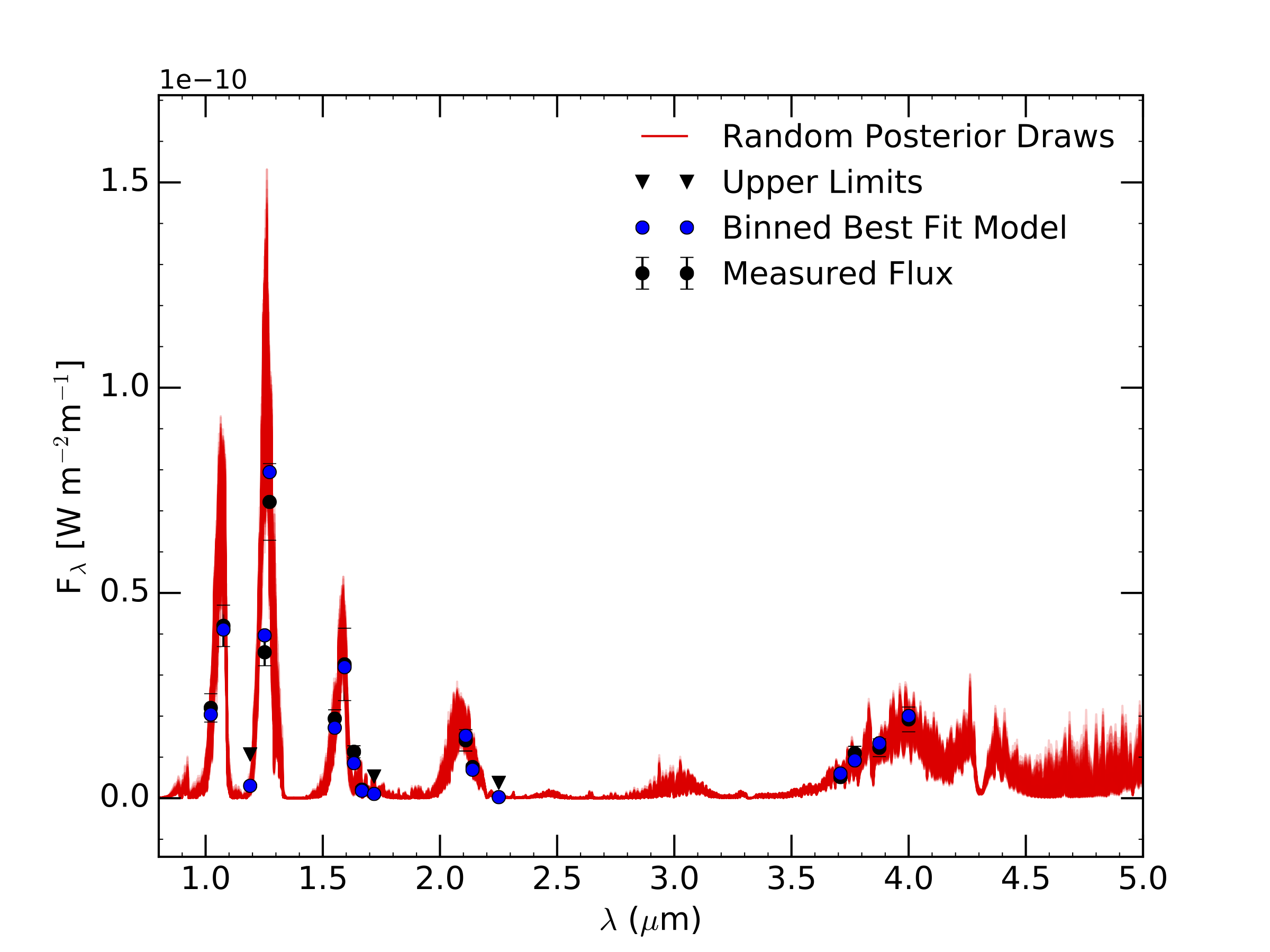}
  \caption{Best-fitting spectrum found with the  forward modeling of GJ~504b SED with cloud-free models exploring the effect of different C/O ratios. }
  \label{fig:BestfitLine}
\end{figure}

	We then considered  a grid of forward cloud-free models (see Appendix \ref{App:D} for the details) exploring different C/O ratios  in addition to $\mathrm{T_{eff}}$, log g, $\mathrm{[M/H]}$, $\mathbf{K_{zz}}$, and R. We used the \texttt{MULTINEST} Bayesian inference tool \citep{2009MNRAS.398.1601F} which implements the Nested Sampling method \citep{Skilling06}. \texttt{MULTINEST} allows for an efficient sampling of multimodal posterior distributions and avoids the convergence issues that can arise in MCMC runs. The upper limits were taken into account using the method of \cite{2012PASP..124.1208S}. We report the posterior distributions in Fig. \ref{fig:NSLine} and the best-fitting spectrum in Fig. \ref{fig:BestfitLine}. The posteriors yield constraints on the $\mathrm{T_{eff}}$ and log g values  which are compatible with those inferred from the model grids not accounting for nonsolar C/O.  The metallicity distribution  points toward values compatible with those reported in \cite{2016ApJ...817..166S}. The C/O ratio is below solar (C/O=$0.20^{+0.09}_{-0.06}$) and not correlated with the $\mathrm{[M/H]}$ value. However we find a strong correlation with the $K_{zz}$ values which is loosely constrained, but points toward lower values than those inferred with other atmospheric models. We refrained from using the C/O ratio value to discuss the formation mode of GJ~504b since our estimate does not account for possible model-to-model uncertainties.

In summary, the Bayesian analysis confirms the $\mathrm{T_{eff}=550\pm50}$ K  found in Section \ref{subsubsec:resultsatmo}. We adopt this value in the following analysis. We do not reproduce the posterior distribution on [M/H] found by \cite{2016ApJ...817..166S} with the full set of photometric points, or restraining the fit to the subset of data used in  \cite{2016ApJ...817..166S}. The metallicity determination is limited by model-to-model systematic error and degeneracies with the cloud properties and log g. The different [M/H] values may be due in part to the prior choices and the reference solar abundances considered in each model\footnote{The \texttt{petitCODE}  models consider the reference solar  abundances reported in \cite{2009ARA&A..47..481A} while  the Morley models consider those of \cite{2010ASSP...16..379L}. There are some notable differences in the two sets of reference abundances, in particular for C, Mg, and Fe.} and/or to the way the clouds are handled. The posteriors points toward a low surface gravity in agreement with the young-age scenario. Nevertheless, the log g determination is degenerate with  [M/H] and the cloud properties (for models with clouds). The C/O ratio can be determined accurately for cold objects such as GJ~504b using the forward modeling approach. It does not seem to affect the other parameter determination considered for the demonstration ([M/H], log g, $\mathrm{T_{eff}}$). However, a more robust determination could be achieved with additional datapoints (or spectra) and better accounting for model-to-model uncertainties.  

\begin{figure}
  \centering
  \includegraphics[width=\columnwidth]{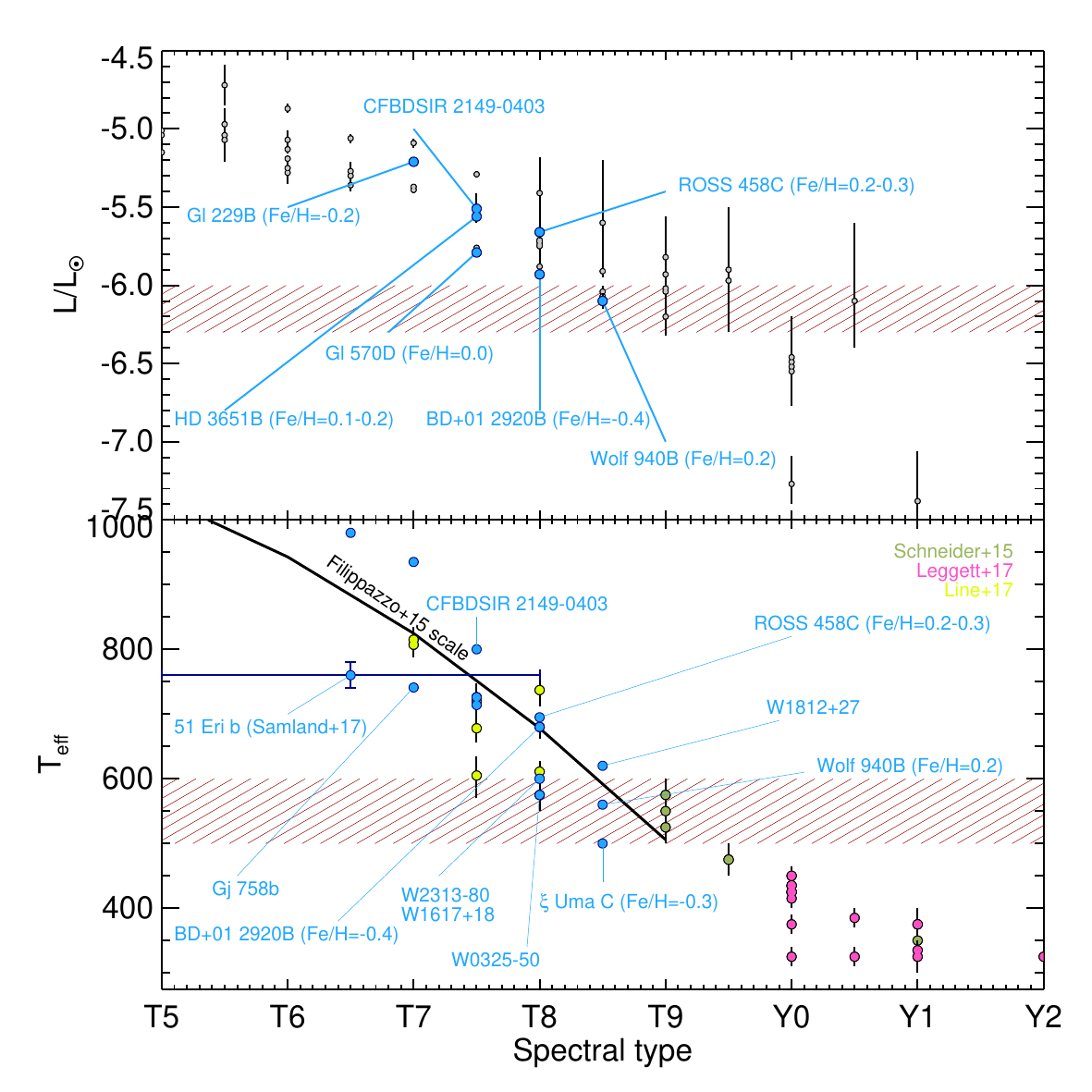}
  \caption{Comparison of the final $\mathrm{T_{eff}}$ and bolometric luminosity of GJ~504b (dashed zone) to those of late-T and early-Y dwarfs. The bolometric luminosity values are taken from \cite{2013Sci...341.1492D} and \cite{2017A&A...602A..82D}. The temperatures and luminosity of benchmark companions are taken from Table \ref{Tab:AppB}. We added the $\mathrm{T_{eff}}$ determined by \cite{2017ApJ...842..118L},  \cite{2017ApJ...848...83L}, and \cite{2015ApJ...804...92S} using atmospheric models and report the $\mathrm{T_{eff}}$/spectral type conversion scale of \cite{2015ApJ...810..158F}.}
  \label{fig:finalTeffLum}
\end{figure}

We adopt a $\mathrm{log(L/L_{\odot})=-6.15\pm0.15}$ for GJ~504b based on the values derived from the empirical analysis and confirmed by  various modelings with synthetic spectra. Both the $\mathrm{T_{eff}}$ and luminosity estimates are in good agreement with  those of T8-T9.5 dwarfs (Fig. \ref{fig:finalTeffLum}).

\section{Mass estimates}
\label{sec:mass}
\begin{figure}
  \centering
  \includegraphics[width=\columnwidth]{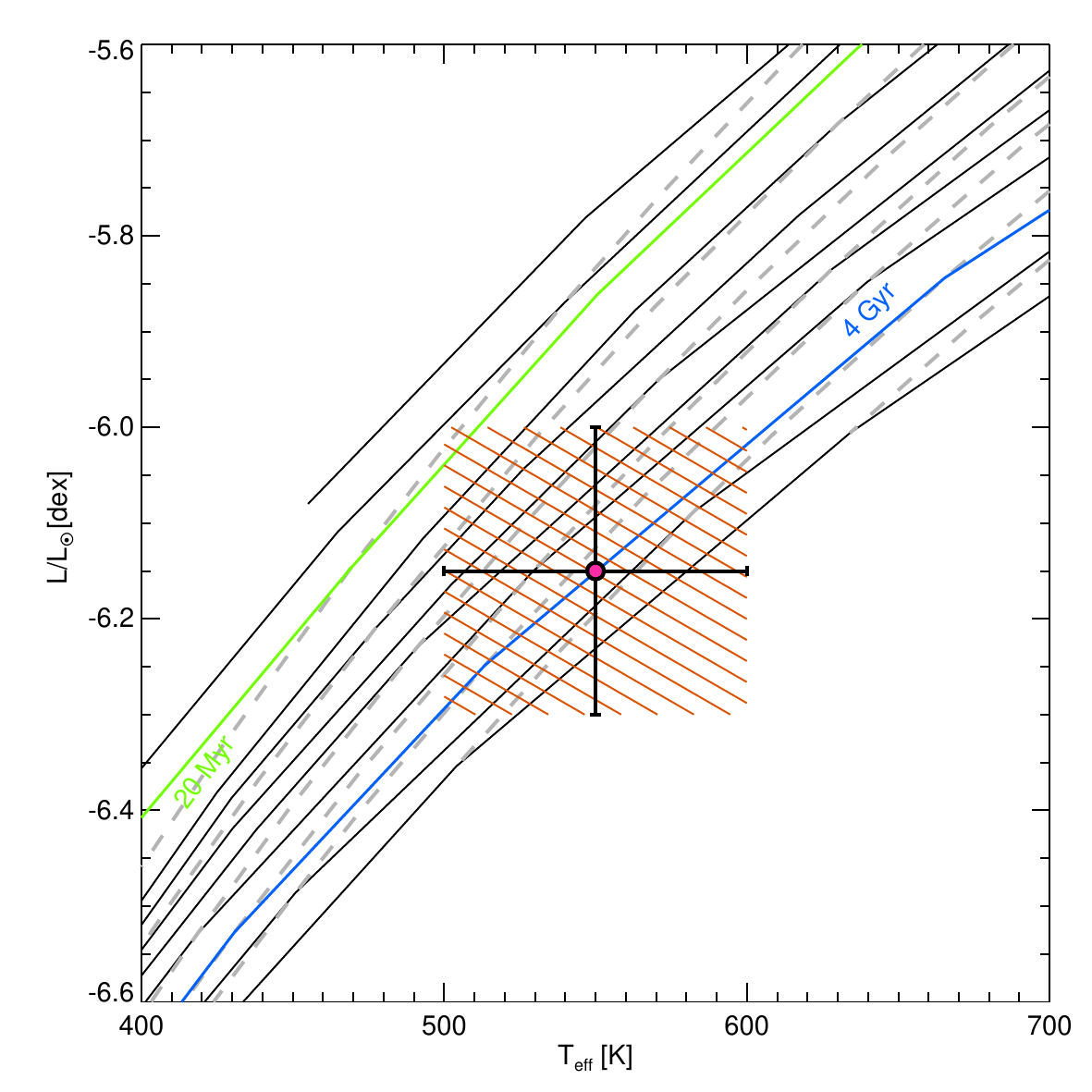}
  \caption{Luminosity and $\mathrm{T_{eff}}$ of GJ~504b compared to the COND03 ("hot-start") evolutionary tracks. The solid lines correspond to the 5, 10, 20, 100, 300, 600 Myr and 1, 2, 4, 6, and 10 Gyr isochrones (from top to bottom). The dashed lines correspond to the model predictions for masses of 1, 5, 10, 15, 20, 30, and 40 $M_{Jup}$ (from top to bottom).}
  \label{fig:hotstartlumteff}
\end{figure}

Table \ref{Tab:masses} reports the masses predicted by the "hot-start" COND models \citep{2003A&A...402..701B}. The masses predicted from the temperature and luminosity agree with each other. The object falls onto the 4 Gyr isochrone in Fig. \ref{fig:hotstartlumteff}. The 20 Myr isochrone is marginally consistent with the object properties.  Conversely, the predicted surface gravities at 21 Myr are in better agreement with those found with the \texttt{BT-SETTL}, \texttt{petitCODE}, \texttt{ATMO}, and Morley atmospheric models, but this parameter can be affected by the degeneracies of the atmospheric model fits discussed above. 

\begin{table*}
\centering
\tiny
\caption{\label{Tab:masses} "Hot-start" evolutionary model predictions}
\begin{tabular}{ll|ccccc|ccccc}
		&				&	\multicolumn{5}{c}{Saumon \& Marley 2008 - no cloud - $[$M/H$]$=+0.3} &	\multicolumn{5}{|c}{COND03 - cloud free - 1x solar}  \\
		\hline
Age	&	Input		&	Mass 	&	R &  $\mathrm{T_{eff}}$	&	log g	 &	 $\mathrm{log(L/L_{\odot})}$ &	Mass 	&	R &  $\mathrm{T_{eff}}$	&	log g	 &	$\mathrm{log(L/L_{\odot})}$  \\
(Gyr)	&				&	$\mathrm{(M_{Jup})}$	 & $\mathrm{(R_{Jup})}$	&	(K)	&	(dex)	& (dex)	&	$\mathrm{(M_{Jup})}$	 & $\mathrm{(R_{Jup})}$	&	(K)	&	(dex) & (dex) \\	
\hline
$0.021\pm0.002$	&	$\mathrm{T_{eff}}$	&	\dots	&	\dots &		\dots & \dots & \dots & $2.5^{+0.6}_{-0.5}$ & $1.24\pm0.01$ & \dots	&	 $3.61\pm0.09$ 	&	$-5.87^{+0.6}_{-0.5}$ \\
$0.021\pm0.002$	&$\mathrm{log(L/L_{\odot})}$	&\dots	&	\dots &	\dots & \dots &\dots & $1.7^{+0.5}_{-0.4}$ & $1.23\pm0.01$& $470^{+43}_{-40}$ & $3.45^{+0.11}_{-0.10}$& \dots \\
\hline		
$4\pm1.8$	&	$\mathrm{T_{eff}}$	&	$23.8^{+7.5}_{-8.1}$	&	$0.94^{+0.07}_{-0.05}$ & \dots & $4.84^{+0.17}_{-0.24}$ &$-6.11\pm0.18$ &  $23.5^{+8.8}_{-6.2}$ & $0.94^{+0.05}_{-0.06}$ & \dots &   $4.83^{+0.20}_{-0.17}$  & $-6.15^{+0.16}_{-0.18}$\\
$4\pm1.8$	&	$\mathrm{log(L/L_{\odot})}$	&	$22.9^{+8.7}_{-8.6}$	&	$0.95^{+0.08}_{-0.06}$& $537^{+68}_{-64}$ & $4.82^{+0.19}_{-0.27}$ &  \dots &$23.5^{+10.2}_{-6.7}$  & $0.94^{+0.05}_{-0.06}$ & $550^{+69}_{-59}$ &  $4.83^{+0.22}_{-0.18}$ & \dots \\
\hline		
 \end{tabular}
\end{table*}

We also report the "hot-start" model predictions for the \cite{2008ApJ...689.1327S} models which account for metal-enriched atmospheres as boundary conditions. The predictions are consistent with those of the COND models for the old age range\footnote{The models do not  make predictions for masses lower than 2 $M_{Jup}$. Therefore, we could not  estimate masses for the young isochronal age interval of GJ~504A.}.

\begin{figure}
\begin{center}
 \includegraphics[width=\columnwidth]{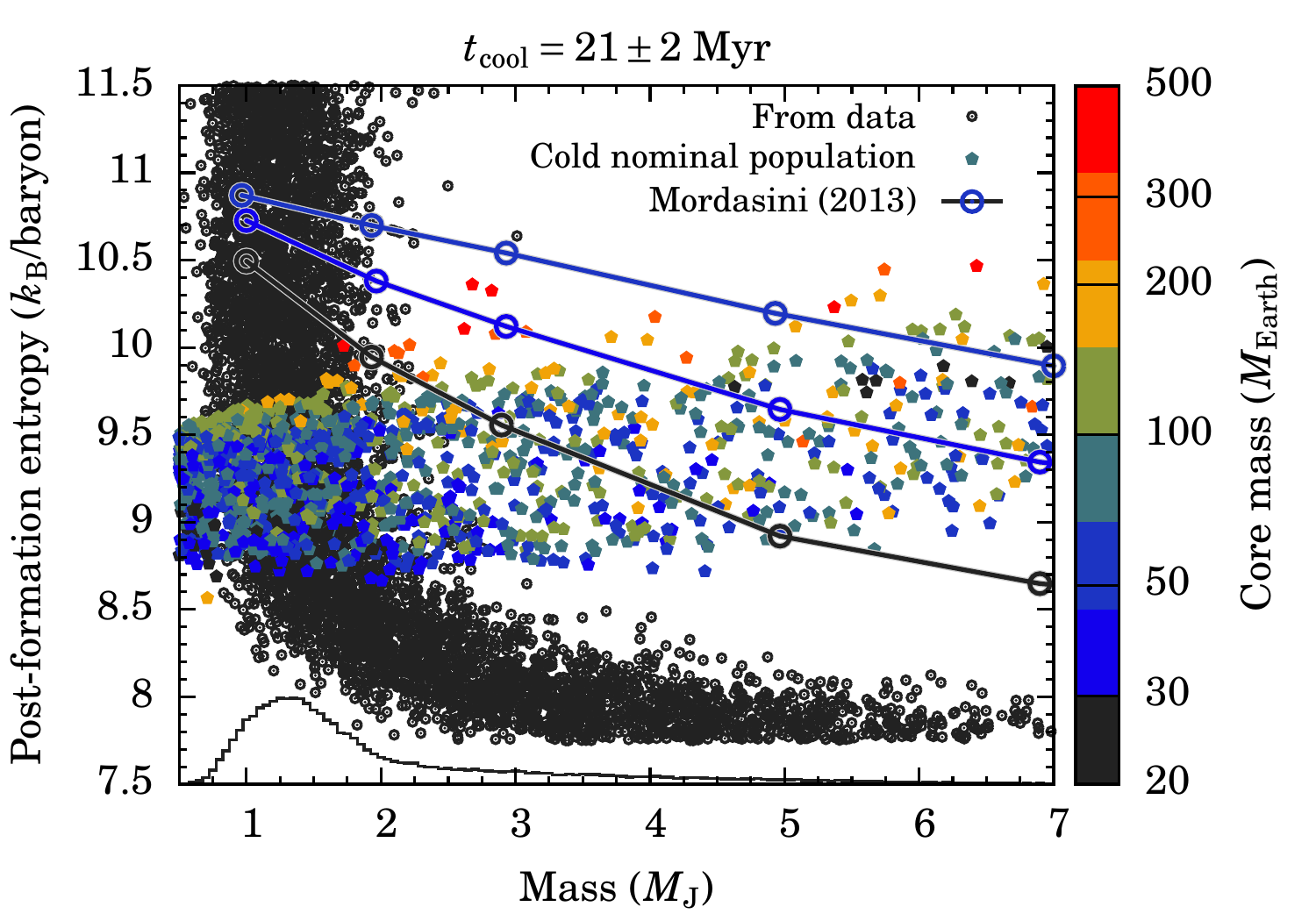}
\end{center}
\caption{
Constraints on the mass and post-formation entropy $\Sinit$ of \gjb\ for
a (cooling) age $\tKuehl=21\pm2$~Myr.
The concave swarm of black points (small open circles) shows all combinations consistent with
the luminosity measurement of $\log L/\Lsun=-6.15$,
following the approach described in detail in \citet{2014MNRAS.437.1378M}
but with an MCMC as in \citet{2014A&A...562A.111B,2014A&A...567L...9B}
and using the \citet{2012ApJ...745..174S} models.
The band of colored symbols (filled pentagons) displays the entropy at the time of disk dispersal
for the cold nominal population of \citet{2017arXiv170800868M}, that is, assuming full radiative losses
at the shock but taking the core-mass effect \citep{2013A&A...558A.113M} into account.
The logarithmic colorscale indicates the core mass $\Mc$.
Shown are also the results of \citet{2013A&A...558A.113M} for core masses of 20, 33, and 49~$\MEarth$
(large open circles connected by lines; bottom to top).
The curve at the bottom of the plot is the marginalized posterior on the mass
for all small black $M$--$\Sinit$ points (without taking the synthesis results into account).
%
}
\label{Abb:MSi von L}
\end{figure}

If GJ~504 is a 21 Myr-old system, the mass predicted by the evolutionary models should be sensitive to the way the companion accreted its forming material  \citep{2007ApJ...655..541M} and to the amount of heavy elements it contains \citep{2013A&A...558A.113M}.  We show in Fig.~\ref{Abb:MSi von L} the joint constraints on the mass and the initial entropy $\Sinit$ of \gjb\ imposed by the bolometric luminosity
for an age of $21\pm2$~Myr (cf.\ \citealp{2014MNRAS.437.1378M}). 

We find that from the luminosity measurement alone, a wide range of masses is possible, from 0.7~$\MJ$ upwards. If we truncate the posterior distribution at 2.5~$\MJ$, we obtain a marginalized 68.3\% confidence interval on the mass of $M=1.3^{+0.6}_{-0.3}~\MJ$ and $M=1.3^{+1}_{-0.6}~\MJ$ at 90\%.
Clearly, higher masses than what is shown here would be found to be consistent with the measurement if the \citet{2012ApJ...745..174S} grid went down to lower initial entropies. 

The locus of possible $M$--$\Sinit$ combinations can however be compared to  planet population synthesis predictions
to derive tighter constraints on both mass and post-formation entropy. While in our core-accretion models no planets are found at the same location in the $a$--$M$ plane as \gjb~(see Fig.~\ref{fig:CA} and Section \ref{Subsec:pathway}), the $M$--$\Sinit$ relation (with its scatter) is relatively universal. We verified this by comparing the output of the population syntheses of \citet{2017arXiv170800868M}, computed for a solar-mass star, to simulations with stellar masses of 1.5 and 2~$\mathrm{M_{\odot}}$ and different migration and planetary growth prescriptions, resulting in very different final $a$--$M$ distributions; the $M$--$\Sinit$ relation in all cases was similar, only with varying amounts of scatter in $\Sinit$ at a given planet mass, which in turn reflects the physics of the core growth.

Comparing the two sets of points in Fig.~\ref{Abb:MSi von L} (inferred from data
and predicted from formation models), it is clear that if \gjb\ formed through
standard core accretion as represented by the ``cold nominal'' population of \citet{2017arXiv170800868M},
its post-formation entropy is 8.7--$8.6<\Sinit<9.6$--9.8 in units of $\mathrm{\kB/\textrm{baryon}}$,
with the bounds slightly depending on the stellar mass (from low to high, respectively). This a priori on $\Sinit$ leads to  $\mathrm{M=1.3\pm0.4~\MJ}$. 

\section{Architecture}
\label{Section:archi}
\subsection{Companion orbit}
\label{Section:orbit}

 \begin{figure*}
  \centering
  \includegraphics[width=18cm]{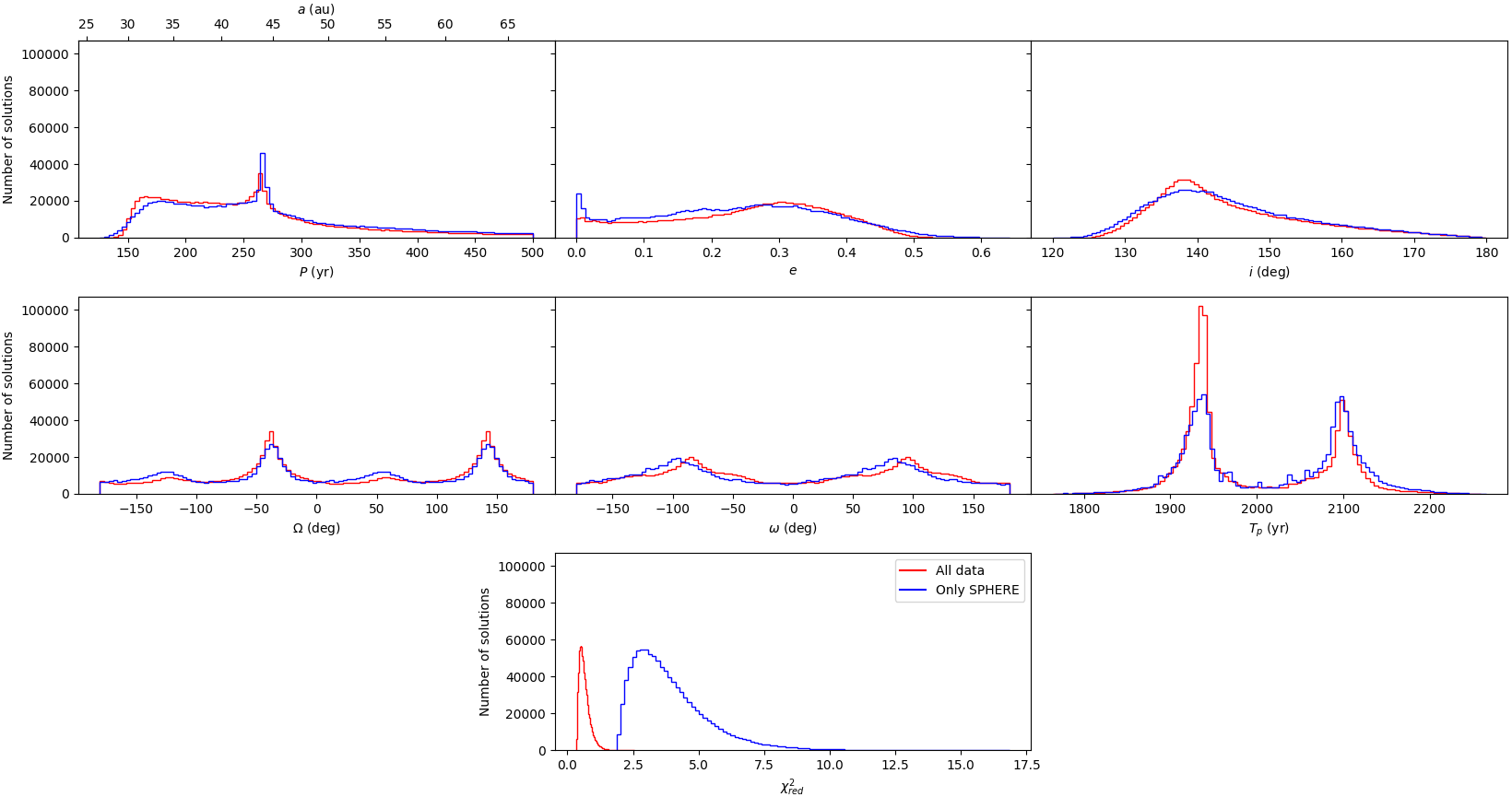}
  \caption{Posterior distributions on the orbital parameters of GJ~504b using all the astrometric epochs as input (in red) or the SPHERE epochs only (in blue).}
  \label{fig:MCMC_orbit_all}
\end{figure*}

 \begin{figure*}
  \centering
    \includegraphics[width=14cm]{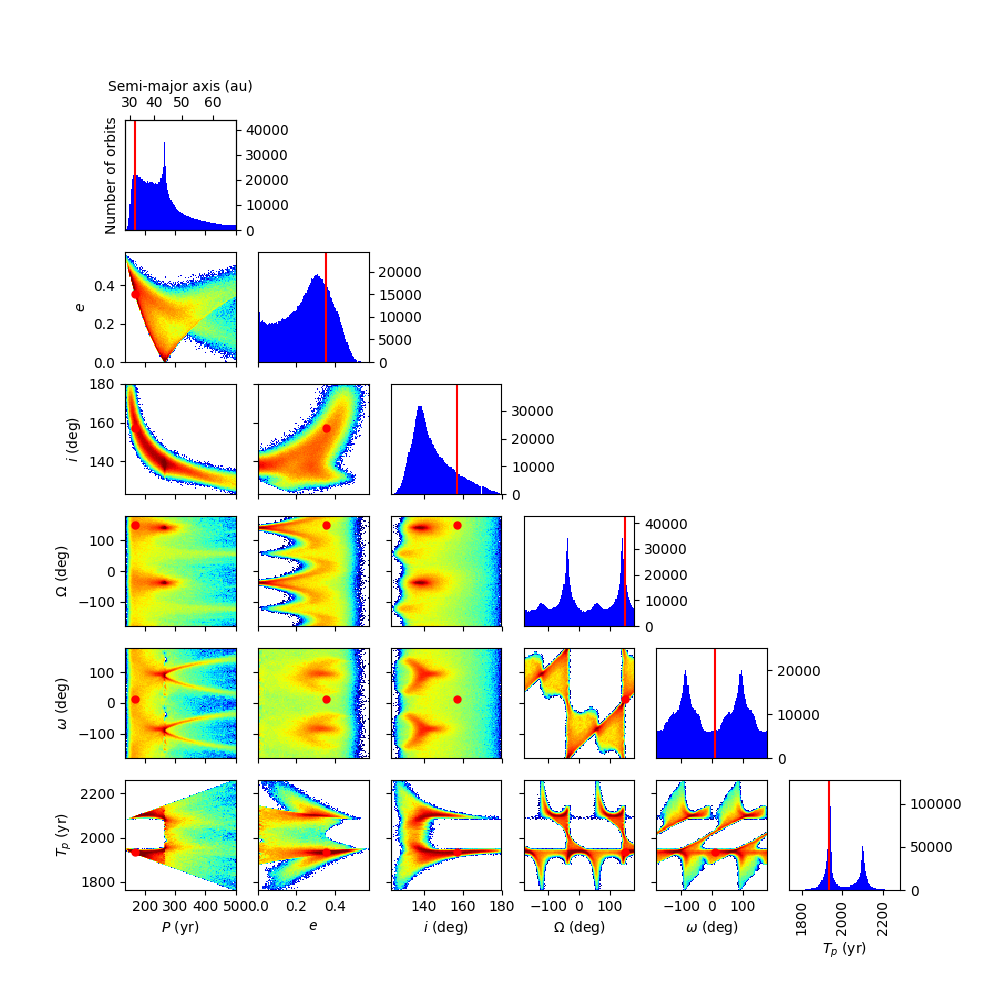} 
  \caption{Posteriors on the orbital elements of GJ~504b when considering the astrometry from 2011 to 2017. The red lines and dots depict the best fitting orbit (better $\chi^2$). The color scale is logarithmic, blue corresponds to 1 orbit and red to 1000.}
  \label{fig:MCMC_GJ504borbit_all_cor}
\end{figure*}

 \begin{figure*}
  \centering
\begin{tabular}{cc}
    \includegraphics[width=9cm]{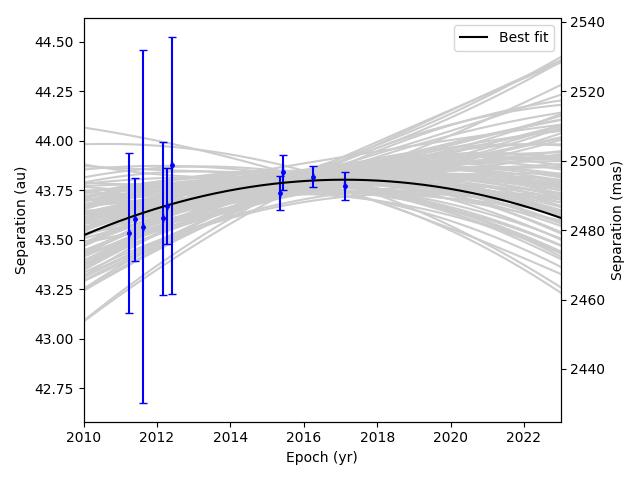}  &     \includegraphics[width=9cm]{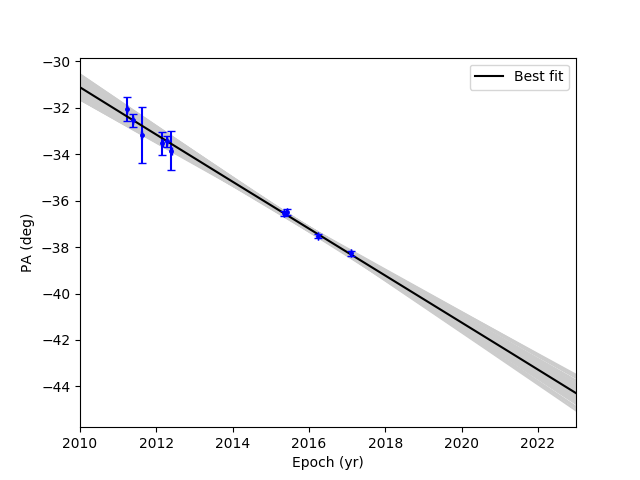}  \\
\end{tabular} 
  \caption{Sample of 100 orbits obtained with the MCMC algorithm applied to GJ~504b astrometry (blue points).}
  \label{fig:GJ504orbitvis}
\end{figure*}

We considered the astrometry reported in Table \ref{Tab:Astrometry} as input of our MCMC orbit fitting packages to set constraints on the orbital parameters of GJ~504b. The code was developed for $\beta$ Pictoris b and Fomalhaut b’s orbits \citep{2012A&A...542A..41C}. We considered a mass of 1.2 $M_{\odot}$  for GJ~504A. We assume  flat priors on  log(P), $e$ , cos(i) , $\Omega + \omega$ , $\omega - \Omega$; and $T_{p}$ following \cite{2006ApJ...642..505F}. We ran ten chains in parallel and used the Gelman-Rubin statistics as convergence criterion \cite[see details in][]{2006ApJ...642..505F}. 

The fit was performed on the whole set of epochs. We neglected the epoch from August 15, 2011 reported in \cite{Kuzuhara2013} for which the data were taken under poor conditions and the astrometry appears to be deviant. However, it is still possible that some systematic angular offsets between each instrument could have biased our analysis. We then also modeled  the SPHERE epochs only, for comparison. The posteriors are shown in Figure \ref{fig:MCMC_orbit_all} for the two data sets.  Figure \ref{fig:MCMC_GJ504borbit_all_cor} shows the correlation between the  different posterior distributions of orbital parameters of GJ~504b when all the astrometric epochs are accounted for in the fit. 

The posterior distributions do not change significantly when considering the homogeneous SPHERE data, or the data from all instruments. The accuracy of the SPHERE astrometry yields the most constraints on the orbital parameters and is therefore not heavily influenced by putative systematic errors on the  HiCIAO and IRCS astrometry. We therefore considered the results from the whole set of epochs in the following. A sample of corresponding orbits is shown in Figure \ref{fig:GJ504orbitvis}. This shows that no curvature can be detected with the present astrometric monitoring.

The posterior on the semi-major axis points at 44 au which corresponds to the  companion projected separation with 68\% of the solutions in the range $44 \pm 11$ au. The fit excludes orbits with a semi-major axis shorter than $\sim$27.8 au. The periods are significantly longer than the time span of the Lick and SOPHIE radial velocities and are likely to prevent us from obtaining constraints on the dynamical mass of GJ~504b.\\

The eccentricity is lower than 0.55 and peaks at 0.31 ($e=0.31\pm0.15$; 68\% solutions). Our new data and fit do not yield solutions at higher eccentricity found by \cite{Kuzuhara2013}. We find an inclination of $137.8^{+12.9}_{-4.6}$ degrees. There is no solution for $i<120^{\circ}$ as found by \cite{Kuzuhara2013}\footnote{We consider that by definition our inclination is larger than 90$^\circ$, since the planet is in a retrograde (i.e., clockwise) orbit.}, but that might be related to our priors which favor small semi-major axis and large inclinations. 

We ran the same analysis considering masses of 1.10 and 1.25 M$_{\odot}$ for GJ~504A. The posteriors  are marginally affected by this change.

\subsection{A spin-orbit misalignment?}
\label{subsec:stellobli}
The radius $R_{\bigstar}$  of GJ~504A  determined  in Section \ref{subsec:interf}  can be used to derive the line-of-sight inclination of the star $i_{\bigstar}$ following:

\begin{equation}
i_{\bigstar}=sin^{-1}\Big(\frac{v_{p} \times P_{\bigstar}}{2\pi \times R_{\bigstar}}\Big)
\end{equation} 

where  $v_{p}=v \cdot sin\:i$ and  $P_{\bigstar}$ is the rotation period. We measure $v_{p}=6.5\pm1.0km.s^{-1}$ from the set of SOPHIE data.  This value is consistent with the one reported in \cite{2017A&A...598A..19D}. We used $P_{\bigstar}=3.33^{+0.08}_{-0.10}$ \citep{1996ApJ...466..384D}

We considered Gaussian  distributions on each measurement to find a resulting probability ditribution corresponding to  $i_{\bigstar}=162.4_{-4.3}^{+3.8}$ degrees or $18.6_{-3.8}^{+4.3}$ degrees. The two values are due to the $[0,\pi]$ degeneracy of the $sin$ function. 

The posterior on the orbit inclination $i_{c}$ of GJ~504b inferred from the MCMC analysis (Section \ref{Section:orbit}) can be used to derive the relative orientation of the stellar spin axis and orbital angular momentum vector $\varphi$, or true obliquity:

\begin{equation}
\varphi = cos^{-1}\Big( cos\:i_{\bigstar}\:cos\:i_{p} + sin\:i_{\bigstar}sin\:i_{p}\:cos\:\lambda\Big)
\end{equation}

where $\lambda$ is the  projected obliquity\footnote{A sketch representing $\varphi$ and $\lambda$ can be found in Fig 3 of \cite{2015ApJ...814...67A}}. In our case, $\lambda$ is unknown, but as explained in \cite{2017AJ....154..165B}, the lower limit on $\varphi$ can be inferred from the absolute difference between $i_{c}$ and $i_{\bigstar}$: $\varphi \geq  \Delta i \equiv \vert i_{\bigstar} -  i_{c} \vert$. Therefore, a system with a posterior probability function on $\Delta i$ extending to $0^{\circ}$ can still have a non-zero true obliquity, and therefore a spin-orbit misalignment. 

\begin{figure}
\centering
\begin{tabular}{c}
\includegraphics[width=\columnwidth]{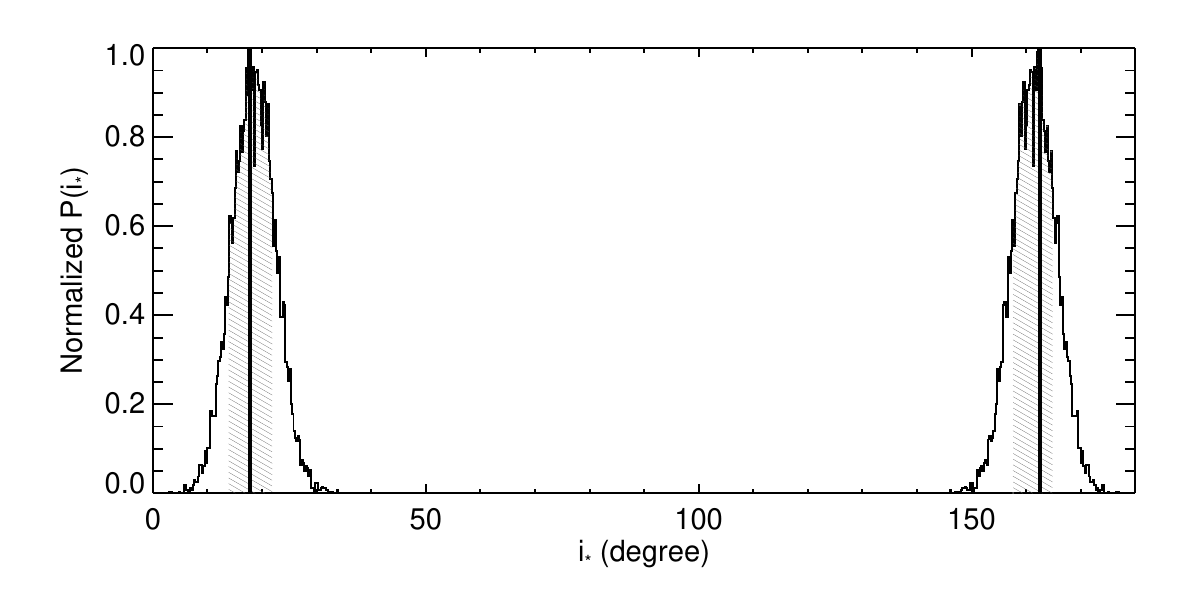} \\
\includegraphics[width=\columnwidth]{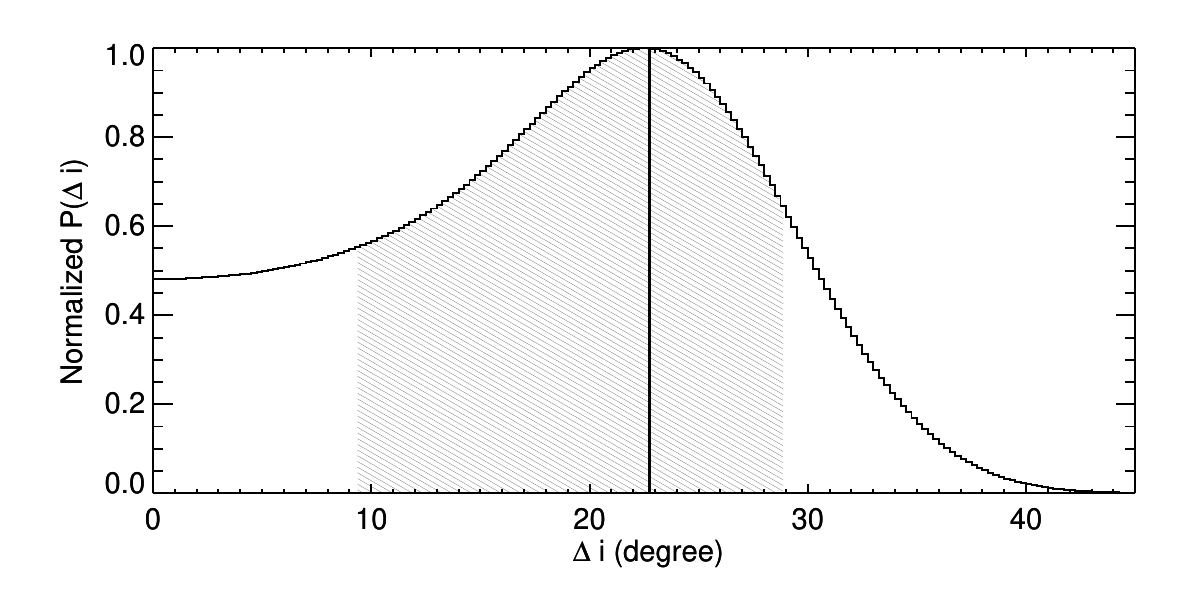} 
\end{tabular}
 \caption{From top to bottom: line-of-sight inclination $i_{\bigstar}$ of GJ~504, and absolute difference between $i_{c}$ and $i_{\bigstar}$ when only $i_{\bigstar}\ge90^{\circ}$ are considered. The dashed zones correspond to 68.28\% of the solutions.} 
\label{fig: obliq}
\end{figure}

We show in Fig. \ref{fig: obliq} the posteriors on $i_{\bigstar}$ and  $\Delta i $. The probability that $\Delta i$ is greater than $10^{\circ}$ is 78.1\%. This probability is conservative given that our priors on the orbit fit of GJ~504b favor large inclinations. In addition, this represents the minimum values of $\varphi$ in this system. Therefore, GJ~504A and b may have a spin-orbit misalignment. Further astrometric monitoring will help to consolidate this result.  

\subsection{Constraints on additional companions}
\label{subsec:addcomp}
The joint use of the radial velocities (RVs) of GJ~504A and contrast limits at each on-sky projected separation inferred from direct imaging data taken at multiple epochs allows  limits to be placed on the mass of undetected companions from the star up to semi-major axis corresponding to the field-of-view coverage of the imaging cameras. 

The \texttt{MESS2} Monte Carlo simulation code generates synthetic planet populations and compares their  RV signals and projected separation  at each epoch to the data \citep{2017A&A...603A..54L} to evaluate detection probabilities. We applied \texttt{MESS2} to RV data obtained with SOPHIE (listed in Appendix \ref{App:F}) and at the Lick Observatory independently.   Potential offsets between the two sets of data may indeed exist because of the different methods used to derive the RVs.  

We used the  local power analysis \citep[LPA; ][]{2012A&A...545A..87M} to interpret these RV data. The LPA method generates periodograms of RV time series for each synthetic planet and compares them to the periodogram of the observed RV data within given orbital periods. We used the IRDIS detection limits obtained from May 2015 to February 2017 that we converted to masses, and those derived from the IFS data from March 2016 and June 2015. We considered, in addition,  the detection limits inferred from  the HiCIAO and IRCS data obtained as part of the SEEDS survey (March 26 	and May 22, 2011; April 12 and May 25, 2012 data) and reprocessed for this study using the ACORNS pipeline \citep{2013ApJ...764..183B}. Those datasets allow for an improved coverage of the orbit of putative companions. All the detection limits inferred from the imaging data assume "hot-start" formation conditions \citep{2003A&A...402..701B}. The predictions should indeed not be heavily sensitive to the initial conditions at an age of 4 Gyr. In addition, the difference in luminosity  predictions  between the "hot" and "cold" stars tends to decrease with the planet mass. 

\begin{figure*}
  \centering
  \includegraphics[width=16cm]{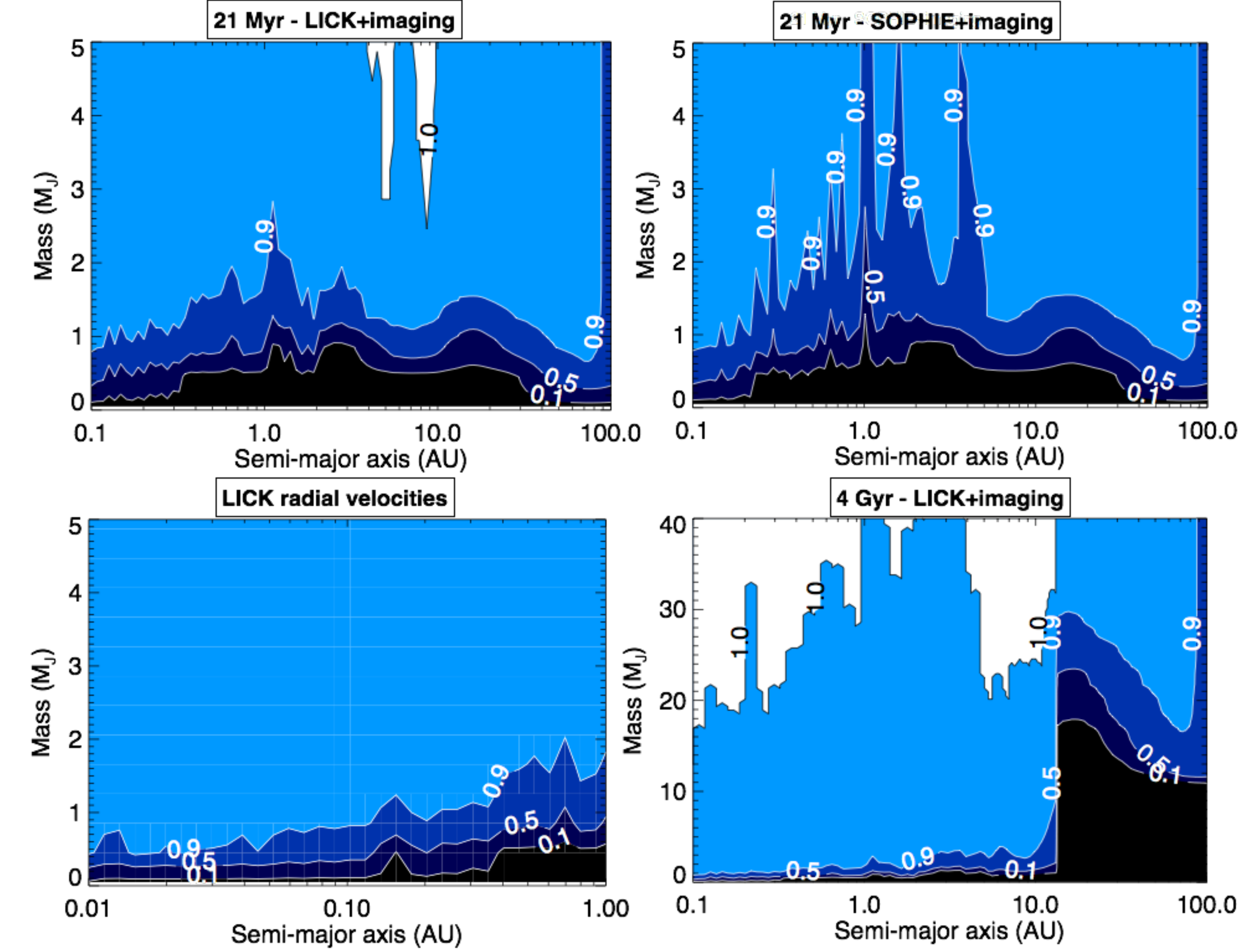}
  \caption{Companion detection probability (white text and isocontours) when combining the sensitivity maps of multiple epochs of imaging data and the Lick  or SOPHIE radial velocities of GJ~504A for the two possible age ranges for the system.}
  \label{fig:MESS2}
\end{figure*}

 The detection probability curves inferred from the  Lick and imaging data are shown in Fig. \ref{fig:MESS2} for the two isochronal age ranges of GJ~504A. When considering the old isochronal age, 90\% of the objects more massive than 30 $\mathrm{M_{Jup}}$ would have been detected from 0.01 to 80 au.  \texttt{MESS2} does not presently enable simulation of the RV signals of planets whose semi-period exceeds the time span of the data. This  explains the sudden loss of sensitivity at $\sim$15 au.  An upgrade of \texttt{MESS2} would allow us, in the near-future, to handle non-detection of planets with longer periods than those set by this observation threshold. 

 No object more massive than 2.5$\mathrm{M_{Jup}}$  (apart GJ~504b)  exists in the system  assuming the young isochronal age.  Our simulations reveal in addition that the Lick data (21.6 years span) enable a more in-depth exploration of the separations from 0.2 to 6 au than the SOPHIE data (3.2 years span). Both of the data sets give comparable constraints from 0.01 to 0.1 au.


\section{Discussion}
\label{sec:discussion}
\subsection{Conflicting age indicators}
\label{subsec:agerec}



\subsubsection{The planet engulfment scenario}


 \cite{Fuhrmann2015} proposed that the engulfment of a jovian planet (2.7 $M_{Jup}$) could have sped up GJ~504A's rotation velocity. \cite{2017A&A...598A..19D}  estimate that the engulfment should have  occurred no more than 200 Myr ago for the system to keep a sufficient imprint of the event on the star rotation speed.  Such an engulfment may also enrich the host star in metals \citep{2012ApJ...757..109C, 2017A&A...604L...4S}.

 In that case, what could have triggered the engulfment long after the dispersal of the circumstellar disk? Our detection limits indicate that no other companions more massive than the proposed engulfed planet are presently located in the first astronomical unit around GJ~504A. GJ~504b is likely the most massive object in the system, and therefore a good candidate perturber.  The Lidov-Kozai mechanism  \citep{1962AJ.....67..591K, 1962P&SS....9..719L} invoked by \cite{Fuhrmann2015} and \cite{2017A&A...598A..19D} could only operate in the system  if  the obliquity $\varphi$ of GJ~504b were higher than at least 40 deg. Additional astrometric monitoring of the companion is required to carve the distribution of relative inclinations $\Delta i$ and provide a lower limit on $\varphi$.  Two known systems have  recently been  discovered with close-in low-mass planets  on eccentric orbits and more massive companions on wide-orbits:  HD 219828 \citep{2016A&A...592A..13S} and  HD 4113 \citep{2017arXiv171205217C}. These systems might then be good proxies of the architecture of the GJ~504 system prior to the putative engulfment.

\subsubsection{Effect of polar spots}
\label{subsub:polspot}

Because GJ~504A is active and  seen close to pole-on, high-latitude spots may be affecting the luminosity and $\mathrm{T_{eff}}$ estimates used for comparison to the tracks. These spots are predicted to occur on rapid rotators such as GJ~504A and young stars \citep{1992A&A...264L..13S, 1997ApJ...484..855B,  2001ApJ...551.1099S, 2006MNRAS.369.1703H, 2015A&A...573A..68Y}. Observations of polar spots on active G-type stars might have been observed \cite[e.g.,][]{2005MNRAS.359..711M, 2006MNRAS.370..468M, 2011MNRAS.413.1922M, 2011MNRAS.413.1949W, 2017MNRAS.465.2076W}. The polar spots (or cap) can fill up to 50\% of the stellar surface and have lifetimes of about a decade. 

 Given a spot filling factor $p$ (defined as  $\mathrm{R_{s} /R_{\star}^{2}}$ , where $\mathrm{R_{s}}$ is the spot radius), the observed luminosity $L_{obs}$ of GJ~504A relative to the photosphere luminosity $L_{phot}$ is:\\
\begin{equation}
L_{obs}/L_{phot} = \frac{p \times T_{spot}^{4} + (1-p) \times T_{phot}^{4}}{T_{phot}^{4}},
\end{equation}

where $T_{spot}$ is the spot temperature and $T_{phot}$ the photosphere temperature. 

The disk-averaged temperature measurement $T_{obs}$ can be influenced by spots in a similar way: 

\begin{equation}
T_{obs}=T_{phot} - ( T_{phot} - T_{spot})\times p
\end{equation}

  We used the SAFIR simulator described in \cite{2007A&A...473..983D} to evaluate the filling factor and inclination of a putative long-lived spot responsible for the RV variations measured with SOPHIE.  Assuming a 1.2 $\mathrm{M_{\odot}}$, 1.3 $\mathrm{R_{\odot}}$ 6205K star inclined by $\sim$17 degrees with respect to edge-on, the observed amplitude of the variations can be reproduced with a $2000$K cooler  dark spot \citep{2005LRSP....2....8B}  inclined by 5 degrees with respect to the spin axis and covering $p\sim6$\% of the star surface. 

The PARSEC evolutionary models indicate that the age predictions should be more sensitive to a bias on the $\mathrm{T_{eff}}$ measurement. 
 We find that a  spot with only $p=7$\% would lower the $\mathrm{T_{eff}}$ determination by 150K and lead to intermediate ages in marginal agreement with the upper limits derived from the gyrochronology (up to 220 Myr) if we assume that the luminosity is not affected by spots at the time of measurements. Conversely,  $p=22$\% would be required to  sufficiently bias the isochronal age based on the luminosity estimate. The data used to compute the luminosity of GJ~504A  (Appendix \ref{sec:AppA}) were  taken on individual nights from 1980 to 2014. Such a large spot would have implied $\mathrm{T_{eff}}$ differences of 440K at least which would have been noticed in the star's SED fit. Therefore, we are confident in the  isochronal age derived from the  luminosity. 

Spots may explain in part the $\sim$200K scatter on the $\mathrm{T_{eff}}$ values found in the literature \citep[see Tab. 2 of][]{2017A&A...598A..19D} and the slight difference on the old isochronal age range ($2.5_{-0.4}^{+1.0}$ Gyr) that can be inferred from the  $\mathrm{T_{eff}}$ value of \cite{2017A&A...598A..19D} and the interferometric radius. However, the $\mathrm{T_{eff}}$ value derived in \cite{2017A&A...598A..19D} is one of the highest reported in the literature; furthermore, it  is  inferred from the excitation balance of 100 Fe lines which form at different optical depths within the star's photosphere. The  lack of a significant scatter in the abundances derived from the individual lines suggests that spots  have not   significantly biased the $\mathrm{T_{eff}}$ determination at the time of the observations. 

Stellar activity is also known to influence the interferometric observables \citep[see,][]{Chiavassa2014, Ligi2014, Ligi2015}.  We have therefore verified whether a spot could have biased our visibility measurements using the \texttt{COMETS} code \citep{Ligi2015} to model the visibility of a star with a spot at its surface. We considered two filling factors $p=7\%$ and $p=22\%$ and a spot temperature of 4205~K, as above. Spots were placed at the edge of the stellar disk, with position angles ranging from 0 to $2\pi$ with $\pi/4$ increments  (that is, all around the stellar disk). Due to symmetry effects, this leads to only three different visibility curves.  The visibilities were computed for the three different wavelengths (550 nm, 710 nm, and 730 nm) used for measuring the angular size of GJ~504A.

\begin{figure*}
  \centering
  \begin{tabular}{cc}
    \includegraphics[width=8cm]{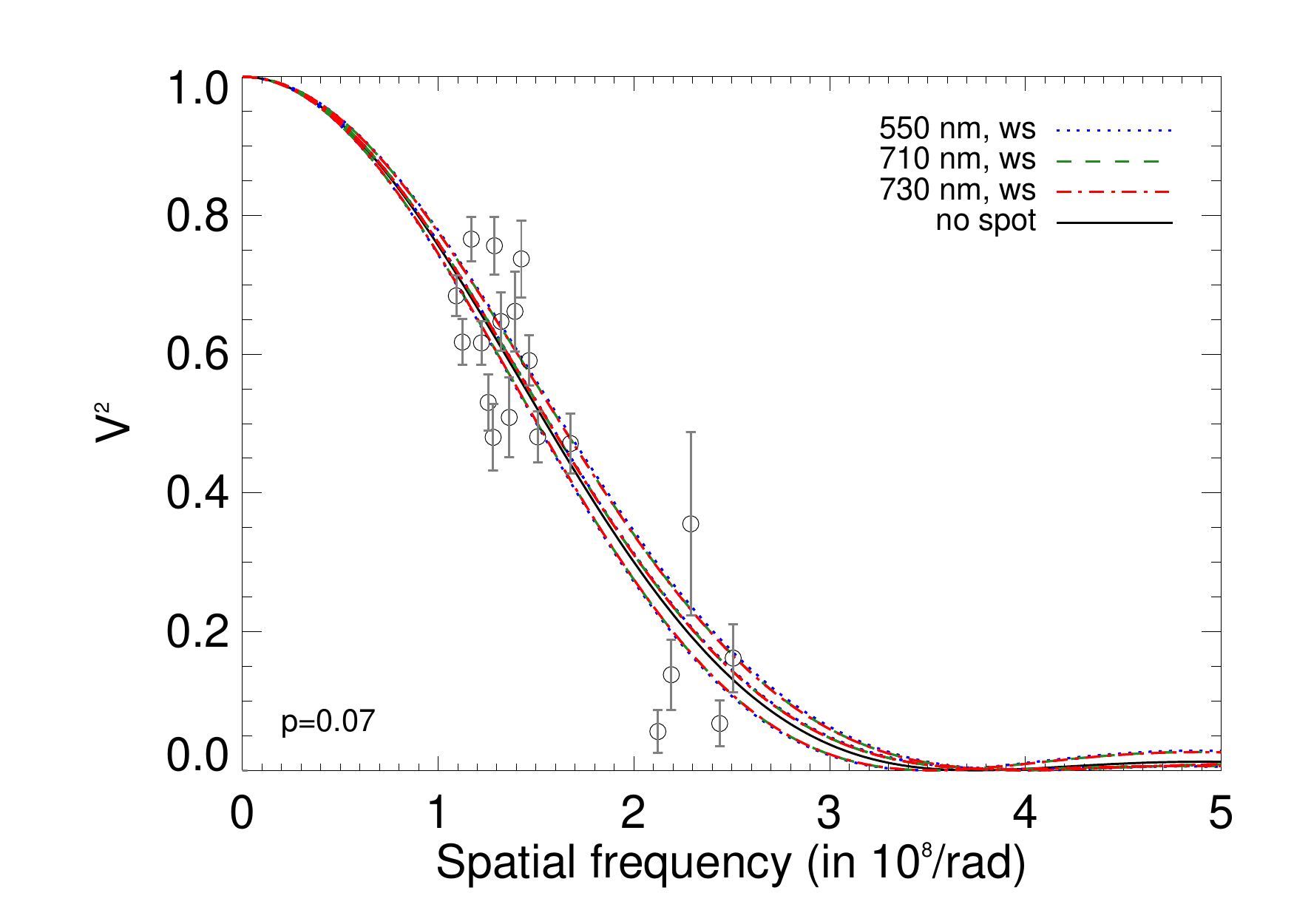} &
    \includegraphics[width=8cm]{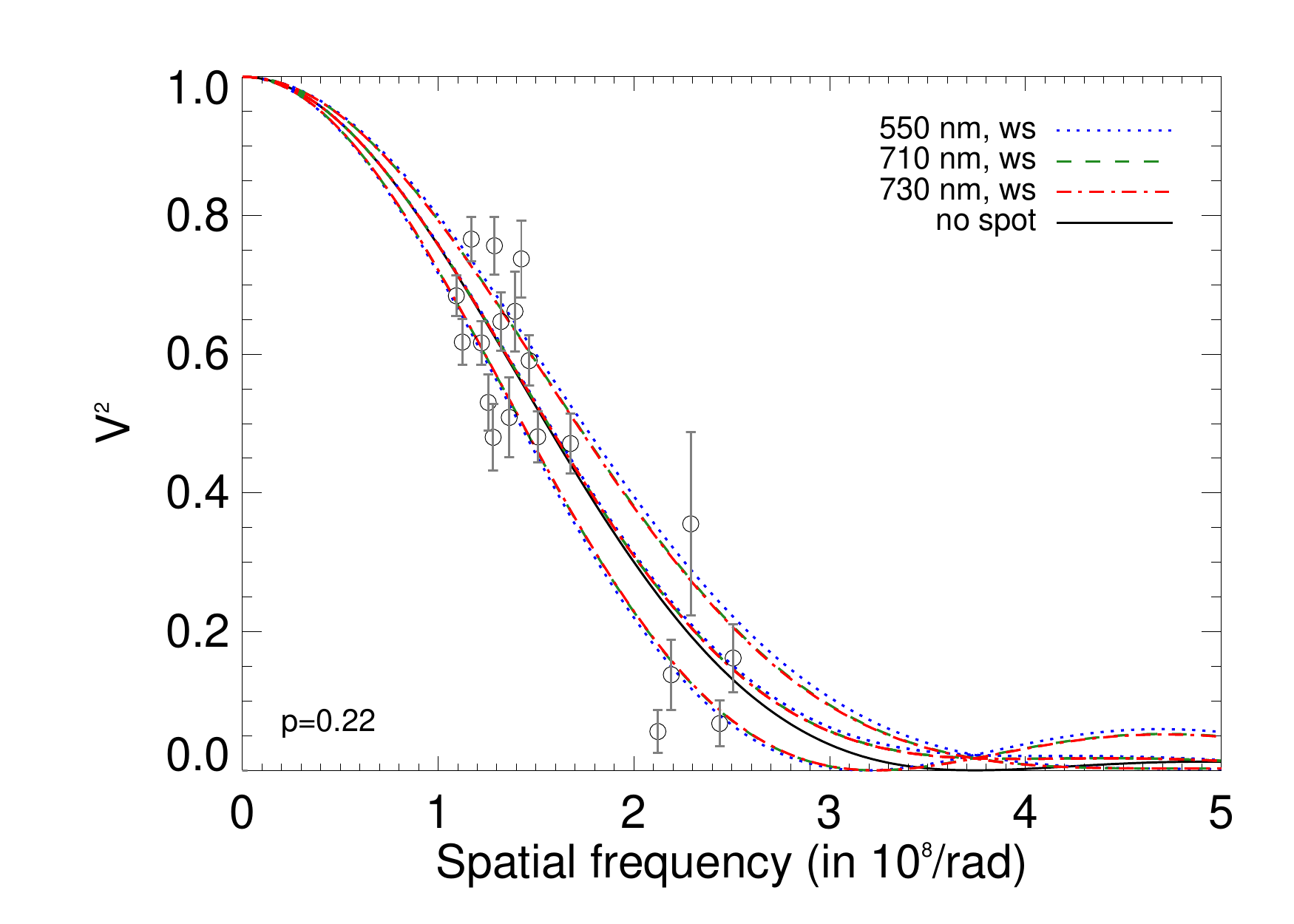} \\
  \end{tabular}
  \caption{Theoretical squared visibilities of a star without a spot (solid black line), and of a star with a spot (ws) as modelised with \texttt{COMETS} \citep{Ligi2015}. The different solid color lines represent the squared visibilities at different wavelengths and positions. The black circles represent the actual interferometric measurements (with error bars) of GJ504 performed with VEGA/CHARA. Left: A spot with a filling factor p=$7\%$. Right: The same but for p= 22$\%$.}
  \label{fig:spotv1}
\end{figure*}

Figure \ref{fig:spotv1} shows the squared visibilities measured with VEGA/CHARA (black circles), that led to an angular diameter of 0.71 mas. The solid black line represents the fit corresponding to this angular diameter (with a limb-darkened model as explained in Sect. \ref{subsec:interf}). The different colored curves represent the theoretical visibilities of a star with a spot as modeled with \texttt{COMETS} at the  different wavelengths and positions tested. 

We find that a spot  with $p=7$ or $22\%$  induces a change in the visibility curve which is still  within the dispersion of measured values. Therefore, spots such as those considered here are not likely to have significantly biased  GJ~504A's angular diameter measurement. 

We therefore conclude that while spots may indeed be affecting the RVs, luminosity,  radius , and  $\mathrm{T_{eff}}$ estimates of GJ~504A, their effect is unlikely to bias all those quantities together by sufficient amounts and change the isochronal age estimates of the star. 
 
\subsection{Disentangling the atmospheric model solutions}
\label{subsec:JWST}

\begin{figure}
  \centering
  \includegraphics[width=\columnwidth]{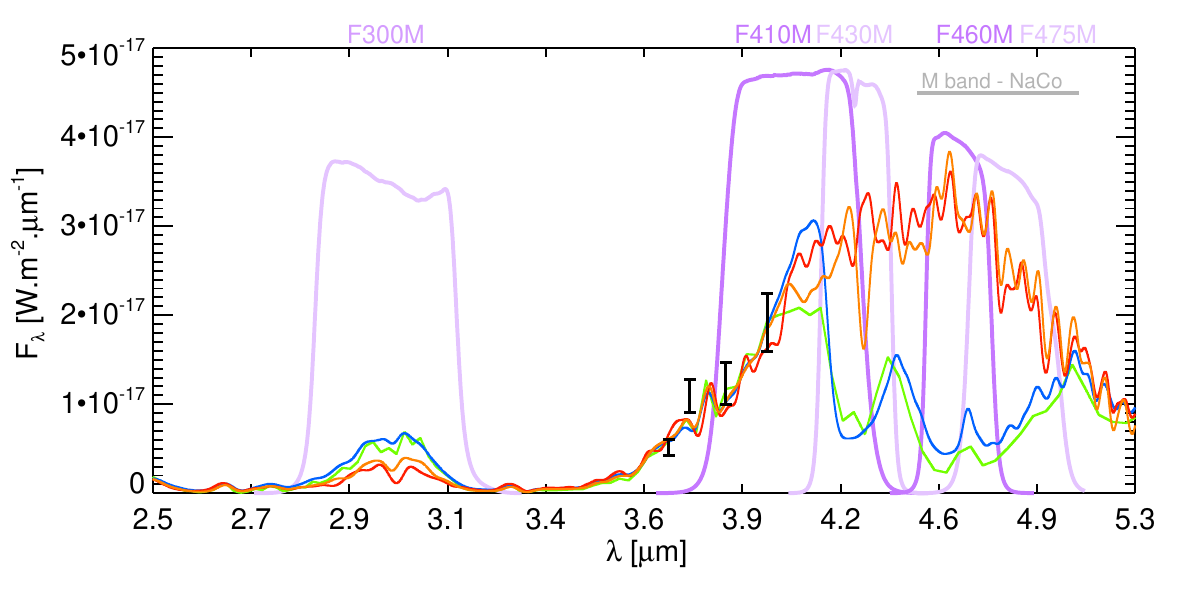}
  \includegraphics[width=\columnwidth]{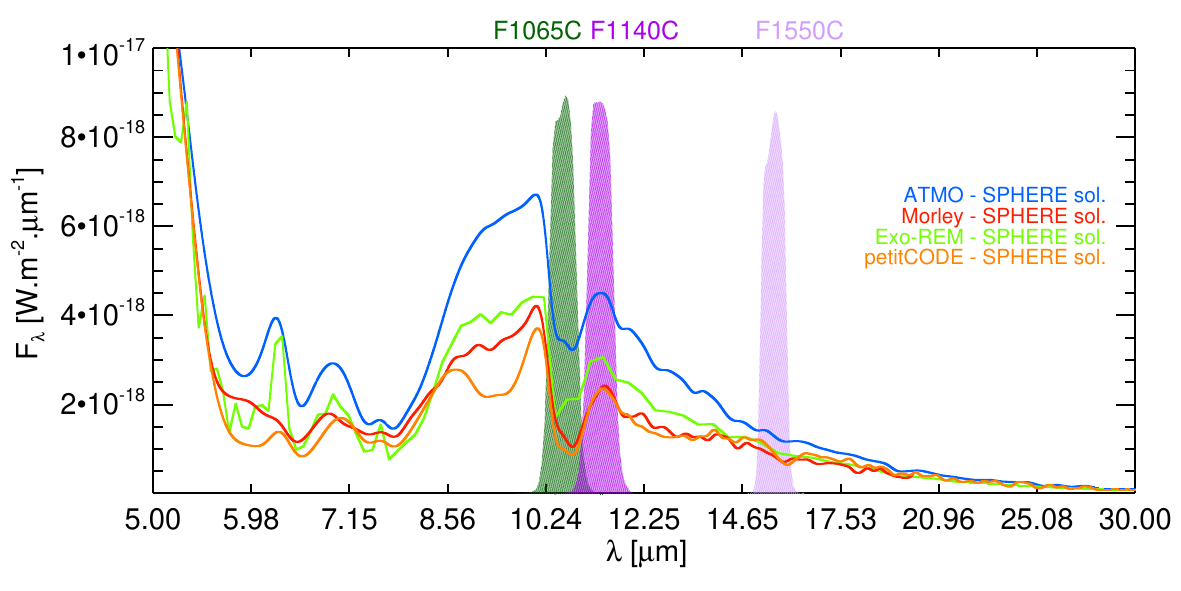}
    \caption{Predicted apparent fluxes of GJ~504b in the near- and mid-infrared  corresponding to the best-fitting synthetic spectra found in Section \ref{Section:atmomodels} with the G statistics and some knowledge of the object radius. The transmission of some key filters of JWST instruments are overlaid. We report the  L-band  photometry (Subaru/IRCS, LBTI/LMIRcam) of GJ~504b (black).}
  \label{fig:JWST}
\end{figure}

We show in Fig. \ref{fig:JWST} the synthetic spectra in the L-M band and in the mid-infrared corresponding to the models fitting the presently available photometry of GJ~504b (Table \ref{Tab:atmopar}; solutions with some pre-requisite on the companion radius).  The \texttt{ATMO} and \texttt{Exo-REM} models predict very similar spectra distinctive from those corresponding to the Morley and \texttt{petitCODE} solutions. The difference arises from the non-equilibrium chemistry which is not considered in the case of the two latter models and modulates the strength of the CO$_{2}$ and CO absorption bands centered around 4.3 and 4.7 $\mu$m, respectively. Adaptive-optics M-band imaging from the ground should already tell whether the non-equilibrium chemistry is a pre-requisite for modeling the companion emission flux (model-to-model contrast between 1.16 and 1.48 mag in the M-band filter of the VLT/NaCo instrument).  Coronographic imaging with the F430M and F460M filters of the Near Infrared Camera (NIRCam) on the James Webb Space Telescope (JWST) should also better constrain the shape of the 3.7-5$\mu$m pseudo-continuum  and could disentangle the \texttt{ATMO} and \texttt{Exo-REM} solutions.

Observations at longer wavelengths will be a niche for the Mid-Infrared Instrument (MIRI) of JWST. We can estimate that the contrast between GJ~504b and GJ~504A should range between 4$\times10^{-6}$ and 2.5$\times10^{-4}$ from 5 to 28.5 $\mu$m using the  set of atmospheric models considered above and the SED of GJ~504A (Appendix \ref{sec:AppA}). The use of the four-quadrant phase masks together with ADI will  be mandatory  to reach GJ~504b contrasts and avoid saturation\ \citep{2015PASP..127..633B}. The four-quadrant phase masks can only be used jointly with the F1065C ($\lambda_{c}=10.575 \mu m, \Delta \lambda=0.75\mu m$), F1140C ($\lambda_{c}=11.40 \mu m, \Delta \lambda=0.80\mu m$), and F1550C ($\lambda_{c}=15.50 \mu m, \Delta \lambda=0.90\mu m$) filters.  The MIRI photometry should enable to distinguish between the  \texttt{ATMO} and \texttt{Exo-REM} solutions. The  \texttt{Exo-REM} models indicate that the spectral slope between 11  and 15 microns probed by the F1140C- F1550C color should also be a good indicator of the percentage of the disk surface covered by clouds. 

To conclude, we considered two representative solutions probing the  log g/[M/H] degeneracy in the posterior distributions shown in Figs. \ref{fig:MCMCmorley} and \ref{fig:MCMCBACON} at $\mathrm{T_{eff}}$=550K.  The spectra indicate that narrow and broad band photometry with JWST longward of 3$\mu$m should not break the log g/[M/H] degeneracy for all but the \texttt{ATMO} solutions. The MIRI data coupled to the SPHERE data points should nonetheless set stringent constraints ($<100K$)  on the $\mathrm{T_{eff}}$ based on our simulations and should allows for reduction of the error bar on the  luminosity. The comparison of that $\mathrm{T_{eff}}$ and luminosity  to evolutionary tracks (Fig. \ref{fig:hotstartlumteff}) is a way to investigate the system age independently from GJ~504A. 

\subsection{Formation pathway}
\label{Subsec:pathway}
If confirmed, a spin-orbit misalignment of GJ~504b (Section \ref{subsec:stellobli}) would be a remarkable property of the system, in particular with respect to the solar system planets.  Such misaligments are indirectly observed on protostar pairs  \citep[e.g.,][]{2016ApJ...827L..11O, 2016ApJ...830L..16B} and are consistent with a stellar-like  formation scenario \citep[e.g.,][]{1979ApJ...234..289B}.  Spin-orbit misalignments are  also clearly established for close-in planets with orbital periods ranging from 0.73 \citep[55 Cnc e;][]{2014A&A...569A..65B}  to 207.6 days \citep[Kepler-462 c; ][]{2015ApJ...814...67A}. Dynamical interactions between planets is a possible cause of those misalignments \citep[e.g., ][]{2008ApJ...686..580C}, but other processes such as the magnetic interactions between the inner disk and the star \citep{2011MNRAS.412.2790L, 2014ApJ...790...42S} or disk-warping \citep[e.g.,][]{2013MNRAS.435..798T} have also been proposed. \cite{2017AJ....154..165B} reported a likely nonzero obliquity for the  ROXs 12 system composed of a  $\mathrm{17.5\pm1.5 M_{Jup}}$ companion ("hot-star" mass) at a projected separation of 240 au from a low-mass ($0.65^{+0.05}_{-0.09} M_{\odot}$) young ($6^{+4}_{-2}$ Myr) star. This is  to our knowledge the only other measurement of the obliquity of a wide-orbit ($>$ 10 au) companion less massive than 30 $\mathrm{M_{Jup}}$. \cite{2017AJ....154..165B} also show that this system has a tertiary stellar component at a projected separation  of 5100 au, which makes the system's architecture  different from that of GJ~504.  

Some other properties of the GJ~504 system may also be informative. The companion is in a mass range either below, or right inside the so-called "brown-dwarf desert" observed at short separations for solar-type stars \citep[e.g.,][]{2011A&A...525A..95S, 2014MNRAS.439.2781M} and likely existing at larger separations \citep{2016A&A...586A.147R}. The companion mass ratio $q$ with GJ~504A is $1.9_{-0.7}^{+1.1}$\%  or $0.11_{-0.03}^{+0.07}$\%, depending on the isochronal age range considered. If the system is 4 Gyr old, GJ~504b still belongs to a very short list of objects resolved at projected separations smaller than 50 au \citep[HD 206893, HR 2562B, HIP 73990B \& C;][]{2017A&A...597L...2M, 2017A&A...608A..79D, 2015ApJ...806L...9H, 2017arXiv171205828M} with such extreme $q$ values. All those companions are found around debris disk stars. GJ~504b's semi-major axis is  probably lower than 200 au in contrast to the 20-30 $M_{Jup}$  G-type star companions \citep[for example, HN Peg b, HD~203030B ;][]{2006ApJ...651.1166M, 2007ApJ...654..570L} found at large (> 500 au) projected separations; e.g., beyond the typical size of circumstellar disks of T-Tauri stars \citep[e.g.,][]{2014A&A...564A..95P, 2017A&A...606A..88T, 2017ApJ...845...44T}. 

If GJ~504b is a 14-33  $M_{Jup}$ object, its orbital properties and mass ratio should still be compatible with a stellar-like formation mechanism \citep[e.g.,][]{2005MNRAS.364L..91L}.  \cite{2009ApJS..181...62M} argue that the companion mass function follows the same universal form over the entire range between 0 and 1590 au in orbital semimajor axis. They predict a peak in semi-major axes for brown dwarfs at $\sim$30 au which is broadly compatible with the present constraints on the semi-major axis of GJ~504b. Most orbital solutions of GJ~504b  also correspond to periods close to the most frequent values found for solar-type star binaries \citep[P=293.6 days; see Fig. 13 of][]{2010ApJS..190....1R}.

\begin{figure}
  \centering
  \includegraphics[width=\columnwidth]{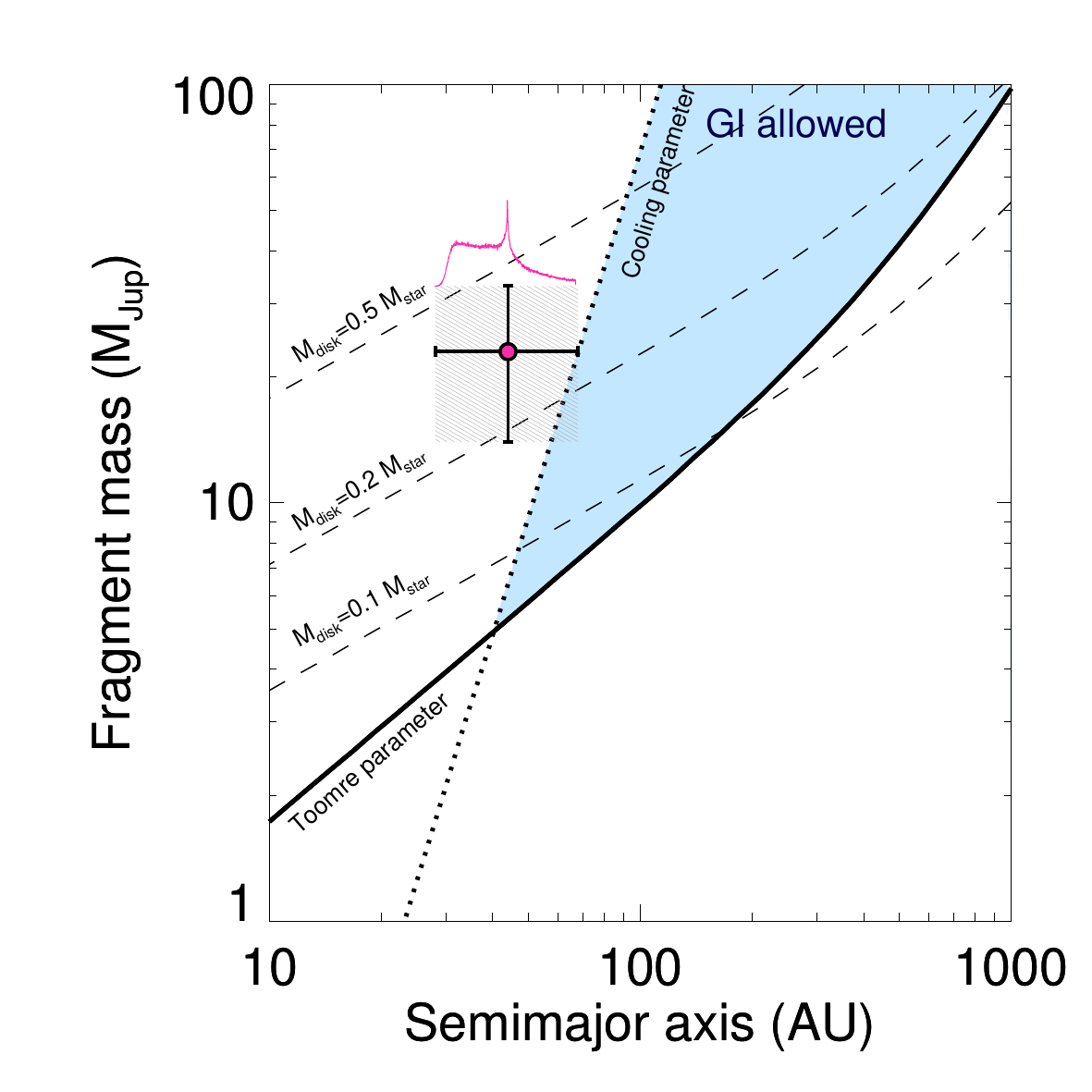}
  \caption{Gravitational instability model adapted to the case of GJ~504. Fragments are allowed to form if they respect the Toomre and cooling criteria. GJ~504b properties are reported. The pink curve corresponds to the posterior distribution of the companion semi-major axis found with our MCMC orbit-fitting package (Section \ref{Section:orbit}). The dashed lines correspond to the disk mass distribution for a different hypothesis on the initial disk mass.}
  \label{fig:GI}
\end{figure}

We also investigated whether the companion could have formed in a disk. We generated a disk instability model \citep[Klahr et al. in prep; see also ][]{2011ApJ...736...89J} adapted to the case of GJ~504 (stellar luminosity and metallicity). The model  predicts the range of semi-major axis and clump masses allowed to form and cool down more rapidly than the local Keplerian timescale in Toomre-unstable disks \citep{1981seng.proc..111T}.  The result is shown in Fig. \ref{fig:GI}.  Clumps with masses in agreement with the companion properties can form if we adopt the old isochronal age for the system. However, the allowed fragmentation zone  is predicted to be at larger semi-major axis than most solutions found from the MCMC orbital fits.  This can be explained if the disk opacity is lowered, and therefore not scaled on the stellar metallicity (this would be the case if GJ~504A was initially a solar-metallicity star that was later enriched by a planet engulfment event; see Section \ref{subsec:agerec}). In such a case, clumps can cool down sufficiently rapidly at shorter separations. The companion may have alternatively been formed at larger separation subsequently undergoing inward disk-induced migration \citep[for instance through the Type II process which allows for clump survival;][]{2015ApJ...810L..11S, 2017MNRAS.470.2387N}.  This formation at a wider distance would also allow for a lower disk mass. 

	The model cannot account for GJ~504b if it is a $1.3^{+0.6}_{-0.3} M_{Jup}$ 21 Myr-old planet. However, more complex models allowing for a more detailed investigation of the free parameters in the GI models \citep[e.g., ][]{2017ApJ...836...53B} and subsequent planet embryo evolution \citep[protoplanet migration, clump-clump dynamical interactions, "tidal downsizing", etc; e.g.,][]{2013MNRAS.432.3168F, 2017MNRAS.470.2517H, 2018arXiv180103384M}  may lead to different conclusions.

\begin{figure}
  \centering
  \includegraphics[width=\columnwidth]{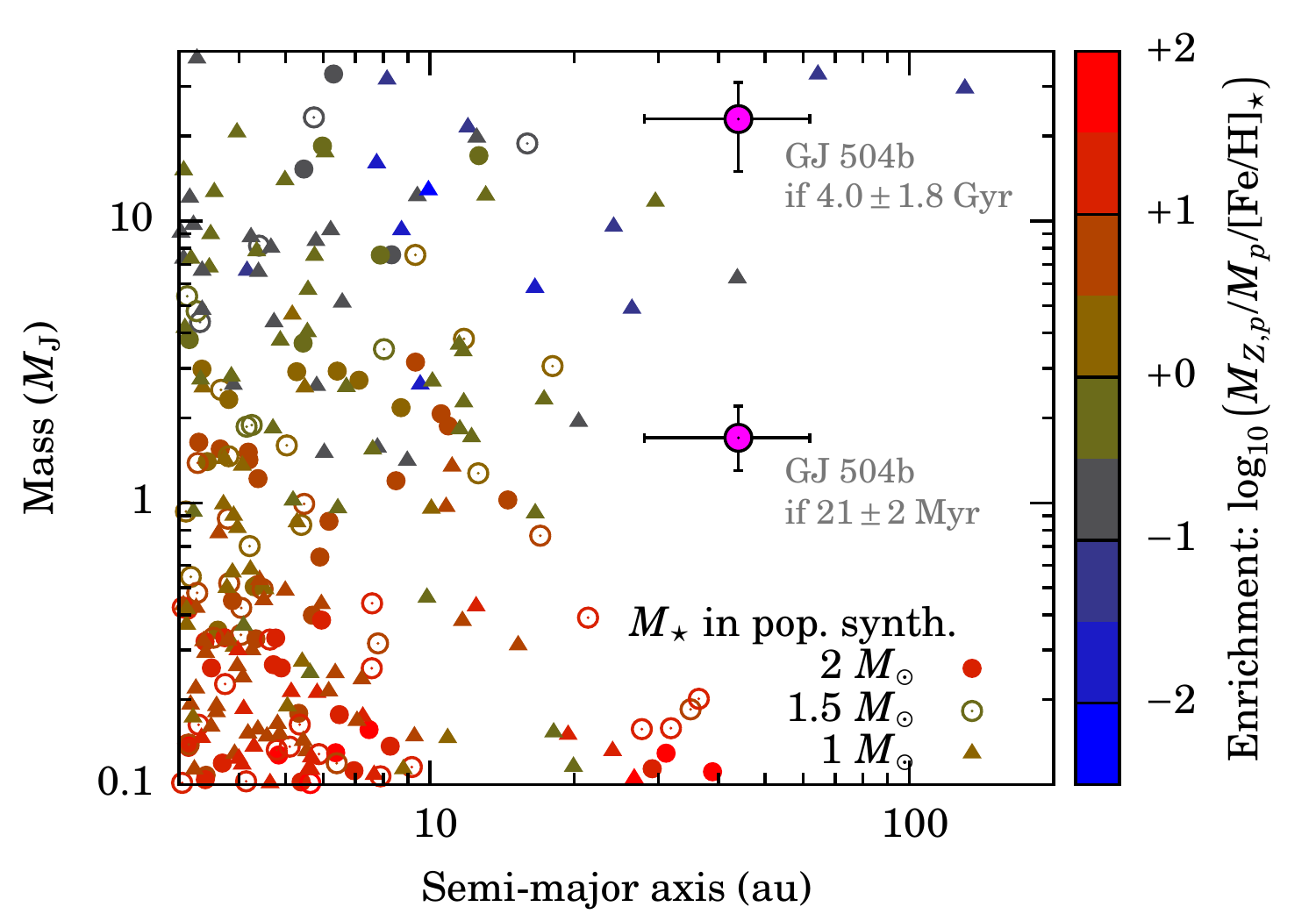}
  \caption{Population synthesis at 20 Myr for core-accretion models including Type I and II migration and dynamical scattering between multiple planet embryos in the disk. We considered the case of 1, 1.5, and 2 $\mathrm{M_{\odot}}$ central stars. The color shows the enrichment relative to the star.}
  \label{fig:CA}
\end{figure}

We compare GJ504b to the Bern core-accretion population synthesis results in Fig. \ref{fig:CA}. The model considers the formation of multiple planet embryos per disk (50, 20, and 10 embryos per disk  for the simulations with 1, 1.5, and 2 $\mathrm{M_{\odot}}$ central objects, respectively) and Type I and II migration \citep{2005A&A...434..343A, 2012A&A...547A.111M, 2013A&A...558A.109A}.  The bulk enrichment in solids of each final planet is reported in the figure. With a lower limit of 27.8 au on its semi-major axis, GJ~504b appears as an outlier of the population for the two possible age ranges. The models can however still form a few objects as massive and distant as the companion. The simulations indicate that all  planets more massive than 10$\mathrm{M_{Jup}}$ should not be significantly metal-enriched with respect to their host stars. This is in good agreement with the atmospheric metallicity found with the Morley and petitCODE models and the MCMC method.

\subsection{Finding analogs of GJ~504b with VLT/SPHERE}
  Most of the SHINE observations  are performed with  the IRDIFS mode of the instrument. The H-band observations ensure good AO performance, an optimal use of the apodized Lyot coronograph, and low background emission. The  IFS can  distinguish  cool companions in the first 0.8-1.2'' from hotter background objects through the detection of characteristic spectral features. That is also one of the best characterized modes for the  astrometric monitoring. The unusual colors of GJ~504b call however for a re-investigation of the detection capabilities of ultracool companions with the various instrument modes of SPHERE.

We estimated the absolute magnitude and colors\footnote{We caution that our predictions do not account for the feedback of the atmosphere on the object evolution. They should not be used for the characterization of individual objects.} of planets and brown dwarfs for three characteristic ages in the field pass-bands using the \texttt{Exo-REM} atmospheric models as boundary conditions (see Appendix \ref{App:G}). 

The Exo-REM models predict a strong sensitivity of the absolute magnitudes to the cloud coverage and metallicity, in particular for the lowest masses (and $\mathrm{T_{eff}}$).  The models also show that the companions have a higher or similar brightness in the J3 filter. This adds to the fact that the typical stars observed with SPHERE have J-H$>$0 or  J-K$>$0, therefore leading to more favorable predicted contrasts at J3. In some cases, DBI imaging with the J2J3 filter could therefore become advantageous for the detection of cool companions. This can be illustrated when considering the  G0 star HIP 19148. The star is  a member of  the 625 Myr-old metal-enriched  Hyades cluster \citep{2001A&A...367..111D}. Using the ESO exposure time calculator\footnote{https://www.eso.org/observing/etc/bin/simu/sphere} (version P101.3), we could generate contrast curves for the J2J3 and H2H3 bands considering median observing conditions (seeing of 0.8-1.0"), 64s exposures to minimize the read-out noise, and the ADI performance reached during a 1.5hr sequence of coronographic exposures. We used the 2MASS J and H magnitudes and the GAIA-DR1 distance \citep{2016A&A...595A...1G}  to compute the sensitivity  and compare it to the predicted magnitudes  of 5, 8, and 15M$_{Jup}$ objects. The simulation predicts that we would miss a 8$M_{Jup}$ object in the H2 band at the physical projected separation of GJ~504b while it would be comfortably detected in the J3 channel, in particular if the object is metal-enriched.

\begin{figure}
  \centering
  \includegraphics[width=\columnwidth]{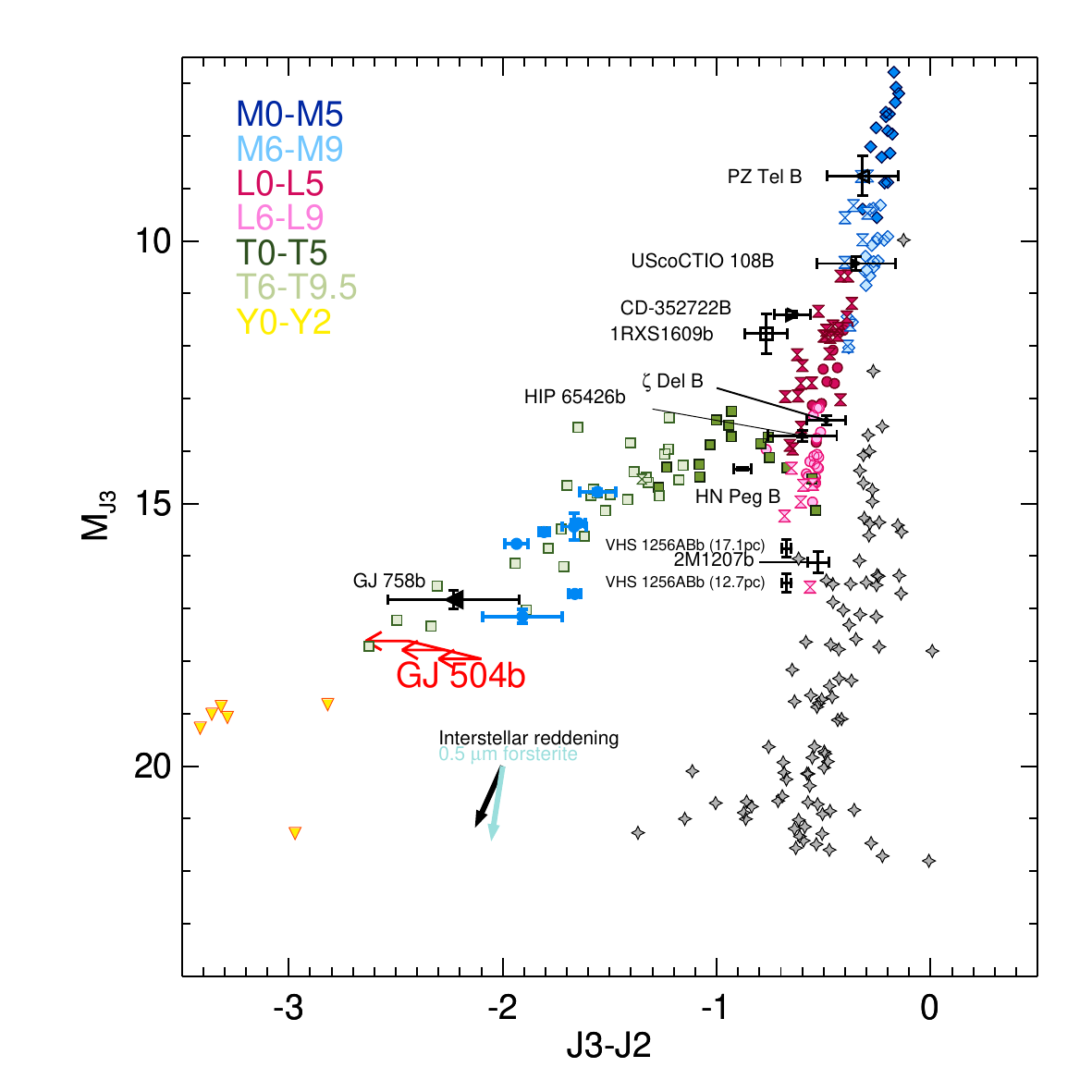}
  \caption{Color-magnitude diagram exploiting the J2 and J3 photometry of SPHERE. We report the photometry of candidate companions detected around three SHINE targets with the DB\_J23 filter of IRDIS (grey stars symbols). The reddening vector of 0.5 $\mu$m forsterite grains and the one corresponding to the interstellar extinction are overlaid \citep[see][for the details on how the vectors are computed]{2016A&A...587A..58B}.}
  \label{fig:CMDJ2J3}
\end{figure}

The J2J3 mode offers a second advantage. Observations of stars in the galactic plane usually lead to the detection of numerous background stars with IRDIS. When reported in color-magnitude diagrams and assuming they are at the same distance of the target, those point sources line up and form a locus. This locus has the same colors as K and early M stars but is spread in luminosity and does not necessarily share the same colors as cool companions.   Therefore, the placement of candidate point sources into those diagrams offers a simple way to disentangle background stars from bound companions.  When considering the H2H3 mode, the locus intersects the sequence of cool objects at the L/T transition (where companions such as HN Peg b or HR8799bcde lies) and falls close to the sequence of late-M dwarfs (Langlois et al., in prep).  It is therefore not always possible to determine whether the object is a background star or a substellar companion.  We build up in Fig. \ref{fig:CMDJ2J3} a locus of contaminants from the  J2J3 observations of HIP~67497, HD~115600, and HIP 92984 obtained as part of SHINE (SHINE collaboration; priv.  com.). The point sources draw a locus distinct from the sequence of young and old late-M, L, and T dwarfs. The faintest contaminants have a 1.5mag color difference with known Y dwarfs.  Their colors follow the interstellar reddenning vector. Cool companions such as GJ~758b or GJ~504b would easily be identified from the method. Dusty L-type planets such as HIP~65426b would also be discriminated from the locus of contaminants. Therefore, DBI observations with the J2J3 filters may not necessarily require  a follow-up to confirm that the point sources do not share the target proper motion. 

\section{Conclusion}
\label{sec:conclusion}
Because it is a nearby bright star, GJ~504 can be observed with a variety of techniques. This work presents new interferometric, radial-velocity, and high-contrast imaging observations  that shed a new light on the system. Two isochronal age ranges ($21\pm2$ Myr and $4.0 \pm1.8$ Gyr) are compatible with the interferometric radius of GJ~504A. The conflicting conclusions from the various alternative age indicators do not allow us to firmly choose one age or the other. The known companion is a T8-T9.5 object with  a peculiar SED from 1 to 2.5 $\mu$m. The SED is compatible with a low surface gravity and/or super-solar metallicity atmosphere. The metallicity determination is limited by systematic errors between atmospheric models and degeneracies with the surface gravity. Our analysis also reveals that the metallicity is not degenerate with the carbon-to-oxygen ratio.  The surface gravity is consistent with the young isochronal age of the system. We estimate a mass of $\mathrm{M = 1.3^{+0.6}_{-0.3} M_{Jup}}$ and $\mathrm{M = 23^{+10}_{-9}M_{Jup}}$ for GJ~504b for the young and old isochronal system ages, respectively. These masses account for a wide range of plausible initial conditions and rely on the bolometric luminosity inferred independently from the empirical and atmospheric model analysis of the companion SED. The orbit of GJ~504b has a semi-major axis larger than 27.8 au, an eccentricity lower than 0.55, and an inclination in the range $[125-180]$ degrees. The interferometric radius of the star and its $v \cdot sin\:i$  allows determinations of the line-of-sight inclination. A comparison with the inclination of the orbit of GJ~504b suggests that the system could have a spin-orbit misalignment. The radial velocity and imaging data allow exclusion of companions more massive than 2.5 and 30 $M_{Jup}$ from 0.01 to 80 au assuming the young and old age range, respectively.   

If GJ~504b is a brown-dwarf in an old system, we show that gravitational instability models possibly coupled to inward migration might explain its properties. Population synthesis models confirm that the core-accretion models can form such a massive object, but preferentially at semi-major axis shorter than 30 au. Both formation models would be challenged if the object is a planet with $\mathrm{M = 1.3^{+0.6}_{-0.3} M_{Jup}}$. 

Additional key measurements could be obtained in the near  future to  better constrain the origins of the GJ 504 system. Additional astrometric monitoring of GJ~504b is crucially needed to  1) tighten down the posteriors on the inclination of GJ~504b orbit and confirm the spin-axis misaligment, and 2) constrain better the eccentricity. The latter could be related to the formation mechanism \citep[see][]{2014MNRAS.439.2781M}. JWST photometry and spectra  of GJ~504b should yield the first robust constraints on the C/O, O/H, and C/H (or metallicity) ratios and on the importance of nonequilibrium chemistry in the atmosphere of GJ~504b. It will then become possible to compare the abundances to those of brown dwarfs \citep{2017ApJ...848...83L} and  planets \citep[e.g. ][]{2015arXiv150407655B, 2017AJ....154...91L}. Deeper imaging data as gathered with the JWST should set stringent constraints on the probability of detection of companions beyond 80 au.  Conversely, additional monitoring with SPHERE may carve the planet detection probability parameter space in the [15-30] au range where companions slightly more massive than GJ~504b may still reside if the system is old. Asteroseismology of the host star might enable us to close the debate on the system age. The more accurate luminosity and surface gravity of GJ~504b gathered by JWST might also enable  the two possible isochronal ages for the system to be disentangled. GAIA may detect the wobble induced by GJ~504b over the duration of its nominal mission (5 years) which could be used to exclude some of our orbital solutions and set upper limits on the companion mass.

To conclude, the J2J3 DBI filter of SPHERE offers good prospects for the detection and follow-up strategy of analogs of GJ~504b.  Direct imaging surveys of nearby metal-rich G-type stars using this pair of filters would be of value to constrain the formation models. 

\begin{acknowledgements}
The authors thank the ESO staff, the CHARA team, and the VEGA team for support in conducting the observations. The first author is grateful to B. Burningham, G. Mace, D. Cushing, M. Liu, R. D. Rosa, D. Lafrenière, D. Kirkpatrick, A. Burgasser, J. Patience, B. Bowler, D. Lachapelle, E. Wright, S. Leggett for providing the spectra of benchmark objects. We thank K. Rice, D. Forgan,and/or with and K. Todorov for fruitful discussions.This research has benefitted from the SpeX Prism Library (and/or SpeX Prism Library Analysis Toolkit), maintained by Adam Burgasser at http://www.browndwarfs.org/spexprism and from the Y Dwarf Compendium maintained by Michael Cushing at https://sites.google.com/view/ydwarfcompendium/. This research has made use of the SIMBAD database and VizieR catalogue access tool (operated at CDS, Strasbourg, France). This research has made use of NASA's Astrophysics Data System and of the Extrasolar Planet Encyclopedia (http://exoplanet.eu/). Part of the planet population of Fig. \ref{fig:CA} is publicly available on DACE, the Data Analysis Centre for Exoplanets of the NCCR PlanetS reachable at https://dace.unige.ch. Additional populations will be added in future. DACE yields both interactive snapshots of the entire population at a given moment in time like the population-wide $a$--$M$ or $M$--$R$ diagrams as well as formation tracks of individual planets (e.g., M(t), R(t), etc.) for all synthetic planets.  This publication makes use of VOSA, developed under the Spanish Virtual Observatory project supported from the Spanish MICINN through grant AyA2011-24052. This work is partly based on data products produced at the SPHERE Data Centre hosted at OSUG/IPAG, Grenoble. Data analysis was partially carried out on the open use data analysis computer system at the Astronomy Data Center, ADC, of the National Astronomical Observatory of Japan. SPHERE is an instrument designed and built by a consortium consisting of IPAG (Grenoble, France), MPIA (Heidelberg, Germany), LAM (Marseille, France), LESIA (Paris, France), Laboratoire Lagrange (Nice, France), INAF Osservatorio di Padova (Italy), Observatoire de Genève (Switzerland), ETH Zurich (Switzerland), NOVA (Nether- lands), ONERA (France) and ASTRON (Netherlands) in collaboration with ESO. SPHERE was funded by ESO, with additional contributions from CNRS (France), MPIA (Germany), INAF (Italy), FINES (Switzerland) and NOVA (Netherlands). SPHERE also received funding from the European Commission Sixth and Seventh Framework Programmes as part of the Optical Infrared Co- ordination Network for Astronomy (OPTICON) under grant number RII3-Ct- 2004-001566 for FP6 (2004-2008), grant number 226604 for FP7 (2009-2012) and grant number 312430 for FP7 (2013-2016). We also acknowledge financial support from the Programme National de Plan\'{e}tologie (PNP) and the Programme National de Physique Stellaire (PNPS) of CNRS-INSU in France. This work has also been supported by a grant from the French Labex OSUG@2020 (Investissements d’avenir – ANR10 LABX56) and from the Agence Nationale de la Recherche (grant ANR-14-CE33-0018). GDM and CM acknowledge the support from the Swiss National Science Foundation under grant BSSGI0$\_$155816 ``PlanetsInTime''. Parts of this work have been carried out within the frame of the National Center for Competence in Research PlanetS supported by the SNSF.  This work is based upon observations obtained
with the Georgia State University Center for High Angular Resolution Astronomy Array at Mount Wilson Observatory. The CHARA Array is supported by the National Science Foundation under Grant No. AST-1211929.  J.H. is supported by the Swiss National Science Foundation
(SNSF P2GEP2\_151842). A.Z. acknowledges support from the CONICYT + PAI/ Convocatoria nacional subvenci\'on a la instalaci\'on en la academia, convocatoria 2017 + Folio PAI77170087.
\end{acknowledgements}
  
  \bibliographystyle{aa}
\bibliography{biblio}

\begin{appendix}
\section{GJ504A SED and luminosity}
\label{sec:AppA}
\begin{figure}
  \centering
  \begin{tabular}{c}
    \includegraphics[width=\columnwidth]{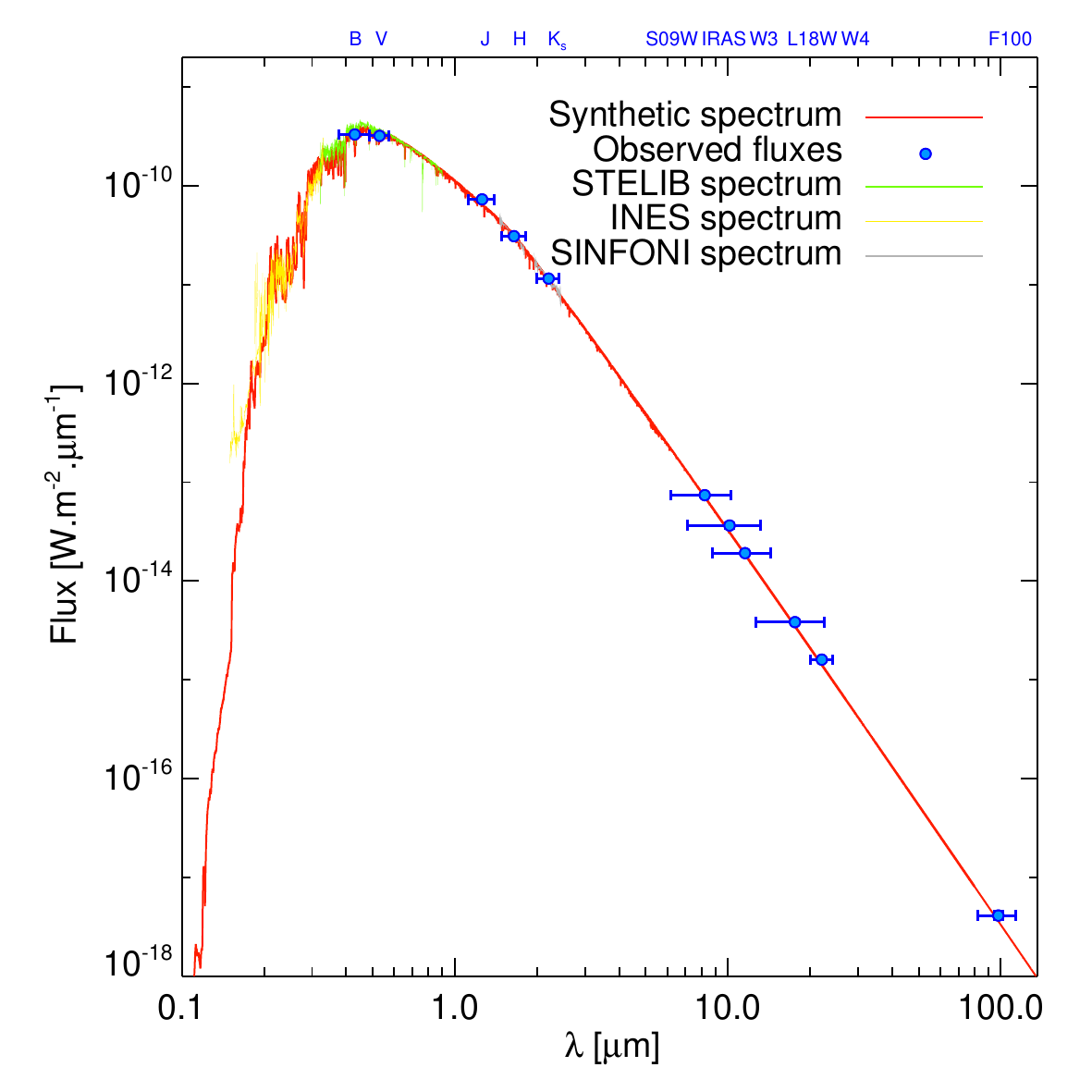} \\
    \includegraphics[width=\columnwidth]{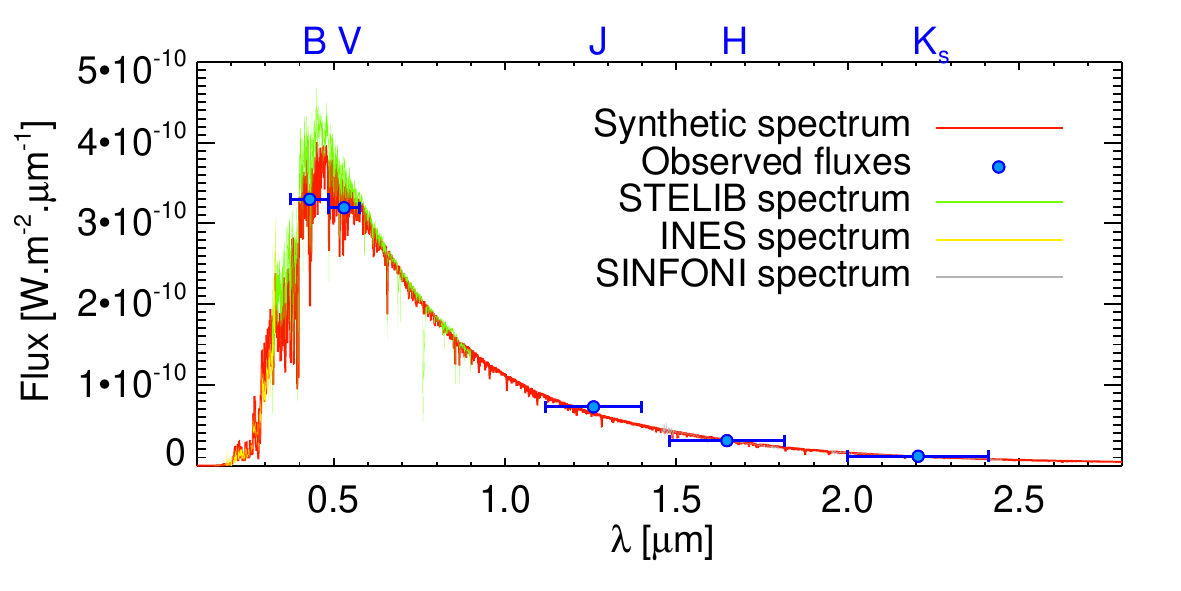} \\
    \includegraphics[width=\columnwidth]{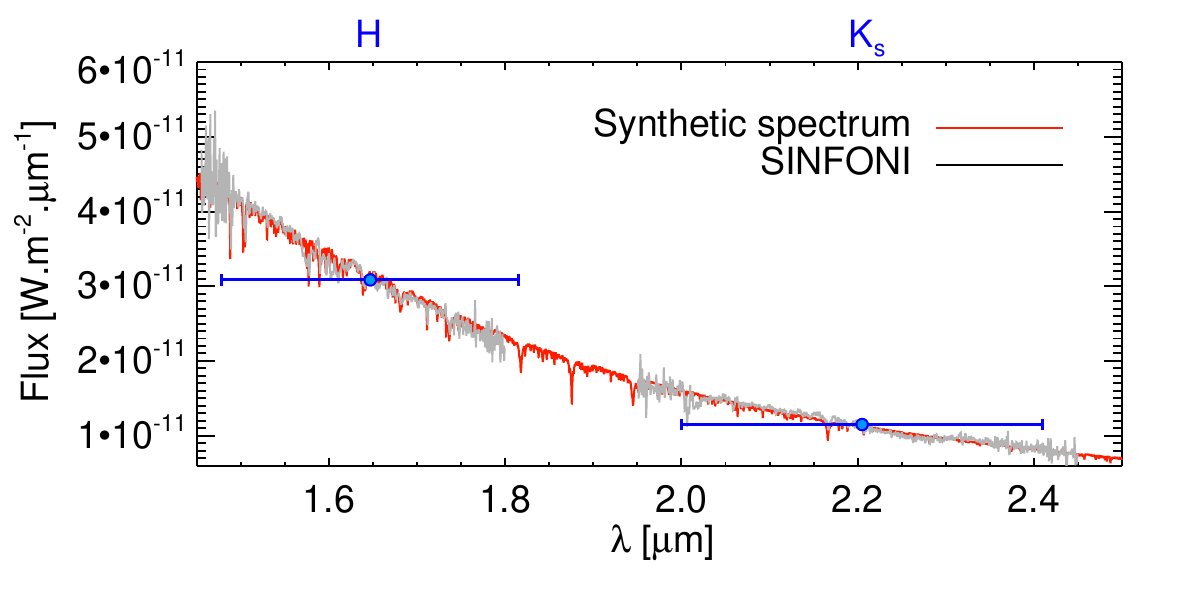} \\
  \end{tabular}

  \caption{Photometry of GJ~504A (blue dots) compared to a BT-NEXTGEN synthetic spectrum (red line) at $\mathrm{T_{eff}=6200}$ K, log g=4.5, and M/H=+0.3 scaled in flux. The flux-calibrated INES, STELIB, and SINFONI spectra (yellow, green, and grey lines, respectively) of the star are compatible with the flux-calibrated model-spectrum.}
  \label{fig:SEDstar}
\end{figure}

		The magnitude of GJ~504A  is unknown in the SPHERE passband. We therefore built a model of the star SED from the Johnson V and B band \citep{2009yCat..35040681K},  J, H, and K band  \citep{2003AJ....125.3311K},   AKARI S09W and L18W   \citep[][]{2010A&A...514A...1I}, IRAS 12 $\mu$m \citep{1989ifss.book.....M},  WISE W3 and W4 \cite{2013yCat.2328....0C},  and Herschel/PACS 100 $\mu$m \citep{2015ApJ...801..143M} photometry\footnote{The source is saturated in the WISE W1 and 2 images. The published \textit{Spitzer} 70 $\mu$m photometry has large error bars \citep{2014ApJ...785...33S} and was not considered in the fit. It confirms the lack of excess emission at 70 $\mu$m}.  That SED is well reproduced by a BT-NEXTGEN synthetic spectra \citep{2012IAUS..282..235A}  with $\mathrm{T_{eff}}$=6200 K, log g=4.5, and M/H=0.3. Those parameters are the closest ones of  the solution found by \cite{2017A&A...598A..19D} using high-resolution spectra. We confirm that no excess can be found up to 100 $\mu$m with our fitting solution. 
		The flux-calibrated model spectrum reproduces equally well (Fig. \ref{fig:SEDstar}) the shape and flux of the STELIB medium-resolution (R$\sim$2000) optical spectrum (320-989 nm) of the star \citep{2003A&A...402..433L} obtained in April  1994. We collected and averaged archival flux-calibrated UV spectra of the star from the "IUE Newly Extracted Spectra" (INES) database\footnote{http://sdc.cab.inta-csic.es/ines/}. The spectra  were collected with the LWR and SWP camera of the International Ultraviolet Explorer \citep{1999A&AS..139..183R, 2000A&AS..141..331C, 2000A&AS..141..343G, 2001A&A...373..730G} and  have a reliable flux in the interval 150-331nm.  We also reduced  data of GJ~504A gathered with the SINFONI NIR integral field spectrograph \citep{2003SPIE.4841.1548E, 2004Msngr.117...17B} on June 9, 2014 (PI CACERES; Program 093.C-0500).  The data were acquired with the H+K mode of the instrument yielding continuous medium-resolution (R$\sim$1500) spectra from 1.45 to 2.45 $\mu$m. The SINFONI data were reduced with the ESO data handling pipeline version 3.0.0  through the \texttt{Reflex} environment. Only one datacube, corresponding to a science exposure obtained at  02h41m04s UT, contained the star in the field of view.  The star spectrum was extracted over a circular aperture with a radius of  325mas. The spectrum was corrected from telluric absorption using the observation of the B9V star HD 141327 observed before GJ~504A. The 1.8-1.95 $\mu$m range was affected by telluric line residuals and was not considered any further. We flux-calibrated the spectrum using the K band flux from \cite{2003AJ....125.3311K}. The H and K-band SINFONI spectrum  is well reproduced by the BT-NEXTGEN model (Fig. \ref{fig:SEDstar}) and can be used to derive reliable IRDIS magnitudes of GJ~504A in the H2H3 and K1K2 channels (see below). 	
		We replaced the  BT-NEXTGEN spectrum with the INES, STELLIB, and SINFONI spectra of GJ~504A and integrated the SED to estimate a log $L/L_{\odot}=0.35\pm0.01$ dex. The error accounts for an uncertainty of 100K on the $ T_{eff}$ of the BT-NEXTGEN model fit and for the uncertainty on the distance \citep[0.08 pc;][]{2007A&A...474..653V}. The value is in good agreement with the one ($\log\:L/L_\odot$~= 0.35~$\pm$~0.05) derived by \cite{Fuhrmann2015} from a V-band bolometric correction. 	
	
	  We used the spectrum  of GJ~504A  considered for the bolometric luminosity estimate and a spectrum of Vega  \citep{1985A&A...151..399M, 1985IAUS..111..225H}  to compute the photometric shifts between the J, H, and K photometry of GJ~504A and the SPHERE DBI filters of IRDIS.  We also re-derived the CH$_{4}$S and CH$_{4}$L photometry of GJ~504A from the published H-band magnitude,  taking into account the SINFONI spectrum of the star. The resulting magnitudes for GJ~504A are reported in Table \ref{Tab:photSPHEREGJ504A}. 
	
\begin{table}
\label{Tab:photSPHEREGJ504A}
\caption{Apparent magnitude of GJ~504A in the IRDIS and HiCIAO CH$_{4}$ filters.}
\centering
\begin{tabular}{ccc}
\hline
Band &Mag & Error \\
\hline \hline
Y2 &	4.32 &   0.03 \\
Y3 &	4.29 &	0.03 \\
J2	 & 4.18  &	0.03 \\
J3	 & 4.07  &	0.03 \\
H2 &	3.87 &	0.03 \\
H3&  3.85  &	0.03 \\
K1&	3.79 &	0.03 \\			
K2&	3.83 &	0.03 \\
CH$_{4}$S	& 3.87	&	0.03 \\
CH$_{4}$L		& 3.86	& 0.03	\\
\hline
\end{tabular}
\label{Tab:photSPHEREGJ504A}
\end{table}

\section{Benchmark late-T objects}
\label{Append:B}
We report in Table \ref{Tab:AppB} the properties of the benchmark late-T companions mentioned in Section \ref{sec:emp_analysis}. 

\begin{sidewaystable*}
\label{Tab:AppB}
\centering
\tiny
\caption{Properties of late-T benchmark companions and peculiar isolated objects}
\begin{tabular}{cccccccccccc}
\hline \hline
Name	&	Sp. type	&	Distance 	&	 Separation$^{b}$ &	Mass		&	[Fe/H] star	&	[Fe/H] comp. &	log g	&	$\mathrm{T_{eff}}$ &	Age	&	log ($\mathrm{L_{bol}}$) & References \\
			&						&	(pc)			&	(a.u.)					&	($\mathrm{M_{Jup}}$) 	&	(dex)	 &  (dex)	&   (dex)	&  (K)	& (Gyr)	& (dex)	& 	\\
\hline
51 Eri b & $\mathrm{T6.5\pm1.5}$ 	& $29.4\pm0.3$	& 13	&	2-10	&	$-0.02\pm0.08$	&	$1.0\pm0.1$	&	$4.26\pm25$	&	$760\pm20$	&	$0.026\pm0.003$	&	$-5.40\pm0.07$ & 15, 16, 17, 18 \\
GJ~758b	&	$\mathrm{T7\pm1}$	&	$15.76\pm0.09$	&	29 	&	30-50	&	$0.18\pm0.05$	&	\dots	&	$4.3\pm0.5$	&	$741\pm25$	&	1-6	&	\dots	&	30, 31, 32 \\
G 204-39B	 &	T6.5 &	$14.1\pm0.4$ & 2685	&	20-35	&	$+0.04\pm0.08$	&	0.	&	4.7-3.9 &	960-1000 & 0.5-1.5	&	$-5.18\pm0.06$ & 1,2, 3, 4, 5, 6 \\
HD 3651B	&	T7.5	&	$11.06\pm0.04$	&	480	&	$66^{+12}_{-21}$	&	+0.09--0.16	&-0.01-0.13 	&	$5.12^{+0.09}_{-0.17}$	&	$726^{+22}_{-21}$ &	3-12	&	$-5.60\pm0.05$	&	7, 8, 9, 10, 11, 12, 13, 14, 20 \\
Gl~570D &	 T7.5	&	$5.84\pm0.03$	&	1525	&	$31^{+27}_{-16}$	&	$-0.05\pm0.17$ &	-0.29 to +0.20 & $4.76^{+0.27}_{-0.28}$	& $714^{+20}_{-23}$ &	1-5	&	$-5.55\pm0.05$	&  14, 19, 20, 21 \\
BD+01\_2920B & T8	& 	 $17.2\pm0.2$	&	2630	&	19-47	&	$-0.38 \pm 0.06$	&	\dots	&$5.0\pm0.3$	&	$680\pm55$	&	2.3-14.4	 & $-5.83 \pm 0.05$	&	24 \\
ROSS~458C	& T8	&	$11.7\pm0.2$	&	1100	&	5-20	&	+0.2-0.3	&	+0.2	& 4.0-4.7	&	$695\pm60$ & 0.15-0.8 & $-5.62\pm0.03$	& 25, 26, 27 \\
Wolf~940B & T8.5	&	$12.5\pm0.8$	&	400	&	24-45 &	 $+0.24\pm0.09$	& 0.0-0.2 & $5.0\pm0.3$	&	$560\pm30$	&	3-10 &  $-6.01\pm0.05$ & 28, 29, 41 \\
Gl~229B	& T7	&		$5.79\pm0.01$	&	45	&	40-55	&	$-0.2\pm0.4$	&	-0.5-0.1	&	4.5-5.5 &	840-1030	&	0.3-3	&	 $-5.21\pm0.04$ & 33, 34, 35, 36, 37 \\
$\xi$ Ursae Majoris C	&	T8.5	&	$8.29\pm0.15$	&	4000	&	14-38	&	$-0.32\pm0.05$	&	\dots	&	5.0	&	500	&	2-8	&	$-6.11\pm0.05$	&	42, 43 \\
Wolf 1130B	&	sdT8	&	$15.8\pm1.0$	 & 3000	&	\dots	&	$-0.64\pm0.17$		&	-0.5	&	5.0-5.5	&	600-900	&	2-15	&		\dots & 45 \\	
HD 4113C & T9 & $41.7\pm0.9$ & 23 & 61-71	&	$0.20\pm0.04$ & \dots & 4.5-5 & 500-600 & $5.0^{+1.3}_{-1.7}$ & \dots & 46 \\
\hline
CFBDSIR2149-0403	& T7.5	&	$54.6\pm5.4$	&	\dots	&	2-40	&	\dots	& 0.0-0.3	&	3.5-5.0	&	700-900		&	$\leq$3 &	$-5.51^{+0.10}_{-0.09}$	&	39, 40 \\ 
WISEPC J231336.41-803701.4	&	T8		&	$\mathrm{9.3\pm0.4}^{a}$	& \dots	&	$7\pm4$	&	\dots	&	\dots	&	$4.0\pm0.3$	&	$600\pm30$	&	$0.3\pm0.4$	&	\dots	&	41 \\
WISEPC J161705.75+180714.0	&	T8	&	$\mathrm{13.1\pm0.6}^{a}$	& \dots	&	$7\pm3$	&	\dots	&	\dots	&		$4.0\pm0.3$	&	$600\pm30$	&	$0.2\pm0.3$	&	\dots &	41 \\	
WISEA J032504.52-504403.0	&	T8	&	$36.0 \pm 2.4$	&	\dots	&	\dots	&	\dots	&	\dots	&	4.0	&	550-600		&	0.08-0.3 	&	\dots & 44 \\	
WISEP J181210.85+272144.3	&	T8.5:	&	$19\pm3^{a}$	 & \dots	&	$13\pm7$	&	\dots	&	\dots	&	$4.3 \pm 0.3$		&	$620\pm30$	&	$0.9\pm1.3$	&	\dots	&	41 \\
\hline
\end{tabular}
\tablefoot{References: [1] - \cite{2004AJ....127.3553K}, [2] - \cite{2009AJ....137....1F}, [3] - \cite{2010AJ....139..176F}, [4] - \cite{2015ApJ...804...64M}, [5] - \cite{2006ApJ...639.1095B}, [6] - \cite{2007A&A...474..653V}, [7] - \cite{2006MNRAS.373L..31M}, [8] - \cite{2007ApJ...654..570L}, [9] - \cite{2007ApJ...658..617B}, [10] - \cite{2007ApJ...660.1507L}, [11] - \cite{2003AJ....126.2048G}, [12] - \cite{2004A&A...415.1153S}, [13] - \cite{2005ApJS..159..141V}, [14] - \cite{2007ApJ...667..537L}, [15] - \cite{2015Sci...350...64M}, [16] - \cite{2017A&A...603A..57S}, [17] - \cite{2017AJ....154...10R}, [18] - \cite{2012A&A...538A.143K}, [19] - \cite{2010ApJ...720.1290G},  [20] - \cite{0004-637X-807-2-183}, [21] - \cite{2014ApJ...796...39T}, [22] - \cite{2006ApJ...637.1067B}, [23] - \cite{2000ApJ...531L..57B}, [24] - \cite{doi:10.1111/j.1365-2966.2012.20549.x} and ref. therein, [25] - \cite{2010MNRAS.405.1140G}, [26] - \cite{2010ApJ...725.1405B}, [27] - \cite{2011MNRAS.414.3590B}, 28 - \cite{2009MNRAS.395.1237B}, 29 - \cite{2010ApJ...720..252L}, 30 - \cite{2009ApJ...707L.123T}, 31 - \cite{2016A&A...587A..55V}, 32 - \cite{2017ApJ...838...64N}, 33 - \cite{2000ApJ...541..374S}, 34 - \cite{2015AJ....150...53N}, 35 - \cite{1997ApJ...484..499S}, 36 - \cite{2002MNRAS.332...78L}, 37 - \cite{1995Natur.378..463N}, 38 - \cite{1996ApJ...465L.123A}, 39 - \cite{2012A&A...548A..26D}, 40 - \cite{2017A&A...602A..82D}, 41 - \cite{2011ApJ...735..116B}, 42 - \cite{2013AJ....145...84W}, 43 - \cite{1994A&A...291..505C}, 44 - \cite{2015ApJ...804...92S}, 45 - \cite{2013ApJ...777...36M}, 46 - Cheetham et al. 2018 \newline $^{a}$ model distance $^{b}$ projected separation, apart for HD 4113C}
\end{sidewaystable*}

\section{Details on the color-magnitude and color-color diagrams}
\label{App:C}
This appendix describes the way the color-magnitude and color-diagrams shown in Section \ref{sec:emp_analysis} are built.

We used spectra of M, L, and T dwarfs from the SpeXPrism library \citep{2014ASInC..11....7B} and from \cite{2000ApJ...535..965L} and \cite{2015ApJ...804...92S} to generate synthetic photometry in the SPHERE filter passbands. The zero points were computed using a flux-calibrated spectrum of Vega  \citep{1985IAUS..111..225H, 1985A&A...151..399M}. We also considered the spectra of young and/or dusty free-floating objects from \cite{2013ApJ...777L..20L}, \cite{2013ApJS..205....6M}, \cite{2015ApJ...799..203G}, and of young companions   \citep{2011ApJ...729..139W, 2015ApJ...804...96G, 2016ApJ...818L..12S, 2014MNRAS.445.3694D, 2015ApJ...802...61L, 2014ApJ...780L...4B, 2017AJ....154...10R, 2014A&A...562A.127B, 2010A&A...517A..76P, 2010ApJ...719..497L}. The colors and absolute fluxes of the benchmark companions  and isolated T-type objects are generated from the distance and spectra of those objects (See Appendix \ref{Tab:AppB} for the details.). To conclude, we used the spectra of Y dwarfs published in \cite{2015ApJ...804...92S},  \cite{2007MNRAS.381.1400W}, \cite{2008A&A...482..961D}, \cite{2008MNRAS.391..320B}, \cite{2010MNRAS.408L..56L}, \cite{2012ApJ...753..156K}, and \cite{2013ApJS..205....6M} to extend the diagrams in the late-T and early Y-dwarf domain. 

	 We used  the distances of the field dwarfs reported in \cite{2000AJ....120..447K}, \cite{2012ApJ...752...56F}, \cite{2013Sci...341.1492D},  \cite{2014ApJ...796...39T}, \cite{2014ApJ...783...68B}, and \cite{2016AJ....152...78L}. We considered those reported in  \cite{2011ApJS..197...19K}, \cite{2012ApJ...752...56F}, \cite{2014A&A...568A...6Z}, and \cite{2016ApJ...833...96L} for the dusty dwarfs.  The companion distances are taken from \cite{2007A&A...474..653V} and \cite{2014A&A...563A.121D}.

\section{Forward models exploring different C/O ratio}
\label{App:D}
The models exploring different C/O ratios treat the gaseous opacity with the k-correlated method  \citep{amundsen:2017aa}. They account for the CIA of H$_{2}$-H$_{2}$/He \citep{2012JQSRT.113.1276R}, H$_{2}$O, CH$_{4}$, CO , CO$_{2}$, NH$_{3}$, H$_{2}$S, PH$_{3}$ \citep[ExoMol and][]{2014ApJS..214...25F}, Na, and K \citep{2003ApJ...583..985B}. The chemistry is computed using the NASA CEA2 routine\footnote{https://www.grc.nasa.gov/www/CEAWeb/} but with "rain-out" condensation implemented. Disequilibrium chemistry of  NH$_{3}$, N$_{2}$, CO, CH$_{4}$, and H$_{2}$O is implemented following the \cite{2014ApJ...797...41Z}  analytic timescale approach.

Grids of synthetic spectra at R=1000 are generated from $300 K \leq \mathrm{T_{eff}} \leq 950 K$ in 50K intervals,  $3.0 \leq \mathrm{log\:g} \leq 5.5$ in 0.5 dex steps,   $-1.0 \leq \mathrm{[M/H]} \leq\mathbf{1.0}$ in 0.5 dex intervals, $\mathbf{-0.2 \leq log(K_{zz} \leq 8)}$ in 0.5 dex steps, and 6 C/O points between 0.1 and 0.85.  They are convolved with the filter passbands corresponding to GJ~504b photometry to generate the synthetic fluxes.

\section{SOPHIE radial-velocity measurements}
\label{App:F}
We report in Table \ref{Tab:RVmsmt} the radial-velocity measurements of GJ~504A used in Section \ref{subsec:addcomp}.

\begin{table}
\centering
\small
\caption{\label{Tab:RVmsmt} SOPHIE radial-velocity measurements.}
\begin{tabular}{ccc}
\hline
MJD - 2 450 000 &	RV (km.s$^{-1}$)	& error (km.s$^{-1}$) \\
\hline
6383.53	&	-0.040	&	0.005\\
6383.53	&	-0.038	&	0.006\\
6385.52	&	0.020	&	0.005\\
6385.53	&	0.016	&	0.006\\
6386.47	&	-0.048	&	0.005\\
6386.47	&	-0.047	&	0.005\\
6388.50	&	0.018	&	0.005\\
6388.50	&	0.017	&	0.005\\
6390.49	&	-0.036	&	0.006\\
6766.53	&	0.075	&	0.006\\
6766.53	&	0.068	&	0.006\\
6767.51	&	-0.061	&	0.006\\
6767.52	&	-0.051	&	0.006\\
7060.60	&	0.015	&	0.005\\
7060.61	&	0.011	&	0.006\\
7061.67	&	0.020	&	0.006\\
7061.67	&	0.018	&	0.006\\
7099.65	&	-0.012	&	0.005\\
7099.65	&	-0.009	&	0.005\\
7100.56	&	0.010	&	0.005\\
7100.56	&	0.010	&	0.005\\
7101.48	&	0.033	&	0.006\\
7101.48	&	0.031	&	0.006\\
7104.57	&	-0.001	&	0.005\\
7104.58	&	0.004	&	0.005\\
7444.64	&	0.013	&	0.006\\
7444.64	&	0.010	&	0.006\\
7447.67	&	-0.019	&	0.007\\
7447.68	&	-0.022	&	0.006\\
7448.67	&	0.002	&	0.007\\
7490.57	&	-0.032	&	0.005\\
7490.57	&	-0.030	&	0.006\\
7491.52	&	-0.025	&	0.006\\
7491.52	&	-0.027	&	0.006\\
7494.47	&	-0.022	&	0.006\\
7494.48	&	-0.021	&	0.006\\
7532.46	&	0.005	&	0.006\\
7532.46	&	0.016	&	0.006\\
\hline
\end{tabular}
\end{table}

\section{Magnitudes of ultracool companions predicted by Exo-REM}
\label{App:G}

\begin{table*}
\centering
\small
\caption{\label{Tab:predcol} Absolute magnitude predictions synthetized from the COND tracks and  the  \texttt{Exo-REM} model atmospheres}
\begin{tabular}{cccccccccccc}
\hline		
age	&	mass	&	models	& $[M/H]$	&	Y2	& Y3	& J2	&	J3	&	H2	&H3	&K1	& K2 \\
(Gyr)	&	$\mathrm{M_{Jup}}$	&	&(dex)	&	(mag)	&(mag)	&(mag)	&(mag)	&(mag)	&(mag)	&(mag)	&(mag)	\\
\hline		
4		&	40		&	Exo-REM/$\mathrm{f_{cloud}=0.75}$ &  0	&		16.96	& 16.69		&16.68		&15.32		&15.48		&16.76	&	15.84	&	17.63 \\
4		&	40		&	Exo-REM/$\mathrm{f_{cloud}=0.75}$ &  +0.5	&		17.03	&	16.68	&	16.74	&	15.18	&	15.38	&	16.72	&	15.18	&	17.44 \\
4		&	40		&	Exo-REM/$\mathrm{f_{cloud}=0}$ &  0	&		16.87	&	16.59	&	16.64	&	15.12	&	15.51	&	16.85	&	15.96	&	17.82 \\
4		&	40		&	Exo-REM/$\mathrm{f_{cloud}=0}$ &  +0.5	&		16.98	&	16.62	&	16.70	&	14.98	&	15.39	&	16.84	&	15.26	&	17.64 \\
4		&	20		&	Exo-REM/$\mathrm{f_{cloud}=0.75}$ &  0	&		20.78 &	20.51	& 20.89	 & 18.76 &	18.65 & 	20.69	& 19.48	& 21.99 \\
4		&	20		&	Exo-REM/$\mathrm{f_{cloud}=0.75}$ &  +0.5	&		20.54 & 20.10	& 20.76	& 18.27	& 18.25	& 21.07	& 18.25	& 21.65 \\
4		&	20		&	Exo-REM/$\mathrm{f_{cloud}=0}$ &  0	&		19.68 &	19.47 &	20.10	& 17.80	& 18.17	& 20.78	& 20.00	& 22.63 \\
4		&	20		&	Exo-REM/$\mathrm{f_{cloud}=0}$ &  +0.5	&		19.60	& 19.17	& 20.13 &	17.39 &	17.78&	21.32&	18.68&	22.30 \\
4		&	15		&	Exo-REM/$\mathrm{f_{cloud}=0.75}$ &  0	&		21.50 &	21.20	&22.11&	19.53&	19.45&	22.02&	20.45&	23.15 \\
4		&	15		&	Exo-REM/$\mathrm{f_{cloud}=0.75}$ &  +0.5	&		21.27&	20.78	&21.93&	19.07&	19.11&	22.47&	19.17&	22.82 \\
4		&	15		&	Exo-REM/$\mathrm{f_{cloud}=0}$ &  0	&		20.33 & 20.10 &	21.31 &	18.51 &	18.83 &	22.07 &	20.96 &	23.77 \\
4		&	15		&	Exo-REM/$\mathrm{f_{cloud}=0}$ &  +0.5	&		20.17&	19.68	&21.15&	18.02&	18.49	&22.60&	19.55&	23.41 \\

\hline		
0.6		&	15		&	Exo-REM/$\mathrm{f_{cloud}=0.75}$ &  0	&		17.50 &	17.20	&17.41&	15.77&	15.91&	17.21&	15.69&	17.55 \\
0.6		&	15		&	Exo-REM/$\mathrm{f_{cloud}=0.75}$ &  +0.5	&		17.85&	17.47&	17.57&	15.86&	15.95&	16.93&	15.15&	17.09 \\
0.6		&	15		&	Exo-REM/$\mathrm{f_{cloud}=0}$ &  0	&		16.97	& 16.68&	17.02	&15.22&	15.58&	17.36&	16.14&	18.40 \\
0.6		&	15		&	Exo-REM/$\mathrm{f_{cloud}=0}$ &  +0.5	&	17.12&	16.74&	17.03&	15.11&	15.51&	17.17&	15.48	&18.03 \\
0.6		&	8		&	Exo-REM/$\mathrm{f_{cloud}=0.75}$ &  0	&		20.43	& 20.08	& 20.91	& 18.47	& 18.43	& 20.94	& 18.68	& 21.32 \\
0.6		&	8		&	Exo-REM/$\mathrm{f_{cloud}=0.75}$ &  +0.5	&		20.34	&	19.86	&	20.84	&	18.17		&18.25	&	21.07	&	17.82	&	21.09 \\
0.6		&	8		&	Exo-REM/$\mathrm{f_{cloud}=0}$ &  0	&		19.26&	18.93&	20.02&	17.32&	17.67&	20.92&	19.21&	22.09 \\
0.6		&	8		&	Exo-REM/$\mathrm{f_{cloud}=0}$ &  +0.5	&	19.28	&18.79	&20.04	&17.06&	17.60&	21.34&	18.20&	21.86 \\
0.6		&	5		&	Exo-REM/$\mathrm{f_{cloud}=0.75}$ &  0	&		22.13	&21.76	&23.32&	20.19	&20.14	&23.46	&21.38	&24.29 \\
0.6		&	5		&	Exo-REM/$\mathrm{f_{cloud}=0.75}$ &  +0.5	&		21.98&	21.37&	23.11	&19.74	&19.92&	23.85&	20.11&	23.84 \\
0.6		&	5		&	Exo-REM/$\mathrm{f_{cloud}=0}$ &  0	&		21.09&	20.73&	22.52&	19.20&	19.35&	23.29&	21.79&	24.74 \\
0.6		&	5		&	Exo-REM/$\mathrm{f_{cloud}=0}$ &  +0.5	&	20.95&	20.32&	22.33&	18.76&	19.29&	23.87&	20.49&	24.45 \\
\hline
0.02		&	5		&	Exo-REM/$\mathrm{f_{cloud}=0.75}$ &  0	& 15.68	&15.53&	15.32&	14.24	&14.08&	14.10&	13.72&	14.18\\
0.02		&	5		&	Exo-REM/$\mathrm{f_{cloud}=0.75}$ &  +0.5	&		15.86&	15.64&	15.53&	14.34&	14.10&	13.85&	13.43	&13.66 \\
0.02		&	5		&	Exo-REM/$\mathrm{f_{cloud}=0}$ &  0	&		15.02&	14.84&	14.82&	13.50&	13.89&	14.26&	14.21&	15.09 \\
0.02		&	5		&	Exo-REM/$\mathrm{f_{cloud}=0}$ &  +0.5	&	15.19&	14.93&	15.04&	13.59&	13.86&	13.73&	13.69&	14.32 \\
0.02		&	3		&	Exo-REM/$\mathrm{f_{cloud}=0.75}$ &  0	&		16.76	&16.51	&16.60&	15.17&	15.26&	15.97&	14.98&	16.06 \\
0.02		&	3		&	Exo-REM/$\mathrm{f_{cloud}=0.75}$ &  +0.5	&		17.14&	16.80	&16.93&	15.41&	15.40&	15.50&	14.59&	15.26 \\
0.02		&	3		&	Exo-REM/$\mathrm{f_{cloud}=0}$ &  0	&		16.14	&15.85	&16.16&	14.48&	14.87&	16.07&	15.31&	17.06 \\
0.02		&	3		&	Exo-REM/$\mathrm{f_{cloud}=0}$ &  +0.5	&	16.35&	15.98&	16.25&	14.51&	14.83&	15.36&	14.78&	16.23 \\
0.02		&	1.5		&	Exo-REM/$\mathrm{f_{cloud}=0.75}$ &  0	&		19.33	&18.96&	19.80&	17.46&	17.63&	19.66&	17.43&	19.58 \\
0.02		&	1.5		&	Exo-REM/$\mathrm{f_{cloud}=0.75}$ &  +0.5	&		19.54	&19.09	&19.79	&17.44	&17.57	&19.38	&16.89&	19.18 \\
0.02		&	1.5		&	Exo-REM/$\mathrm{f_{cloud}=0}$ &  0	&		18.27	&17.89&	19.06&	16.38&	16.87&	19.89&	17.90&	20.61 \\
0.02		&	1.5		&	Exo-REM/$\mathrm{f_{cloud}=0}$ &  +0.5	&	18.50&	18.05&	18.94&	16.30	&16.83&	19.37&	17.21&	19.97 \\
\hline
 \end{tabular}
\end{table*}

\end{appendix}

\end{document}